\documentclass[graybox,envcountchap]{svmono}

\usepackage{type1cm}         
\usepackage{makeidx}         
\usepackage{graphicx}        
\usepackage{multicol}        
\usepackage[bottom]{footmisc}
\usepackage{newtxtext}       %
\usepackage{newtxmath}       

\makeindex             
\graphicspath{{./}{images/}}

\begin{document}

\author{Zoran Levnaji\'c}
\title{Network model selection: A review of methods}
\subtitle{SpringerBriefs in Complexity}
\maketitle

\thispagestyle{empty}
\noindent \textit{Author:} \\
Zoran Levnajić \\ 
Complex systems and Data science Lab \\ 
Faculty of information studies in Novo mesto  \\ 
Ljubljanska cesta 31A, 8000 Novo mesto, Slovenia \\ 
\texttt{zoran.levnajic@fis.unm.si}

\frontmatter
\tableofcontents

\preface

Someone told me that Preface is a place where I articulate the frustration that drove me to write the book. So, here it is: It bothers me that in the science of complex networks all we produce are methods. We don't seem to be interested in applying them, i.e., in actually analyzing complex networks. Almost as if the only goal of network science was to invent as many methods as possible, without asking if there is a need for them, and with no pressure to use them. I find this unbearable. Just imagine astrobiology insisting that its mission is to keep improving methods for detecting life on other planets, but not to apply them. But make no mistake: I am fully complicit in the trend I just called unbearable.

Having reached a certain level of career, I felt it was time to make a concrete contribution to my scientific field. Obviously, this wasn't to be another method. Instead, I decided that the most meaningful way to contribute is to make a review of existing methods that concern one aspect of network science. I eventually settled on methods for network model selection: given a studied (target) network, find the network model (e.g. a formation/growth model) that best describes it, whereby the selection is done from a list of candidate network models. In other words, find the network model which is most likely to be responsible for growth/evolution of the studied network, usually a real empirical one. I find this an interesting problem, for which several nice (but fragmented) methods are already available, with no review paper (or book) to consolidate them. So, this book is a review of methods for network model selection, whose overarching aim is to encourage research towards unifying them into a single best method. Such ultimate method could finally reveal the true formation mechanism underpinning real networks.

I will not dive further into this frustration of mine since that's what most of the Chapter 1 is devoted to. Namely, I felt the need to describe this issue in more detail, not just to explain my motivation for the book, but to make a statement about what do I see as a problem with our scientific field. Of course, I wholeheartedly welcome any disagreement about this. I encourage readers to reach out to me with comments and criticism.

This book is intended for anyone doing research in network science. This scientific field is also called complex networks, network analysis, and social network analysis. My guess is that readers have to be at least graduate (doctoral) students. Knowledge and background required to grasp the book is stated in Chapter 1. The same chapter also includes details about me and my work, literature search behind my review of methods for network model selection, and book's organization. 

Many thanks for my colleague Prof Ljupčo Todorovski for very useful discussions and feedback. He helped me organize these pages into a coherent flow. I am grateful to my faculty and in particular to our librarian Nina Malovrh for assistance with publishing questions. I devote this book to my family, which had to put up with me during all these years. 

I declare that did not use Artificial Intelligence. Literature search is mine, preparing the sample of papers/methods to review is mine, reading these papers and studying methods is mine, writing and polishing the entire text is mine.

As for financial support, this work was partially supported by the Slovenian Research and Innovation Agency (ARIS) through the program P1-0383 entitled ``Kompleksna omrežja'' (Complex Networks).

\vspace{\baselineskip}
\begin{flushright}\noindent
Novo mesto,\hfill {\it Zoran Levnajić}\\
April 2026 \hfill {\it $\;\;\;$}\\
\end{flushright}

%
%
%

\begin{dedication}
To my family
\end{dedication}



\mainmatter
\chapter{Why I wrote this book and how}  \label{chapter-why}

\begin{svgraybox}
\abstract{This book is a review of methods for finding which network model -- among a set candidate models -- best fits a given (target) network. In other words: given some (real, empirical) target network whose structure is known to us, and several candidate network models, the problem is to select which model is most likely to have generated the target network and/or best explains its structure. Before I start, I want first to explain a bit about me and about my motives for writing this book. I felt the urge to do this, because my motives contain a criticism of network science -- the field this book and I belong to. Of course, I will wholeheartedly welcome any disagreement on this from my colleagues. Further in this chapter I describe how I gathered, filtered, summarized, and grouped all these methods that this book is about. The chapter finishes with an outline of the rest of the book and of the knowledge required to understand it.}
\end{svgraybox}

To put everything in context, I will first say a bit about my scientific career and experience. I will also explain how I see network science and why I find it exciting. Then I will move on to explain my motivations for writing this book. Actually, my motivation is two-fold. I will first explain my main (nominal) motivation which is nothing unusual for a network scientist. I will then proceed to explain my other motivations for writing this book. These will contain a critical note about our field, or at least a worrisome trend that I see and feel should be addressed. The final sections in this chapter will be devoted to detailing the methodology I used to put together all these papers on model selection in networks and summarize all these methods. I will finish the chapter with information about how the rest of the book is organized and what is good to know before continuing to read it.


\section{A bit about me and my work}

I started my career in science as an undergraduate student of physics back in 1996. I was always fascinated by it's theoretical aspects, specifically by how we could describe nature through mathematics. I went on to do my Master's degree in dynamical systems, which also relied on math and computing. It was then when I first got in touch with computer science and programming. Except dynamical systems and chaos, I was always fascinated by complex systems, i.e., the systems composed of many interacting units, which jointly display emergent collective behaviors that no unit in isolation can ever display. I then started my PhD on network analysis, which I saw as a way to access complex systems. My thesis had to do with modeling the emergence of collective chaotic dynamics on large networks, including on networks of biological interest (gene regulation). It was based on constructing and examining synthetic network models, without relying on empirical data. I defended my PhD in 2009, and from this point on, network science became the main field of my work (although not the only field, as I will explain below). After getting my PhD, I did three years of postdoc where I touched upon diverse aspects of synchronization in complex networks, specifically, the aspects of modeling dynamics of coupled neuronal cells through networks. This is where I got in touch with the neuroscience part of the field. Eventually, I became assistant professor and then associate professor. Since my institution is (also) social science oriented, this introduced me to social science aspects of network science and complex systems. Moreover, while still collaborating with physicists, I had a chance to work also with medical scientists and psychologists. So, can say I collaborate with a diverse range of scientific fields, but network science always remains my primary interest and passion.

Perhaps now it's a good point to spend a few words on the defining what do I mean by network science. I mean the scientific field that deals with complex networks, which are the mathematical or computational representation of systems composed of many interacting units. The traditional examples are social networks where nodes represent individuals, while links or edges represent their interactions such as love or friendship. Other examples include networks of biological entities such as proteins, transportation networks involving roads or airline routes, neuroscience networks where one studies connectivity among brain regions, trade the networks which capture the business relations between the involved stakeholders, etc. The field originated many decades ago primarily through social sciences. Sociologists were looking at social networks and over time developed methodologies for studying them systematically. They primarily used mathematical graph theory and statistics. Later, with blossom of computer science, the field of social network analysis benefited from various algorithms, many of which were at this point developed specifically for analyzing social networks. Towards the end of 1990s, physicists start joining in through pioneering works by Barabasi, Albert, Watts, and Strogatz. Physicists realized that some of the properties found in statistical physics of critical phenomena -- quite unexpectedly -- are present also in real complex networks found in nature, society, and technology. This triggered the development of many new approaches and study of many novel problems. Last two decades saw further expansion of network science into completely new territories, including neuroscience, psychology, biology, engineering, economics, and many others, including even linguistics and sports. So, what I call network science is the confluence of all these research efforts and directions \cite{Estrada2015,Zweig2016,Yang2016,Latora2017,Newman2018,Menczer2020,Coscia2021,Dorogovtsev2022,Izenman2023}. Common to them is that one studies (complex) networks of varying origin and in various contexts, relying on analysis methods from mathematics/graph theory/statistics, computer science, and (statistical) physics. This endeavor is so interdisciplinary that is has been accused of being fragmented \cite{Hidalgo2016}. 

When I say that I consider myself a network scientist, what I mean is that over the last 20 years I made a few contributions to this field. They start with the papers I published during (and after) my PhD on modeling collective phenomena on networks, especially in the context of biological networks \cite{Levnajic2008,Levnajic2010a}. During my postdoc time I worked on various aspects of synchronization of phase (and other) oscillators on complex networks \cite{levnajic2010b,levnajic2011b,faggian2019}. A particular topic I was very interested in during and after my postdoc was networks reconstruction (or inference): how to discover the structure of a (hidden) network from the data describing the dynamics exhibited by its nodes (time series). Examples include, at least in principle, experimentally measured/observed time series of neuron or gene activity, i.e., time resolved data on neuronal spiking or gene activation/repression. I was involved in designing several new methods along those lines in various settings \cite{tokuda2019, grau2019a, grau2019b, simidjievski2018, grau2017, levnajic2014, levnajic2013, levnajic2011a}. I was also interested in ideas related to designing networks with pre-specified functions \cite{levnajic2012}. More recently I worked on an interesting new problem: how to use the collective dynamics of coupled phase oscillators to check if a given network can be colored with a certain number of colors (in the context of graph coloring problem) \cite{crnkic2020}. 

In parallel with these theoretical efforts, I also researched more applied aspects of network science. One of them is network neuroscience, where I examined the (dis)similarities between brain functional networks \cite{tadic2016}. Another one revolves around scientometrics and bibliometrics \cite{luzar2014, subelj2015}. Naturally, this entire time I was involved with computer science too. I collaborated on a paper where we proposed a new way to compare networks and measure their similarity \cite{Yaveroglu2014}. Based on it we developed a method for network model selection, applied it to networks of international trade, and found what models underlie them. This paper will be presented in detail in later chapters: It is one of the methods of model selection for complex networks that I will review.

However, my career did not involve only network science. Actually, even before starting my PhD I worked with chaotic dynamical systems and ergodic theory \cite{levnajic2010d, levnajic2015}. My research in this direction even involved quantum chaotic dynamics at one point \cite{levnajic2010c}. Later, as I began mentoring PhD students, I found myself collaborating with a wide range of disciplines and a variety of people. Stemming from my involvement in social network analysis, I had a chance to collaborate with social scientists and psychologists, but on topics not related to network science. They include modeling of crowdsourcing and collective problem solving \cite{Guazzini2015}, dynamics of Wikipedia growth \cite{Ban2017}, psychology of cooperation among human players \cite{Guazzini2019}, and social perception of “Us” vs. “Them” in face of terrorist attacks \cite{Jovic2023}. Along the same lines I worked on problems related to economic and management \cite{Damij2015, Joksimovic2023}, including the issue of adjusting information systems to pandemic conditions \cite{Damij2023}. My works farthest away from network science includes collaborating with neuroscience (studying the dynamics of vagal tone) \cite{Grote2019} and medical sciences/physiology (how to tell if a person is awake or asleep from data on heart activity) \cite{Zorko2020}. At the time of writing this text, I am actively working with not just medical sciences, neuroscience and psychology, but as of recently, also materials science.

\begin{svgraybox}
Collaborating outside network/computer science and theoretical physics taught me a few important lessons. Besides those related to the difference in scientific culture between, for example, theoretical physics and experimental neuroscience, I learned some other lessons. Actually, it is better to say that these other lessons refreshed my knowledge about how science is supposed to work. One such lesson is the distinction between a new method and a new discovery, and the scientific relevance of each. Another such lesson is the difference between an insight obtained from a computational simulation vs. an insight obtained via real experiment. Spending your time in theoretical physics or in computer science won't teach you this, yet clarity on this is crucial for grasping your role in overall scientific endeavor. In what follows I will discuss this further and show how it has to do with the current situation in network science.
\end{svgraybox}


\section{An important habit of mine}

Before proceeding, I must describe a work habit of mine that is very important for this book. I am devoting a considerable amount of time -- on a daily basis -- to carefully following the literature in network science. This means that I check newly published papers in (almost) every issue of all journals relevant to this field: Journal of Complex Networks, Applied Network Science, EPJ Data Science, Physical Review E, Journal of Physics: Complexity, Social Networks, CHAOS, Network Neuroscience, and a few others occasionally. In addition to this, I check (almost) every day new preprints on arXiv. I also follow major conferences in network science, at least by checking the list of published contributions. Naturally, I don't exclusively follow network science. Most often I also look at new issues of Nature, Science, PNAS, Science Advances, Scientific Reports, Royal Society Open Science, Nature Communications, PLoS ONE, Physical Review X, Physical Review Letters, Nature Human Behaviour, and occasionally some other. In them I search not just for papers on networks, but check out interesting new research in physics, astrophysics, social sciences, cognitive science, psychology, neuroscience or even humanities. As a hobby, I sometimes find myself reading about cancer science, paleoanthropology, geophysics, or even cosmology. 

Each time I come across an interesting paper I save the pdf. As I've been doing this for over 10 years, my collection of papers has grown to over 2,300 titles. Probably about $3/4$ of these papers are related to network science or some of the pertaining keywords (network analysis, social network analysis, complex networks, network modeling/simulation, etc) or are associated with the term data science, which for me means it deals with some data that can be approached and analyzed in similar ways. This is not to say that I carefully read each of these papers. But when I'm downloading a paper, I at least take a quick look at its basic ideas, methods and results. And this hold true for all papers in my collection, be it on network science or otherwise. I revisit some papers later and read them more carefully, others I check from time to time depending on what I am working on at the moment. Some papers I just never come back to. I use Mendeley Reference Manager to have a good manage my collection and have an instant overview should I need it. I warmly recommend this habit to anyone working in science. It pays off well when you must, for instance, put together the literature for your next paper, or for a book like this one. 

Browsing through my collection of papers gave me the inspiration for this book, which I will describe in the next section. Having a bird's eye perspective of network science for ten years revealed the field's trends and rhythms, but also deviations and sidetracks. I will speak more about this towards the end of the book, but for now let me just say that I found network science a very fragmented field. There are many interesting results, but they pop up from different places and seldom build on each other. In fact, I find that there is much more need for consolidation of the existing findings than for creating new ones.

Furthermore, since my collection includes papers from other scientific fields, I had a chance to compare what’s happening in network science with the trends in other fields (albeit not perfectly, since I don't have nearly as many papers in other fields, and in some fields I have none). Nevertheless, this comparison taught me several important lessons, which I will touch upon later. Before closing this section let me also mention another detail. I used to volunteer for a few years for the Slovenian media portal \textsc{metinalista.si}. Every month or so I would pick the most relevant recent papers (about 5 of them) published by the Slovenian scientists. I would then prepare a column presenting them to the general public. Naturally, this encompasses the entire science that Slovenian researchers were at the time engaged in (covering all fields of science). My columns can be found here \cite{metinalista}. 

This was a very revealing experience: Trying to present a scientific result to a wider audience (using layman's terms) makes you see it in a much more realistic light. Moreover, I understood much better what kind of scientific results general public (read: taxpayers) understand, are able to digest, and are interested in paying for. It is only against the backdrop of this reality that one can sharply see the place of networks in overall scientific enterprise.


\section{My main motivation for writing this book}

Let me explain now why I decided to write a review about current state of methods for model selection in complex networks. Recall that the problem is to choose which network model -- among a set offered candidates -- best explains a given network. This network I will call target, assuming that it is some real network obtained via some kind of measurement or observation. A good example can be a social network coming from storing the data on proximity/contact among a group of students.

Network models were always among my favorite topics in the field. I am fascinated by how we can transform simple intuitive ideas about how networks came to be into precise growth models that generate networks very similar to the ones we observe in nature, society, and technology. The reason I like this paradigm probably hides in my physics upbringing: Use well-understood first principles to create equations that describe the evolution of the system, but in a way that allows for validation via experiments and/or observations. In my mind, that is what creating network models is supposed to do for the case of network science. Another topic I used to be interested in is network reconstruction or network inference: given some observations of the dynamics or behavior of a given network, find its structure, i.e. determine which pairs of nodes are connected and which are not. As explained above, I contributed several new methods for this purpose. But note that network reconstruction can be seen as an inverse problem. In contrast to analyzing the structure and/or dynamics of a known network, here we work in the opposite direction: study the available data about an unknown network to reveal its structure. 

My passion for model selection comes from combination of the above two interests. In fact, I would argue that model selection is nothing but the inverse of the usual problem of creating network models. Instead of examining what network instances are generated via some network model, we here ask the opposite question: \textit{given a network, what model is most likely to have generated it}? For me, this is a very profound problem, since it has the obvious potential: finding what model(s) are behind the real networks, in principle all of them. Reading \textit{any} book on complex networks, reader will quickly learn that this field is about finding `common universal mechanisms' that underpin the real networks. Well, the best way to find these mechanisms -- at least in my mind -- is to find the generative/growth models that underpin the real complex systems. And finding the best method(s) for this purpose is the first step. This is one of those open issues in network science that I was always fascinated by, but never got a real chance to contribute to (or very little of it). 

Coming back to my curated collection of papers on networks, I found that there are many interesting open problems that many researchers touch upon, but never really finish. What I mean by finishing is coming to some definite and well-agreed-upon solution, so that the problem becomes less and less interesting, researchers eventually abandon it, and move to new problems. Examples and many. Let's take community detection (mathematicians would call it graph partitioning). It's about constructing methods to find communities (densely connected subnetworks) in a studied network. There are probably hundreds of published papers with hundreds of methods, even with books and review papers, but without a good agreement on which method(s) work best and should be used to actually find communities in networks. Meanwhile, new papers with new methods are pouring on weekly basis. It's just that most of these new methods for community detection -- as far as I can tell -- go overlooked by people who are supposed to use them. I will come back to this point later in the book, but for now let me say that model selection is exactly one of such open problems. Quite a few results have accumulated over the last decade, but they are scattered and fragmented, without any common thread, let alone an agreement about what model selection method works best. I believe that at this point in time we we should consolidate the existing methods and not rush to design new ones. That's where this book comes in. 

\begin{svgraybox}
So, my aim with this book is to review the state-of-the-art in methods for model selection in complex networks. I will identify the key existing methods for this purpose, summarize their main ideas, arrange them into groups according to their main principles, and discuss their usefulness in practice. This also includes reviewing which software and/or programming codes are available for immediate use. I hope that this will provide a bird-eye perspective on this topic to network science researchers, which should enable them to do the next two steps. First, implement all these methods (or groups of similar methods) and study which of them are the most efficient and precise in selecting the best models. Second, use this winner method(s) to actually find what models underpin real networks. Achieving this second goal for me means a major milestone in network science. 
\end{svgraybox}


\section{Another (much deeper) motivation for writing this book}

I think everyone will agree that science progresses along two main avenues: making \textit{discoveries} about something and making \textit{methods} that are used to make discoveries. Discoveries (or insights) deepen our understanding about some phenomena in nature or society, including its potential applications. Methods are the tools needed to make all this happen, since any result in science is always obtained by applying certain approaches and methods. Even if operating of some methods contains an insight into how the studied system works, the prime purpose of methods is to be applied to make discoveries. For example, establishing that there is (or that there was) life on Mars amounts to making a new discovery. In contrast, creating or improving scientific tools that we can use to establish presence/absence of life on Mars amounts to working on methods. Naturally, both discoveries and methods are essential for progress of science, but we will probably all agree that there should be a balance between how much effort (and funding) we invest in one compared to the other. We should keep in mind that the foremost (albeit not the only) role of methods is to be the tool for discoveries. Methods are necessary (and often very interesting in themselves), but they are not the main goal of science. 

Following network science literature I become convinced of something. We -- the network scientists -- are producing and publishing much more methods than discoveries. It seems that we prefer developing methods much more than using them. The ratio between the number of papers proposing new methods and papers presenting results of new analyses is tilted towards the former, not by a little. What I call discovery in this context are the results of the analysis of some network, obtained e.g. from some empirical data. But such papers have become an oddity. Nowadays, a vast majority of papers is proposing new algorithms for community detection, new centrality indices, new ways to count subgraphs, etc. A na\"ive outsider would conclude that network science desperately needs new methods.

Anyone can easily convince himself or herself of this trend. It is enough to look at the table of contents of Journal of Complex Networks, Physical Review E (section on networks), or even interdisciplinary Scientific Reports (papers on complex networks). The trend is present also in journal Applied Network Science, despite the adjective `applied’ in the title. The trend is not very different in other journals that cover network science, including high-profile venues such as Science Advances or PNAS. I invite the reader of this book to follow the papers published in any of these journals for a few months. Read them and strip their message to the bone. You will realize that, by and large, they propose new ways and approaches to analyze networks and only seldom actually analyze them. When a paper includes some results of network analysis, these results serve only to convince you that the proposed method works properly.

This trend is present across all venues of network science, including most conferences too. Interestingly, I found that it is least pronounced in the oldest network science community: That of social scientist analyzing social networks, gathered around venues such as the journal Social Networks. From my following, they struck a better balance between new methods and new discoveries. The same is true (to a similar degree) for the journal Network Neuroscience, which covers studies such as network analysis of functional brain connectomes. And, (not) surprisingly, I also found that most of the papers that present the analyses of new networks actually rely on the existing, not to say old, methods. They rarely -- if at all -- employ newly proposed methods. Exceptions are the cases where authors use the methods that they developed themselves. In fact, I wonder what fraction of newly proposed methods (e.g. in the last 10 years) is actually even applied to anything. Calculating this could be a very interesting research idea. 

Why this trend? I don’t know, but I could make a few guesses. These guesses would be educated, since I am fully guilty of following this trend. Most of my papers on network science are about new methods. In this regard I am no different from an average network scientist. However, even if educated, my guesses would still be speculative, so let me leave this discussion for another time.

But why is this trend a problem? Why can't we go on inventing better and better methods? To me the answer is clear. Any book on network science will tell you that network science is about \textit{studying networks}. For me, that means new methods are useless unless they are applied to \textit{study some networks}, no matter how intuitive, precise, or computationally efficient they are. Yes, it is also true that a method can be more than just a method: It sometimes contains an intuition about how the underlying systems works. Novel methods mean novel ways of thinking about scientific problems, which can inspire new solutions or new ways to frame the old problems. But coming back to my example of searching for life on Mars: What good is a new method for searching for life on Mars unless we actually apply it?

I argue that what network science needs is not new methods. We have more than enough of them. What our field needs to flourish are the new insights and discoveries about real empirical (complex) networks. In my humble opinion that is what science (and humanity) wants of us. No one needs endless list of unrelated methods developed independently of one another without any central guide for actual application. And that is what we have now. This is true not just for methods of selecting suitable network models, but in virtually any methodological aspect in this field. I claim that much more useful now is to consolidate the existing methods, i.e., make them more applicable to actual real cases of interest. 

\begin{svgraybox}
Hence my second (much deeper) motivation for writing this book: To consolidate and explain in a comprehensive way this group of existing methods, which are devoted to the problem of network model selection. This should allow us to extract the \textit{true scientific value} from all these methods, which will hopefully lead to their application to real networks and new discoveries about them. I find this to be a better usage of my time than looking for a new method, since we have \textit{enough} methods, not just for network model selection. In fact, it could be that consolidating the existing methods, rather than designing new ones from scratch, is the best way of finding the most useful methods.
\end{svgraybox}


\section{How I sampled the papers that are reviewed in this book?} \label{sampling}

As already mentioned, this book is a review of methods that I found in the literature about selecting the most suitable network model to describe a given target network. Specifically, it is about those methods where the selection is done between several candidate network models. Actually, there is another group of methods, those which fit a single prescribed model to the target network (e.g. stochastic block models), which I won't be reviewing here. My review is actually a collection of reviews of individual papers, i.e., methods, and it's presented in Chapter 3. For each of the reviewed papers I decided that it does contain a method of interest for this book, in opposition to papers for which I decided that they don't. Naturally, this obligates me to explain how I made my inclusion and exclusion decisions, i.e., to explain the filtering process that led me to the sample of papers reviewed in Chapter 3. This what I will do in this section. 

The time of COVID pandemic a few years ago gave me an excellent opportunity to reflect upon my work. This also included strolling through my Mendeley collection of papers and thinking about what papers I am revisiting the most. I soon realized that these are the papers that have to do with network models and network modeling in general, for example, studies of mechanisms of network growth or ways of recognizing growth patterns from the data. As I selected these papers out from the rest of my collection, I realized that there were several hundreds of them: They were a broad and heterogeneous sample of papers that all had to do with network models in one way or another. Last few decades indeed saw many new network models being proposed (from various intuitions) and I am certain that this development demands a review of its own. However, I quickly gave up on it (for now) because it was just too huge.

What I did next was to exclude the papers that were only proposing new models as such, without proposing methods to infer them or in some way recognize them in the data. This exclusion step was not exactly simple, since many papers do both, or seek to do both. Still, I went through the papers in more detail and made my inclusion vs. exclusion decisions manually. Eventually, I was left only with the papers that spoke about inferring network models from the data, or (at least) recognizing certain structural/growth patterns from the data.

By then I had already started looking for papers that contained methods of model selection with multiple candidates. This involved excluding the papers which were proposing methods of fitting data to a single model, i.e., finding what choices of parameter values lead to the best fit of some target network to some network model (stochastic block models are an excellent example with a huge volume of literature behind). I will review these methods very briefly in the next Chapter, but they are not the main topic of this book. Again, I wish to stress that they definitely deserve a review. These exclusion choices were somewhat easier since a paper either considers multiple candidate models or just a single one. This borderline was still not perfectly clear, but as I went manually through all the papers again, I eventually made my decisions.

At this point I was left with less than 30 papers in my sample. I read each of them in every possible detail. After few careful considerations I decided to exclude several more papers until I got around 20 papers. This was no small task since many great papers include several results to present, and a new method of model selection is only one of them, sometimes even a side results. 

Along the way I detected another group of similar methods that didn't fit in any of the above categories. In them, the network model for a given target is not selected, but constructed directly piece by piece. This means there are no candidate models to begin with, but one creates the model from scratch. I will review these papers too at the end, but very briefly since they are also not the main focus of this book. Still, being so interesting, I think they merit a proper review of their own (which I may do one day). Nevertheless, for the purposes of this review I excluded these papers.

\begin{svgraybox}
After going through all the exclusion steps described above, I was left with the final sample of 17 papers. This is very coherent sample, because all papers in it deal with the same problem: how to find which network model -- from several candidates -- best explains a given target network. Moreover, I found that each method in the sample is independent from other methods, i.e., each method offers its own approach to selecting the best model. Hence, one could argue that the sample contains \textit{all} existing methods for network model selection, at least at the time of writing this book. Actually, upon compiling the sample, I made various attempts to find additional methods of the same kind in the literature, but I was not successful.
\end{svgraybox}


\section{Organization of the book}

All 17 existing methods for network model selection, covered in 17 papers from my final sample, are summarized and discussed in Chapters 3 -- 6. These four Chapters contain the core material and the main scientific contribution of this book. I found that these 17 methods can be arranged into four groups depending on what basic principle they employ to look for the best network model. Accordingly, I decided to organize these 17 reviews in four separate Chapters. Chapter 3 is devoted to six methods that rely on a chosen set of topological (structural) measures to encode networks and then develop model selection based on such encoding. Similarly, Chapter 4 includes five methods that encode network based on counting the presence of graphlets (small subgraphs) and looking at their statistics. Four methods from Chapter 5 utilize spectral properties of networks' matrices (e.g. adjacency matrix) and various information theory tools. Chapter 6 has only two methods, which are somewhat different from the previous 15 but worth discussing nevertheless. In the rest of this section I explain the content and the role of other Chapters.

This Chapter is, as the reader has noticed, composed of various introductory remarks. I felt it was important to explain my reasons for writing this book. But given that, I also felt I need to put things in context and explain a bit about myself and my career. Naturally, I also had to mention my habit of collecting interesting papers and putting them in my Mendeley collection. That said, I devoted the rest of this Chapter to explaining how I made my sample of papers for review, so that I wouldn't burden Chapter 3 with this. Chapter 1 finishes with the next section where I list the knowledge that is required for a reader to understand the remainder of this book. 

The next Chapter -- Chapter 2 – is where I give some introduction to general topics in network science that are fundamental for understanding the methods presented in Chapters 3 -- 6. First and foremost, I define what do I mean by network model and why we need them in the first place. I also provide a (non-exhaustive) list of network models currently in the market. In the same context I discuss two distinct families of network models, since as it turns out, each of them has a different logic of what does it mean to `model’ a network. With this out of the way, I define precisely the problem treated in this book. Chapter 2 finishes with sections on two important disambiguations. The first is a rough overview of methods that do model selection using a single candidate model. This is important, since it highlights that my main topic is model selection with \textit{multiple} candidate models. Secondly, I felt it necessary to include a section where I seek to prevent the misuse of term `model selection’. Namely, this term comes from statistics and machine learning, where it is used for a somewhat different purpose, which calls for clarification. My intention is that upon covering Chapter 2, the reader should be well prepared to dive into Chapters 3 -- 6.

The book ends with Chapter 7, in which I discuss all 17 methods jointly. This involves discussing common and opposite aspects of diverse methods, confronting them with one another. I also discuss the question of interpreting the results of network model selection and mention alternative routes towards this goal. Chapter 7 (and hence the book) ends with a few concluding remarks.


\section{What is required to understand the remaining chapters} 

I will do my best to make the scientific content of this book as accessible as possible to everyone. Nevertheless, I will not be able to shield the reader from all the technicalities, nor will I be able to explain those technicalities in parallel with the main content. So, I close this Chapter by outlining the technical matters that the reader is expected to be familiar with. There are three kinds of topics where at least some basic familiarity is required. 
\begin{itemize}
   \item This book belongs to the field of computer science/mathematics/statistics. These are the disciplines from which modern network science stems. Therefore, I am expecting the reader to have at least an undergraduate-level understanding of them. In other words, I need the reader to have a basic familiarity with mathematical/statistical and algorithmic thinking. It would be great for the reader to be also familiar with discrete mathematics, more specifically graph theory -- the mathematical underpinning of any and all network science.
   \item Second, this is a book in network science. Recall that what we call networks are graph representations of complex systems found in nature, society, and technology. We make graphs into networks by defining what real-work entities are represented by nodes (vertices) and edges (links). So, a social network is composed of human individuals and the connections among them, whereby we capture individuals as the nodes of a graph and their connections (e.g. friendship) as the edges of that graph (or network). Same logic applies when using graphs to model systems in biology, physics, neuroscience, etc. Along these lines the reader should be familiar with basic concepts in network analysis. They include node degree, node clustering, shortest path, the ideas of directed and non-directed networks, weighted and non-weighted networks, the concept of sub-graphs (sub-networks), adjacency matrix, network Laplacian, etc. These include networks metrics such as various centrality measures, distribution of various quantities (e.g. degree), along with the standard methods (algorithms) for computing them. All this is covered in a huge volume of wonderful literature \cite{Estrada2015,Zweig2016,Yang2016,Latora2017,Newman2018,Menczer2020,Coscia2021,Dorogovtsev2022,Izenman2023}, so I will not be spending more time on it here. In the next section I will discuss network models, assuming the familiarity with the topics necessary for understanding them.  
   \item The third is Machine Learning (ML). Actually, most of the methods I review work via some kind of ML, or at least employ some of ML approaches. Therefore, it's advisable for the reader to be acquainted with the fundamentals of it. This is specially important for readers that don't have computer science training (e.g. physicists, such as myself). You must be familiar with the concepts of supervised and unsupervised learning, classification and clustering, training and testing. So for example, the idea of `training a classifier' must ring a bell. I suggest to start here with this clean and simple webpage \cite{zubarev2024}, from which I learned a bunch. Once done, you may want to move to more grown-up textbooks like these \cite{shalev2014,rogers2016,james2023}, and eventually reach books like \cite{hastie2009}. Pay special attention to the idea of structural features -- they often go under names network metrics and network (graph) indices. In short, these are values you can compute for any network, such as the number of nodes or links, average degree or shortest path, network diameter, etc. Computing many features for the studied network converts its structure into a vector of features. This conversion is called embedding and is used to succinctly describe the network and feed as input to e.g. classification algorithms. This is one -- but not the only -- ML framework that a reader must be familiar with.
\end{itemize}
Familiarity with the ground stated above should guarantee that the reader will grasp the chapters to follow. I will also do my best to back-up all claims with adequate references. Finally, note that the main material of this book are the reviews, meaning that an interested reader can always check the original papers themselves for any additional clarification. 

\chapter{The problem of network model selection} \label{statement} 

\begin{svgraybox}
\abstract{This Chapter covers all the basics necessary for accessing the next four chapters, in which I review the methods for finding which network model among multiple candidates best fits a target network. After explaining the general scientific idea of \textit{a model}, I define what network models are, trace where they came from, how we build them, and to what purpose. Along the same lines I describe a few popular network models and explain their classification in two distinct families: statistical models and mechanistic models. This leads me to precisely formulate the problem treated in this book. Last two sections are devoted to two important clarifications, related to how I think of model selection in network science. The first addresses the difference between model selection and model fitting. The second is the disambiguation of what do I mean by `network model selection'.}
\end{svgraybox}

Our understanding of reality -- be it natural or social -- often relies on models. A \textit{model} is a mathematical or algorithmic representation of a system, process, or a phenomenon, observed in the real world. Building a model amounts to identifying the most relevant features of the studied empirical system and then developing a mathematical construct that captures (mimics) the same features. Modeling is an effort to replicate the intuitions about the studied piece of real world via equations or programming codes \cite{wiki-modelling}. Having a model allows us to understand and visualize the studied natural or social system. Using computers and numerical simulations, we can `play' with a model, i.e., simulate its behavior and evolution in order to understand the functioning and predict the future of the original real-world system. This means leveraging power of (super)computing to confirm the initial intuition or gain new intuition about the studied process or phenomenon. In contrast to the two traditional methodological frameworks in science -- \textit{in vivo} and \textit{in vitro} -- computer models are sometimes said to belong to \textit{in silico} methodological framework. 

For clarity, note that there are two distinct notions of `model' in circulation: model of a (natural or social) phenomenon explained above stems from physics culture (i.e., from the tradition in natural sciences). Another notion is that of \textit{model of data}, commonly used in statistics and machine learning. This refers to looking for patters and regularities in a studied dataset. In this context, a researcher may hypothesize that this or that model best explains the dataset at hand. Two notions are equivalent (or at least similar) in cases where the dataset comes from experimental measurements of some system or phenomenon \cite{stanford-encyclopedia}.

Almost all scientific fields rely on models, especially in natural and computer sciences. Examples are very diverse: Physicists use them to clarify logic behind anything and everything from elementary particles to collisions of galaxies. Chemists use them to capture the dynamics of a reaction. Neurosciences relies on them to study the connectivity among brain regions in various situations. Most of climate science is simulating complicated models to estimate the raise of average temperatures and long-terms changes of climate. Epidemiologist -- such as during COVID-19 crises -- employ models to extrapolate the spread of infection and advise authorities on quarantine measures. Social sciences need models to grasp the complexity of social interactions. There are many more examples \cite{Page2018}, but I can not go through all of them here. 

Of course, in order to build useful models, scientists must rely on simplified versions of reality. A complete and realistic representation of a process in nature (let alone society) is almost never possible. But it makes very good sense to discuss which is the best model for capturing a given real-world process -- for instance, which climate model is better for seasonal forecasting. Of course, all simplified reflections of reality are just (rough) approximations, but those approximations can be extremely useful. Actually, I am a fan of the aphorism attributed to George Box: ``All models are wrong, but some models are useful'' \cite{wiki-wrong}. The aphorism acknowledges that models \textit{always} fall short of including all complexities of the reality, but they are scientifically useful nonetheless. Models provide valuable insights, but only when applied judiciously, i.e., only for as long as they keep in touch with empirical observations. I use to work with Kuramoto model \cite{levnajic2011b,levnajic2012,crnkic2020} -- a simplified but useful model for capturing all kinds of synchronization phenomena, not just on networks \cite{acebron2005,rodrigues2016}.

There are two distinct concepts of how modeling is used in science. \textit{Forward modeling} is when a scientist builds a model as described above, runs it on a computer, and compares the output data with the empirical data. Scientist hopes that patterns found in the synthetic (model-generated) data will be the same (or at least similar) to the patters found in the empirical data. The opposite concept is when a scientist starts from the empirical data and looks for a model that best explains (fits) it. This model is to be selected from diverse candidates or even built from scratch relying on insights about the empirical data. Scientist hopes that at least one of the candidate models will explain the data well. This is called \textit{inverse modeling} and is of interest in this book. This tradition has strong roots in statistics and data mining.


\section{Network models and their purpose}

Similarly to many scientific disciplines, network science also relies on models. It builds them in ways not unlike those described above: From intuitions on why real-world networks look as they do and how they came to be as they are. Historically, development and study of network models follows the insights from systematic observations made about real networks (starting in the early 1990s). Basically all of them showed that there are certain properties -- like heterogeneous degree distribution, very short average paths, unexpectedly large number of triangles, etc. -- , which are universally found in virtually all real-world networks \cite{Estrada2015,Zweig2016,Yang2016,Latora2017,Newman2018,Menczer2020,Coscia2021,Dorogovtsev2022,Izenman2023}. In fact, this universality was so striking that (almost) identical characteristics were revealed in empirical networks ranging from protein-protein, neuron, and gene interactions, via air-traffic and infrastructure networks, to social, linguistic, and trade networks. Naturally, these discoveries beg the obvious question: Where do these common features come from? Do they arise by chance or are there some shared principles and mechanisms on which all real-world networks are built? 

These considerations triggered a rapid development of network models of all kinds. Their primary aim was to generate synthetic networks with properties mimicking those found in the real world. In other words, designing network models meant guessing the mechanisms behind the real networks, formalizing them as iterative processes, and hoping that thus designed artificial (synthetic) networks will exhibit the same (or at least similar) properties to what observed in the real world. Let me define a network model as follows. 
\begin{svgraybox}
Network model is a set of recipes, rules, or instructions that allow us to generate synthetic (artificial) networks with pre-specified structural features and patterns. These recipes involve intuitions, or even hypotheses, about how empirically observed features emerge in real-world networks. For example, such intuitions can be about the probability of a link forming between a given pair of nodes, or about which of the existing nodes will the new incoming node become linked to. By abstracting away certain details, network models focus on the key mechanisms like randomness, attachment, or proximity. They are based on simplified assumptions about how nodes and links assemble into networks. The variety of models enable researchers to study networks in a controlled setting in which parameters can be tuned as desired. This allows us to identify universal patterns, explain emergent phenomena, simulate collective dynamics, test hypotheses, and make predictions about network behavior under various circumstances. While no single model can capture the entire complexity of real-world networks, each model emphasizes its own aspect of their structure, formation, and dynamics. 
\end{svgraybox}

Once we have a network model, we can use it to generate synthetic networks. Synthetic networks are artificially made by iterating mathematical or algorithmic procedure (model) for a certain desired number of iterations. This procedure generates a network instance according to the prescribed recipe, albeit in a probabilistic fashion. As opposed to real-world networks, synthetic networks can be made `clean’ from noise, missing or spurious nodes and links, and other imperfections from the real world. This admittedly makes them unrealistic, but: (\textit{i}) we can generate as many of them as we wish, (\textit{ii}) we can control all the model parameters as we please, including network size, and (\textit{iii}) we can always improve the model, i.e., add new rules and intuitions to make the generated networks closer approximation of the real-world ones. So, synthetic networks are `fake', but there are several excellent reasons why it is interesting to study them nevertheless \cite{Estrada2015,Zweig2016,Yang2016,Latora2017,Newman2018,Menczer2020,Coscia2021,Dorogovtsev2022,Izenman2023}. Let me list some of these reasons:
\begin{itemize}
    \item  To understand the emergent network properties: Reveal if and how large-scale characteristics of real-world networks stem from simple connection rules, i.e., from the best (most suitable) model. This includes identifying universality of such large-scale characteristics across diverse networks;
    \item  To simulate the network evolution and dynamics: Run any dynamical process on a synthetic network and see how it evolves or run a simulation in which the network's structure itself is changing. This process should be, at least in principle, similar to what observed in the real-world networks. This also has to do with being able predicting the behavior of e.g. disease spreading or information diffusion in the real world;
    \item  To test the statistical significance of an empirical fact observed in some real network: Having an ensemble of networks with a certain property in the controlled conditions allows us to calculate the probability that the empirical fact occurred by chance, as opposed through same peculiar mechanism;
    \item  To study and improved the network model itself: examine how networks generated by a model compare to the real-world networks, and not just through the property that the model was supposed to capture. The closer synthetic networks are to their realistic counterparts, the better is the studied model. Gradually improving a model by consistently comparing the synthetic with the real networks is a way to uncover the true mechanisms behind them.
\end{itemize}
The breadth and depth of network models currently circulating in the literature deserves a review of its own. If you ask me, this rapid development of new network models is likely the key driver behind the swift advancement of network science over the past decades. Some network models are general in the sense that they seek to capture fundamental mechanisms underlying all real networks. Other network models are tailored to illustrate certain specific mechanisms found only in specific real networks (for instance, gene regulation network of a certain biological organism). Naturally, some network models prove to be more successful than others in explaining the empirical characteristics of real-world networks. However, to the best of my knowledge, no network model has been entirely successful in explaining all the features of real networks. The search continues, and the researchers constantly introduce newer models, which better and better approximate specific aspects of real networks.

In the next two sections I present some among the most prominent network models. This is essential for readers to grasp Chapters 3 -- 6, which constitute the core of this book. I have divided this overview in two sections because there are two fundamentally distinct families (classes) of network models. The section 2.2 is about Statistical network models: These models are focused on explaining the observed features of static real-world networks, such as the presence of community structure or the shortness of paths. In section 2.3 I discuss Mechanistic network models: They take a different angle and seek to illuminate the growth mechanisms that caused the real networks to look as they do. They are an attempt to explain how nodes and links assemble into large-scale networks. The two classes of models complement each other. Each class has its own advantages, which is why they both deserve careful consideration. 

One last remark before I begin. All network models discussed in what follows refer to ordinary non-directed and non-weighted networks. The development of models for directed/weighted networks is in its infancy (compared to the development for ordinary networks). In the context of model selection, no method that I examined included models for directed or weighted networks. This is actually an interesting problem for future research.


\section{Static/statistical/probabilistic network models}  \label{static-models}

This family of network models has a somewhat longer history and (mostly) originates from the original literature in mathematical graph theory and social sciences. Static network models stem from directly observing real-world networks and trying to spot statistical patterns in them, i.e., in their connectivity structures. This class of models is not (primarily) interested in how these structures came to be. Instead, the focus is on finding the regularities in the connectivity structures in empirical (e.g. social) networks and capturing them via simple statistical or probabilistic rules \cite{Newman2018,Menczer2020,Dorogovtsev2022,Izenman2023,goldenberg2010}. Here is my attempt to define this family of network models more consistently. 

\begin{svgraybox}
Static/statistical/probabilistic network models are focused on mathematically describing network connectivity structure as static, i.e., as it was observed by a researcher at a specific point in time (snapshot). These models aim to capture patterns and regularities like block/community structure, homophily, or role equivalence. They do not (explicitly) examine how such a connectivity structure was formed in the first place or how it changes over time. Mathematically, statistical models are about formulating the likelihood of observing a specific network structure under prescribed constraints, which are defined by the model itself. These models seek the best set of parameters to parsimoniously describe static real-world connectivity patterns. 
\end{svgraybox}

Static network models are useful for several reasons. First, they enable us to examine the dynamics and other processes on networks which are considered static, in the sense that their connectivity does not change with time. These processes include epidemic spreading (contagion), diffusion of information, dynamics of opinions, various synchronization phenomena, etc. Even if a synthetic network (or an ensemble thereof) does not model the real network perfectly, the properties of the dynamical process on it may still be very close to what observed in reality. 

Second, they allow for a better understanding of how centrality and similar network measures relate to node importance or node influence. They are also great in explaining (capturing) well-known empirical properties of real networks, such as high clustering (over-presence of triangles in connectivity) or unexpectedly short average values of shortest paths. Crucially, static network models are an excellent avenue to understand why and how real-world networks are neither completely regular nor completely random, but their observed structures can be very accurately represented by statistical/probabilistic laws \cite{Newman2018}. 

Below is the list of several most prominent static network models. Naturally, the list can not be considered exhaustive or complete. I did my best to explain each model briefly. The reader is referred to cited references for more detail.

\runinhead{Regular graphs.}  The simplest example of a static network model is any regular graph, i.e., a structured graph exhibiting some form of regularity or pattern. Such graphs are normally studied within mathematical graph theory \cite{Izenman2023,Bollobas1998,Frieze2015}. Examples of regular graphs that can serve as network models are many, so let me mention just two most common ones. The first is the lattice graph: nodes are arranged in a regular grid, mimicking crystal structures studied in solid state physics or chemistry. Links connect adjacent nodes. Each node has the same degree, except for the border nodes (in case of a finite lattice). Lattices come in 2, 3, or any other number of dimensions. The second such examples are various versions of the ring (cycle) graph. Nodes are lined up in a circle, links connect adjacent nodes in a sequence, like a necklace. More complex versions of rings include additional links that connect second neighbors, third neighbors, etc. All nodes always have the same degree. Such regular graphs are the first na\"{\i}ve idea of a network model. Actually, the clustering coefficient of nodes in these networks can be adjusted to reproduce the real-world values. However, most other properties observed in the empirical networks -- shortest paths, hubs, power-laws, etc. -- are not immediately reproduced by these models. Nevertheless, these are useful as a starting point. Additionally, they come handy as null models, i.e., for investigating possible over- or under- presence of a certain pattern in the studied network.

\runinhead{Erd\H{o}s--R\'enyi (ER) random graphs.}  Another good starting point for building network models is the ER random graph \cite{Erdos1959}. The study of random graphs goes back a long time in graph theory \cite{Bollobas1998,Frieze2015}. There are two ways to think about an ER random graph: One is to start from $n$ disconnected nodes and put a link between each pair of nodes independently with a probability $p$. The values of $n$ and $p$ immediately define the graph's density. The other way is to use $L$ links to connect $n$ nodes, so that each link connects exactly one pair of nodes. In other words, an instance of graph is chosen uniformly at random from the set of all possible graphs with $n$ nodes and $L$ links. One can show via simple exercise that the two ways of defining an ER random graph are equivalent. ER random graph (network) is exactly the opposite of a regular network -- it contains no deterministic structure at all. Instead, the entire graph is as random as it can be. Of course, this is not realistic as a model for real-world networks. Nevertheless, ER random graph captures an important hallmark of real-world networks -- the shortness of (average) shortest paths \cite{Estrada2015,Menczer2020,Dorogovtsev2022,Izenman2023,barabasi2016book}. ER model cannot reproduce other well-known features of real networks, such as high clustering coefficient or hubs. There are many generalizations of ER random graphs \cite{michel2019}. In fact, many standard network models including mechanistic ones (dealt with in next Section) are referred to as `random' because that include some randomness. ER graphs are also routinely used as null models to test if a given network exhibits structure beyond what would be expected in a simple ER random graph \cite{fosdick2018}.

\runinhead{Small-world model.}  One of the early milestones in network science was the discovery that in almost all real-world networks, after averaging over (pairs of) nodes, we find large values of local clustering and small values of shortest paths. Watts and Strogatz already in 1998 proposed a simple model that reproduces both features \cite{watts1998}. This model is a hybrid between regular and random graphs: Regular graphs can be highly clustered, while ER random graphs naturally have short shortest paths. One starts from a (ring) lattice, where each node is connected to several of its nearest neighbors. Then a fraction of the links is rewired at random with probability $p$. When $p=0$, the network remains the original regular graph (high clustering). When $p=1$, it becomes a random graph (short shortest paths). As it turns out, at intermediate values of $p$ (in particular, for $0 < p \ll 1$), the network simultaneously exhibits high clustering \textit{and} short shortest paths -- two key characteristics of real networks. This is known as the \textit{small-world phenomenon}, from which comes the famous ``six degrees of separation'' -- a hallmark of real-world social networks \cite{Estrada2015,Menczer2020,Coscia2021,barabasi2016book}. The design of this model illustrates the fact that complex systems balance regularity with randomness, making them more efficient and better organized \cite{latora2001,deSantos2014,ferreira2021}. However, Small-world model fails to reproduce other important features of real networks, primarily hubs and hierarchical node connectivity (power-laws). The model is still too rudimentary to fit actual real networks, but has served as a foundation for developing more elaborate models. In fact, most (if not all) modern network models automatically reproduce the small-world effect. Finally, Small-world model can be utilized as a form of null model, in the sense that one can measure the `small-world-ness' of a network \cite{humphries2008}.

\runinhead{(Stochastic) Block models.}  Block models (in their many versions) are the most grown-up family of static network models. They are widely used with real-world data and accurately explain real networks, or at least their static snapshots \cite{Coscia2021,holland1983}. (Stochastic) Block Models (SBMs) originate from the social science literature \cite{doreian2004,lee2019,cugmas2020} and predate modeling concepts developed later in physics and computer science literature. SBMs rely on the idea that nodes or a (social) network can be arranged into several latent groups called \textit{blocks}. The interactions (links) between nodes are determined (probabilistically) by which blocks they belong to. Say we have $K$ blocks denoted as $Z_k$, where $Z_k \in {1, 2, \dots, K}$. Now, the probability of a link between the nodes $i$ and $j$ depends only on their block memberships. So, if node $i$ belongs to block $Z_n$ and node $j$ to block $Z_m$, the probability of a link existing between nodes $i$ and $j$ is given by the matrix element $\Theta_{nm}$. An SBM is defined by the number of blocks $K$, node membership, and the matrix $\Theta_{nm}$, called block-wise connection probability matrix. $\Theta_{nm}$ determines, in a probabilistic (stochastic) fashion, the structure of each network instance (realization) obtained from that SBM. Note that $\Theta_{nm}$ is the same for any pair of nodes as long as the first belongs to the block $Z_n$ and the second to the block $Z_m$. This is a powerful modeling framework, which enables us to capture real networks with clustered, hierarchical, role-based many other types of structures. There are numerous generalizations of SBMs, including degree-corrected SBMs (varying degree distributions), hierarchical SBMs (nested block structures), mixed-membership SBMs (nodes can belong to multiple blocks), etc. There are well-developed and statistically principled methods for fitting a given (target) network to an SBM, i.e., calculating the most suitable number of blocks $K$ and the matrix $\Theta_{nm}$ to represent the target network as an SBM \cite{peixoto2014,peixoto2015,ramirez2022}. Personally, I prefer to call this model fitting, but literature often calls it model selection, albeit with a single candidate model: SBM. I will say more about this later in this chapter. 

\runinhead{Exponential random graphs.} (\textit{not a network model})  There is an inference method heavily used in the social science networks community that goes under the name Exponential random graphs (ERGs) or Exponential random graph models (ERGMs) \cite{snijders2006,robins2007,lusher2013}. This inference method takes a real (observed) network as an input and regards it as one realization from a set of all possible networks with similar features. We postulate that this observed network is the outcome of some (unknown) stochastic process, from which the network instances are drawn. The observed network is characterized by a specific pattern of network features (measures). These features can include the number of nodes or links, degree sequence, number of triangles, etc. The pattern of network features characterizing the observed network can be seen as a particular realization out of a set of all possible patterns of network features, which correspond to all possible network instances. The entirety of all possible networks can be formulated as a probability distribution, by which ERGs enable us to compute the likelihood of seeing a particular network instance picked from this probability distribution. For example, assume we observed some real directed network. If ERGs find a high value of reciprocity (bidirectionality of links), it means that mutual (bidirectional) links are more likely than by chance. Social networks are, for instance, notorious for having more triangles than by chance. Given an interesting property of some network, ERGs can tell us how likely it is to find that property by chance: Obviously, the properties that are ultimately interesting are those that are hard to find by chance \cite{stivala2021,karkavandi2022,setayesh2022}. Famous Configuration model is an example of ERG where the degree sequence is that property. Now, in my mind, ERGs are not a network model in the sense of the work that I use in this book. They are not a recipe for how to create a synthetic network from scratch. Instead, they take an observed network as an input and then carry out a specific kind of analysis on it. Also, I found no method of network model selection that considers ERG as a candidate model. Nevertheless, since ERGs/ERGMs are often brought up in the context of network models, I wish to make this important clarification here.

\runinhead{Random geometric graphs.}  This is another simple model based on the intuition to incorporate geometric thinking into network modeling \cite{Estrada2015,michel2019,penrose2003,dall2002}. We want to connect the nodes based on their geometric (spatial) relationships. Random Geometric Graphs (RGGs) start from $N$ nodes that are randomly placed in a (bounded) $d$-dimensional metric space. The metric is usually assumed to be Euclidean. For example, one can consider a $d$-dimensional cube and throw $N$ points (to become nodes) randomly in it. Alternatively, one can throw them in a cube with some probability density. The dimensionality of space $d$ can be anything we like, but normally 2 and 3 suffice. Next we define a distance $r$, much smaller than the characteristic length of the space, and called it connection radius. Now, a link between a given pair of nodes is established if the two nodes (points) are closer than $r$. Doing this for all pairs of nodes creates our network -- a random geometric graph. Given this set-up, one can vary $d$ and $r$ and obtain a rich variety of synthetic networks. In a RGG the spatial proximity influences the likelihood of connection, capturing the intuition that in many real networks connected nodes are in some sense `close' to each other. RGG are used to model real networks that live in physical/spatial world, such as power grids \cite{fujiwara2011} or grey-matter in human brain \cite{lo2015}. RGGs are not to be confused with the idea of spatial embedding of a network, which is about representing a given network as a set of points in an abstract (latent) space so that the distances between the points (nodes) replace their connectivity \cite{barthelemy2011,boguna2021}. I must also say that I was not sure whether to include RGGs in this section or the next: RGGs are meant to capture the static network structure, but they are `grown' mechanistically by connecting the nodes in a geometric space. Either way, RGGs (sometimes called just Geometric Graphs) are an excellent model candidate to consider in any network model selection framework.

\runinhead{Other models that reproduce specific connectivity patterns.}  Literature is full of network models designed specifically to reproduce a particular connectivity pattern observed in real networks. Such models are usually very successful in reproducing the pattern aimed at, albeit sometimes neglecting other connectivity patterns that are just as present. The best example is the preferential attachment model, which seeks to reproduce the power-law degree distribution (this model will be covered in the next section). Here I will list several such models that can be classified as static because they aim at reproducing the connectivity patterns that were found from statistical/probabilistic analysis of real networks as they are observed (even though some of these models rely on some form of attachment). Let me start with the models that seek to reproduce the community structure (which can be regarded as a special case of block structure). Presence of communities (densely connected groups of nodes) has been found in essentially all real networks, in particular social networks, and has been studied extensively \cite{Newman2018,Menczer2020}. That triggered researchers to create network models that generate synthetic networks with a community structure \cite{Coscia2021}. These models allow us to construct networks that display a community structure with desired properties, such as hierarchical and/or overlapping communities \cite{karrer2011,zuev2015,cherifi2019,stadtfeld2020}. They can also be used to make networks with pre-selected planted communities, which are needed for testing (benchmarking) algorithms for community detection. A classic example is Lancichinetti-Fortunato-Radicchi (LFR) model \cite{lancichinetti2008}. The next family are network models that generate synthetic networks with a core-periphery structure: a highly connected core and a less connected periphery \cite{borgatti2000,verma2016}. In a similar fashion, there are models that create networks with a rich-club, centered around a strongly connected hub core \cite{csigi2017}. A $k$-core of a given network is its maximal subnetwork in which every node has degree at least $k$. There are models that make networks with an arbitrary $k$-core structure \cite{hebertdufresne2013}. Assortativity, found in many real networks, means that nodes of similar degree are more likely to link to one another (think of homophily). Researchers developed models the produce networks with desired level of assortativity \cite{sendina2016}. I could go on like this, but it's too daunting to list all such network models. All these models are (at least in principle) viable candidates for network model selection methods. Diversity of such models also shows that network scientists are quite successful in finding the mechanisms that underlie the connectivity patterns in real networks. \\

This concludes my list of static/statistical/probabilistic network models, though it is by no means exhaustive. I encourage the reader to consult the literature for additional models in this category, including recent advancements such as statistical models for temporal networks. It is important to note that for many models in this group there are well-established methods for model inference/fitting. That is what makes them easy candidates for model selection. A prime example are (Stochastic) Block Models, for which statistically rigorous techniques exist to fit a given network to a general (Stochastic) Block Model in the most parsimonious manner. I will explore this later in the chapter. However, static/statistical/probabilistic models are not concerned (perhaps with some exceptions) with network formation, i.e., do not explain how real networks become what they are. This is where our next family of network models comes into play.


\section{Mechanistic/growth/formation network models}  \label{mechanistic-models}

This is a more recent family of network models. In the 1990s, physicists began to contribute to network science (as I mentioned earlier, this scientific stream is what got me into this field). Physicists brought physical thinking with them and started asking new questions. How a given real network came to be the way it is? What are the attachment/rewiring/deletion mechanisms that produce the connectivity patterns that we routinely observe in the real-world network? Are these mechanisms the same for all networks, or do they depend on the network in question? Do they change throughout the evolution of a network, or are they constant? This line of inquiry gave rise to a new class of network models: rather than capturing static connectivity patterns, the focus shifted to identifying growth/formation processes that generate connectivity patterns resembling those observed in the real networks \cite{Newman2018,Menczer2020,Dorogovtsev2022,Izenman2023,barabasi2016book}. Let me properly define this class of network models.

\begin{svgraybox}
Mechanistic/growth/formation network models assume that all networks grow and change with time, and that this is driven by a well-defined (albeit stochastic) set of rules. These mechanistic rules are typically formalized as equations that encode our understanding of stochastic microscopic processes that govern the network formation and evolution. Such rules may include operations such as attaching new nodes, removing existing nodes, deleting and/or rewiring links. They are generally described using discrete time steps, where one or more operations occur at each time step. Most intriguing mechanistic network models are those that generate synthetic networks which resemble real networks. We can reasonably hypothesize that such mechanistic models reflect the actual real formation/evolution mechanism.
\end{svgraybox}

I feel I can say (with quite some confidence) why physicists love this class of models. It is because the logic of developing mechanistic models is akin to the logic that physicists traditionally rely on when developing, for example, differential equations that model some physical system. Mechanistic models enable us to encode interactions between system's units (in this case nodes and links) that happen on a microscopic scale, and which stem from general understanding of such processes. In other words, we use intuition about what network formation is expected to look like to build mechanistic network models, and hypothesize that such models underpin the formation/evolution of real networks. Once a mechanistic model is available, we can use it to generate synthetic networks (by varying the values of parameters) and compare them with the real networks. This allows us to assess whether we are headed in the right direction. We are optimistic if we find that at least some characteristics of synthetic networks are consistent with their real counterparts. 

On the other hand, contrary to static models, mechanistic models are not (necessarily) concerned with describing snapshot network structures. For this reason they (typically) have intractable likelihoods and do not easily lend themselves to the use of inferential statistics. Nevertheless, mechanistic models are just as valuable candidates for model selection -- in fact, they are the most frequent ones. Below is my incomplete list with short descriptions of prominent mechanistic models (I refer the reader to references for more insight). Since there are so many of them, I did my best to include at least those that appear in the papers that I review in what follows. There are ways to categorize mechanistic models into groups \cite{toivonen2009}, but I will not delve into that.

\runinhead{Preferential attachment (Barab\'asi-Albert) model.}  This is definitely the most celebrated network growth/formation model \cite{Newman2018,Menczer2020,Dorogovtsev2022,barabasi2016book}. It is one of the origins of modern network science \cite{barabasi1999} alongside Watts-Strogatz Small-world model discussed earlier. This model is the key avenue through which physicists entered the field \cite{albert2002}. Barab\'asi-Albert (BA) model relies on `rich-gets-richer' idea to capture network growth in its simplest form: new nodes that join the existing network are more likely to connect to nodes which are already well connected. If you already have a lot of friends you are more likely to make new friends than someone who has only a few friends. If a website has many links pointing to it, it is very likely that newborn websites will also point to it. More technically: we start from a network of few nodes, say $m_0$. At each time step, a new node is added to the network. This new node connects to $m$ existing nodes with probabilities proportional to their degrees. So, an old node with degree 10 is 10 times more likely to receive a link from a new node than an old node with degree 1 -- this is called preferential attachment. The network is then grown following this process until the desired size. The celebration behind this model comes from the fact that it produces a power-law degree distribution, commonly observed in many real networks (hence the term scale-free networks). Actually, the power-law is clearer if we make attachments with probabilities proportional to the old node's degree plus a constant (Attractiveness model) \cite{dorogovtsev2000}. The progenitor of the BA model appeared in the 1970s and is referred to as the Price model \cite{Newman2018,price1976}. It was designed for citation networks: better cited/older papers are more successful in accumulating new citations. The BA model is extremely popular in network science, especially among physicists, and it was employed in a myriad of scientific contexts. So much so that it became somewhat controversial \cite{broido2019,barabasi2018blog,voitalov2019,jacomy2020}. Debates surrounding this issue are relevant for overall network science, but are beyond the scope of this book.

\runinhead{Variations of the Barab\'asi-Albert model.}  Huge popularity of BA model in the early 2000s triggered an avalanche of new network models, each proposing an improvement, generalization, or adjustment of the original version. The aims were to reproduce more features of the real networks accurately and/or to capture the hypothesized mechanisms of network formation more realistically. This trend is still going on today and the diversity of resulting models is impossible to grasp in this book \cite{Menczer2020,barabasi2016book}. However, I will mention some of them, also because they frequently appear as candidates in network model selection methods. The first immediate generalization is to make the preferential attachment depend not on the degree of the existing node, but on some power (exponent) of that degree. It is called Non-linear preferential attachment \cite{krapivsky2000}. Next variation is known as the Fitness model. We assign to each node its own individual appeal, called fitness. The fitness value is an intrinsic characteristic of a node and does not change over time. Now, the attachment probability is proportional to the product of the degree and the fitness of an existing node. So, new nodes attach preferentially to old nodes with high degree and/or high fitness \cite{bianconi2001}. Then there is the Rank model: instead of attaching new nodes according to the degree, we now attach them according to the rank of the degree (or rank of some other node property). So, we first rank all the nodes according to the chosen property (e.g. degree). Then we make attachments with probabilities proportional to some inverse power of these ranks \cite{fortunato2006}. While all these models neatly reproduce power-laws and hubs, they mostly fail to reproduce high values of clustering (triangles) in the resulting synthetic networks. That can be obtained by suitably adjusting the rules of attachment, as done in several other models, one example being \cite{klemm2002}. Another large group of models seeks the attachment mechanisms that would generate networks with specific properties found in life sciences networks, e.g. brain networks. One example is the Affinity model: an extension of BA model that captures the assortative and hierarchical structure known to exist in brain networks \cite{klimm2014}. With the 2010s came the idea to embed networks into geometrical/metric spaces. This led to Geometric preferential attachment models: Nodes are placed in a space (often hyperbolic), so that geometric constraints are included in the attachment rules. Now, the attachment probability depends both on node characteristics (e.g. degree) and spatial proximity \cite{papadopoulos2015}. And so on. There are many more such models, which are (slight) modifications of the original BA model in one form or another. I will stop here and move to describing other growth/formation models that differ from (but often build upon) the original version in a more substantial way. Please note, however, that it is hard to establish a precise border between a variation of BA model and an entirely new model, since many formation models rely on iterative node attachment. 

\runinhead{Random walk attachment model.} As I just mentioned, there are countless versions of growth models relying on some form of attachment. The growth mechanisms are defined by specific micro-rules of new node additions. In general, such models lead to power-law degree distribution. There are constant attempts at making the attachment rules even more realistic. Ideally, in addition to being intuitive, such attachment rules would also generate synthetic networks that resemble their real-world counterparts even more realistically. There are several models referred to as Random walk attachment models. The literature mentions a few versions, so I will mention two of them here. The first one \cite{Menczer2020,vazquez2003} is inspired by the idea that when we meet a new person, we are very likely to also meet his/her friend(s), and possibly their friends as well. So, besides creating random links to new neighbors, we also connect to the new neighbor’s neighbors. The rules are as follows: A new node is first linked to a randomly chosen old node $i$. Then, with probability $p$, the new node is also connected to another old node $j$, which is a randomly chosen neighbor of the node $i$. Otherwise, with probability $1 - p$, the new node is linked to an old node $j$, which may not be a neighbor of $i$, but is chosen randomly from the entire network. Note that this mechanism does not presume that new nodes know the degrees of old nodes. The parameter (probability) $p$ dictates the emergence of triadic closure (average values of local clustering). Another version of Random walk attachment model is that upon getting attached to the old node $i$, the new node gets attached to another old node $j$, which this time is chosen by a random walk of a certain length that starts from the node $i$ \cite{saramaki2004}. The length of this random walk has to do with local clustering and can lead to a community structure. My guess is that there are many more versions of the Random walk attachment model, all very intuitive, but I will skip them here. 

\runinhead{Node copying models.} These models are inspired by the idea that incoming nodes may try to mimic the connectivity of the existing (old) nodes. For example, a scholar publishing a paper might try to put together the list of references for his/her new paper by copying it from an existing paper, or at least partially copying it. A website creator could decide to link a new website to the existing websites based on how this was done by some preferred existing website, which the creator considers to be suitably connected. In its basic form, the model is as follows: a new node chooses one of the old nodes as its `prototype' node. Then, the new incoming node establishes links to some of the neighbors of the prototype node. For each prototype's neighbor, the new node creates a link to it with the probability $p$. Otherwise, with probability $1 - p$, it attaches itself to another node, randomly chosen from the entire network \cite{Newman2018}. So, every new incoming node `copies' a part of the neighborhood of the node it took as the prototype, and then creates the rest of its neighborhood at random. The specific choices of the prototype node from one node addition to the next become less important as the size of the entire network grows. This is a very intuitive model for networked ecosystems such as scientific publishing. Node copying model leads to hubs and power-laws. However, it does not (necessarily) lead to triadic closure and high presence of triangles. As in all other cases, this model also has many and diverse versions \cite{kleinberg1999,krapivsky2005}.

\runinhead{Duplication-divergence models.} As I mentioned at the beginning of this section, an important advantage of mechanistic models is that they allow us to seamlessly convert domain knowledge and intuitions into mathematically sound network models. Duplication-divergence models are a great example of this. They come from the domain of biology, in particular, from the study of gene networks and protein-protein interaction networks. Say we have a gene (node) in an existing genetic network, which undergoes the natural evolutionary process involving replication and mutations. We start from an existing gene network composed of old nodes (genes). We create a new node by replicating an old node and its connections. Hence, the new node initially attaches itself to the nodes that have links to the chosen old node. Now, with some probability $p$, each of thus created new links is either deleted or rewired to randomly chosen nodes. When rewiring, the end of the link attached to the new node is kept (and the other end is rewired). So, the new node (gene) is an imperfect replica of the original node (gene), in a sense that it has some of its links. Note that the old node plays a role similar to the prototype node in the Node copying model. In fact, Duplication-divergence model can be seen as a version of Node copying model, except that it stems from biology of gene networks, not bibliometrics. The evolutionary processes of random replication and mutations are captured by the new node first inheriting links of the original node (duplication) and then losing some of them (divergence). This model produces networks with power-laws etc., but more importantly, it produces networks with properties similar to those empirically found in biological networks. As always, there are many versions in circulation \cite{wagner1994,sole2002,bhan2002,vazquezFlammini2003}.

\runinhead{Forest fire model.}  This model is inspired by the idea of fire that spreads through a forest: a burning tree ignites trees close by and fire spreads. New nodes connect by ``burning’’ the edges in the existing network. The model simulates network growth as a recursive process: A new node connects to a randomly chosen old node and then ``burns’’ (i.e., connects to) a subset of its neighbors. The old node is in this model called `ambassador' node. This model was originally intended for web graphs with hyperlinks, so it is casted as a model for directed networks. The precise mechanism is: The new incoming node randomly chooses its ambassador node to connect to. Then it burns (visits) a subset of the ambassador’s neighbors (both in-links and out-links), chosen with forward and backward burning probabilities, denoted respectively as $p_f$ and $p_b$. This process continues recursively: each newly burned node may ignite its neighbors in the same manner. Once the burning process is over and no additional links are burned, the new node establishes links to nodes visited during the burning process. This model is used for citation networks, social media graphs, and web hyperlink networks. Except power-laws, this model captures two other empirical properties of real networks. The first is densification: the number of links in real networks usually grows super-linearly with the number of nodes. The second is shrinking diameters: as the network grows, the average shortest path in general decreases. As always, there are many versions of the forest fire model \cite{Coscia2021,leskovec2005,bancal2010}. Some versions adjust the burning probabilities, other restrict the burning depth etc. 

\runinhead{Kronecker graphs.}  Real networks, as always, surprise us with a variety of properties they exhibit. Consequently, it is extremely hard -- if not impossible -- to construct a model that reproduces them all. In addition to properties such as power-law and high local clustering, real networks sometimes display self-similar (fractal-like) structure and scalability. These properties are usually not captured by the standard network models discussed so far, such as BA model and other attachment-based models. This could be because it is not immediately clear how to reproduce them via attachment-based models. Kronecker graph is a network model based not on addition/attachment of new nodes, but on another, more mathematical concept. It relies on directly constructing the adjacency matrix of the synthetic (generated) network via iterative matrix operations. We start from some small matrix $A$, for instance $2 \times 2$, called the initiator matrix. We think of matrix $A$ as the adjacency matrix for our initial network, called initiator graph \cite{leskovec2010}. Then we multiply the initiator matrix with itself by applying the Kronecker product. The Kronecker product of two $2 \times 2$ matrices $A$ and $B$ is defined as:
\[     
A \otimes B = \begin{pmatrix} 
a_{11}B & a_{12}B \\ 
a_{21}B & a_{22}B 
\end{pmatrix},    
\]
with obvious generalization to the case of $n \times n$ matrix. After multiplication we are left with a matrix $A \otimes A$ of size $n^2$, which we interpret as the adjacency matrix for the next iteration of our network. This graph has $n$ times the size of the original graph and $n$ times the number of links. We then multiply the matrix $A \otimes A$ with itself again using Kronecker product. The resulting adjacency matrix $A \otimes A \otimes A \otimes A$ has $n^4$ times more nodes and links compared to the initiator graph/matrix $A$. We now keep iterating (Kronecker-multiplying) in the same manner until we reach the desired matrix (network) size: Kronecker graphs grows in a self-similar manner by recursively applying the Kronecker product. After $k$ iterations our synthetic network will have $n^k$ more nodes and links than the initiator graph. The result is the adjacency matrix of the Kronecker graph of order $k$. It can be shown that this network structure exhibits self-similarity and scalability. In addition, Kronecker networks display densification power law and shrinking diameter, another set of important real network properties \cite{leskovec2010}. There are many many variants of Kronecker graphs. Some include extensions for directed and/or weighted graphs, others capture community structure, etc. \cite{Coscia2021,seshadhri2013}. This model complements the rest of formation models by not relying on any attachment process. Yet, it is a viable candidate for any network model selection method.

\runinhead{Other growth/formation models.}  There are an infinite number of formation network models in the literature (not literary, but almost..). Reviewing them all and finding common denominators is an important research task for future work. Here I will limit the discussion to mentioning a few more prominent mechanistic/growth/formation models, primarily those that did or could appear as candidates in network model selection methods. The bulk of these models come from physics and computer science literature. Some of them are constructed by assuming some general growth principles and logic, which are expected to be universal (or close to) for all real networks. These growth models are not (necessarily) intended for any specific situation studied in any domain science. These are some examples \cite{xie2008,akoglu2009,smolyarenko2013,morzy2019,falkenberg2020,levens2022,kharel2022}, but of course, there are many many more. Another framework for designing growth models is to get inspiration from empirical insights, conceptions, or ideas about how real networks should form, while acknowledging that any given growth process depends on the specific context studied within a specific scientific field. In biology, there is a need to grasp the formation and evolution of PPI networks, gene interaction networks etc. One example is the stickiness model \cite{przulj2006}. The behavior of functional or structural brain connectome is of interest in modern neuroscience. A range of models exist to describe this situation \cite{betzel2016}. Formation of social networks is primarily studied in social sciences, but growth models for them have also been developed in physics and computer science literature \cite{antal2005,talaga2020}. Network models have also been proposed to explain growing economic systems, where opportunistic attachment could account for the observed structures \cite{mattsson2025}. Network problems in logistics discuss issues such as formation of traffic (road), airline, railway, or infrastructure networks. For example, one could look for a network that satisfies the passenger and cargo demands while minimizing the overall transport costs \cite{zhang2015}. And so on. There are many more scientific fields and domains where various network growth models have been proposed to describe the empirically observed networks in those contexts. There are also several review papers discussing and comparing various growth/formation models (and static models) \cite{toivonen2009,goldenberg2010,piva2021}. At any rate, an interested reader will face no trouble finding many more growth models (and review papers about them) in the existing literature. The point is, all (or most) of these formation/growth models are suitable candidates for network model selections methods. \\

This section was my modest attempt to list major mechanistic/growth/formation network models. The list is by no means complete or exhaustive. Let me close this section by stressing two things. First, mechanistic models indeed offer a smooth way of translating our intuitions about network growth into algorithms for the purpose of making synthetic networks similar to the real ones. This was one of the key directions of development of network science over the last decades, not just in the physics community.

However, there are little to no statistically principled methods for fitting a given growth/formation model onto a real network coming from empirical data. In other terms: the world of mechanistic/formation/growth models is a world of forward-simulation. There is a lack of attention on the inverse problem: given a target network, which formation model is most likely to have generated it? This is in stark contrast with the static/statistical models, where such methods abound (e.g. SBM). As we will see in chapters 3 -- 6, methods for solving the inverse modeling problem for mechanistic models do exist, but are all heuristic in nature.


\section{The problem of network model selection reviewed in this book}  \label{formulation}

With all of the above said, I am finally in a position to define -- in a very precise manner -- what this book is about. Over the last decade or so, researchers recognized the lack of inference (inverse) methods for mechanistic network models. This lack prevents us from answering questions like: Are preferential attachment models \textit{really} behind real networks? Everyone seems to claim that they are, but concrete evidence is inconclusive. And given the amount of literature produced about preferential attachment, answering questions like this one is paramount. As I emphasized above, unlike static network models, mechanistic models do not simply lend themselves to statistical inference methods. 

So, researchers resorted to various heuristics and proposed several methods for checking if some (mechanistic) network model is behind a given target network. Being anyhow based on heuristics, these methods can, without additional complication, consider several candidate models at once. Importantly, these candidate network models may be static or mechanistic, since methods do not distinguish between them. This framework enables us to answer a broader question: which network model -- from a mixed list of candidates -- is likely to be behind some target network. Such list can include all network models mentioned in both sections above, and many more. From this comes the name \textit{network model selection}. 

\begin{svgraybox}
This book is a review of these methods. I selected them very carefully, making sure that they share the same (or equivalent) definition of the problem. As I explained in section \ref{sampling}, I eventually settled on 17 methods that fulfilled all the criteria. In addition to reviewing all these methods, I did my best to classify them depending on the approach they take and to discuss their practical implementability. The aim of this, at least in my mind, is to show us the way forward: How should we find the most useful method of network model selection and apply it to real networks. This should contribute to (if not settle) the question of what mechanisms \textit{really} underpin real networks.
\end{svgraybox}

As I have stressed earlier, the scientifically meaningful strategy -- in my opinion -- is not to keep creating new methods, but to focus on improving the one(s) that work best, and then apply them. We should not lose sight of the fact that our key aim is to \textit{apply} these methods to the real networks and discover things about them. The goal of network science \textit{should not be} to design methods for its own sake. This is another motivation I have for completing this book.

I will close this Chapter with two short sections devoted to two important clarifications. The first has to do with the issue of model selection with one vs. multiple candidate models. The second is the disambiguation of the term `model selection', which in machine learning/statistics denotes something slightly different.


\section{Network model selection -- one vs. multiple candidate models}

There is a certain confusion in network science among the terms model selection, model inference, and model fitting (and I admit I am not perfectly sure either). As I understand these terms, they anyhow refer to similar problems. For the purposes of this book, I will focus on the following key difference: Are we trying to fit a single network model (perhaps parameterized in some way), or are we choosing the best model among multiple genuinely different candidates? The former is what I would call the problem of model inference or perhaps model fitting. It can be formulated as hypothesis testing: what is the probability that this target network was generated by this particular network model (and for what values of model parameters)? Instead, this book is about the latter formulation: selecting the optimal network model among several different candidates, of which each might have its own set of parameters. This is what I will call network model selection in this book. For me, this framework is superior, because it makes no assumption about which model is behind the target network. Instead, it makes candidate models `compete' so that the best fitting one is selected. Let me explain.

Static/statistical network models have a long tradition of single model inference methods. All of them consider that model as the sole candidate. A prime example are Stochastic Block models (SBM). For them, there are inference methods the work under very general conditions and are mathematically precisely founded (see \cite{doreian2004,lee2019,cugmas2020,peixoto2014,peixoto2015,ramirez2022} and references therein). And indeed, SBM have been amazingly successful in capturing empirical structures of real (mainly social) networks. They are thought in university courses on (social) network analysis, are featured in various software packages, etc.

But I have two issues with the framework of single model inference/fitting. The first is that by using \textit{only one} model, we are not giving the target network a chance to be fitted by another model. Let's say we are analyzing some target network, and we found an excellent fit of our target to some SBM. That is a great result, but what if there is another, perhaps simpler model, which fits equally well (or better)? We will never know unless we try. Yes, in their full generality SBMs encompass a lot, but still: they represent only one specific logic of connectivity among many that can be out there. When we fit a single model, be it SBM or something else, we are implicitly assuming that this is the correct model to fit. But what if it's not? 

My second issue is that SBMs speak only about static network structure and make no attempt at explaining the network growth/formation. So, even if we obtain a perfect fit to an SBM, and even if we have good reasons not to try any other model, we still haven't learned much about how our target network formed in the first place. In other words: We sacrificed the knowledge about target network's growth at the expense of knowledge about its structure. The same is true for all static network models and their inference methods.

This brings me to the problem of model inference for mechanistic network models. As I will show in the chapters that follow, researchers are developing such methods based on diverse heuristics. These approaches are unavoidably approximate and are never mathematically exact like those for SBM. But they offer another advantage: ability to consider multiple candidate models at the same time, so that running the method (algorithm) amounts to identifying the best model. This is why I believe the term `network model selection' is appropriate here. Furthermore, the competing candidate models can now be either static or mechanistic. For the most part, heuristic network model selection methods make no difference between the two. 

However, it must be stressed that finding a good fit to a mechanistic model does not (always) guarantee that the target network actually grew via that mechanistic model. It is hard even to estimate the probability that it did. This is the downside of heuristics, which is (in general) less exact than statistical hypothesis testing that underlies static (SBM) inference methods. More about this later.

To finish this section, let me mention that there are single model inference methods for network models other than SBMs. As I mentioned, they are very well developed for basically all static models (see literature). There are even more general approaches to evaluate the suitability of any static/statistical model with respect to a target network \cite{duvivier2022}. But more interestingly, there are single model inference methods suited for mechanistic models, in particular, for preferential attachment models. The methods \cite{pham2015} checks whether preferential attachment is responsible for generating a given target network. Of course, this depends on how many snapshots of the network are available \cite{pham2021}. There are many remarks worth stressing here, but let me mention just two. First, power-laws do not always come from preferential attachments \cite{tsiotas2019}. And second -- in reality, there is no guarantee that one and the same (mechanistic) model always dictates the network growth. What if the growth mechanism (models) is to change at some point \cite{arnold2021}? 

\begin{svgraybox}
At any rate, as I see it, all single model inference methods -- despite their success -- have important limitations. These limitation come to light when we seek the answers to the most fundamental question in network science: What mechanisms underlie real-world networks? This is why I am devoting this book to reviewing the methods of network model selection \textit{with multiple candidates}.
\end{svgraybox}


\section{Disambiguation of the term `model selection'}

As stated above, in this book I use the term 'network model selection' to denote the problem of determining which network model -- among several candidates, static and/or mechanistic -- best describes a given target network. This is how the term is used in the literature that I will review. In this section I wish to clarify that the term `model selection' has been in use for decades in statistics and machine learning, but to a somewhat different scope. This is an important disambiguation to make.

Originally, the term `model selection' was associated with the broad scientific problem of finding the best model for some (empirical) data. In the simplest version, the problem can be framed as follows. We are given some dataset in the form of a sequence of values (points) defined by their $x$ and $y$ coordinates. The problem is to find the equation (or a function) that best fits these data points. How to find such an equation? We resort to the fact that the data comes from some experimental measurement, which is conducted in a certain scientific context. Now, this context may lead us to hypothesize that there could be several candidate equations. Each equation is rooted in our current understanding and assumptions about the studied scientific context. Each of them is equally well motivated: none of the equations is a-priori more likely than others to be the best one. Furthermore, these candidate equations can be \textit{very} different among them, not just in the number of parameters. In the language of machine learning, each such equation is called a \textit{model} and the list of all candidate equations is called the model set. So, how we select the best among them? We fit each equation (model) to the data points. We quantify the goodness of each fit using one or more measures (e.g. least squares). We also keep track of the model complexity (e.g. the number of parameters). There are various criteria to identify the best model from goodness of fit, model complexity, and possibly other requirements. Finally, the winner model is \textit{selected} as the model that best explains the data. This procedure is the original model selection.

More generally, model selection involves situations where the data can be more complex than a sequence of points with $x-y$ coordinates. Likewise, model can be a more intricate thing than a mere equation. All this complicates the game of defining the ``best'' model, let alone searching for it. Still, looking for the best model usually involves balancing between goodness of fit and model complexity. Without this balance, we risk underfitting (model is too simple) or overfitting (model is too complex) the data. We may evaluate the candidate models based not only on how well they explain the known data, but also by how likely they are to generalize to new, unseen data. This is linked to the classic dilemma of statistics: do we need the model to explain the existing data or to make predictions about the future data? Very often we can not have both, so our selection criteria should reflect our preference. Most common measures of ``best''-ness of a model rely on some information criterion. The most standard among them is the \textit{Akaike Information Criterion} (AIC), which is oriented toward predictive accuracy. Other information criteria penalize model complexity and reward goodness of fit in different ways, hence balancing between predictive and explanatory power. The literature on this is overwhelming: I suggest the reader to start from standard texts on machine learning \cite{zubarev2024,shalev2014,james2023,hastie2009} and then work his/her way up to \cite{burnham2002,kadane2004,nannen2010,ding2018}.

One situation where data is not a sequence of points and model is not a mathematical equation is exactly the situation treated in this book. Here, `data' is replaced by the target network, while `model' is replaced by the candidate network model. The model set is the list of all candidate network models. Apart from this, the game remains the same: how to find the best model to fit the studied data. This is what connects the traditional concept of model selection with the concept employed in this book. However, fitting a network model to network data is more complicated than fitting a sequence of points to an equation. This is why network model selection can not rely on precise measures of ``best-ness'' such as AIC. Nevertheless, research community has developed other, albeit heuristic, approaches that still serve the purpose. Next four chapters are a tour of such approaches. \\ 

\runinhead{An organizational note about the next four chapters.}  This short note is not related to the scientific content of this section, but to the organization of the next four chapters. Namely, while preparing the reviews of 17 methods (papers) on network model selection, I noticed that they can be arranged into four different groups, according to the approach they take to the problem of network model selection. I then decided, in the interest of overall clarity, to organize these 17 reviews in 4 different chapters. I find this arrangement to offer more structure for the reader, as compared to putting all reviews in a single chapter (with many sections and subsections). However, for consistency, I reviewed all 17 papers in all four chapters relying on the same template, which I will present immediately in the next Chapter. Each of four chapters begins with a statement of what is common (in terms of methods' operation) to all methods reviewed in that chapter. Each chapter ends with a brief discussion of methods from that chapter.

\chapter{Methods based on encoding networks via topological measures}
\label{encoding-measures}

\begin{svgraybox}
\abstract{This is the first of four chapters in which I review the literature on network model selection. In this chapter I review the first six (out of 17) methods. These methods begin by choosing a set of \textit{topological (structural) measures}, such as size, density, degree, assortativity, transitivity, etc. All of them can be easily computed for any network, be it real (target) or a synthetic one generated by some candidate model. Upon computing the (averaged) values of these measures (ML literature calls them features), we arrange the obtained values in a vector. This vector represents a network and is often called encoding (embedding) of that network. Once networks are encoded into vectors, we can do all kinds of operations with them as we usually do with vectors, including e.g. comparison. This framework is what six methods reviewed in this chapter rely on to carry out network model selection. After reviewing them, I end the chapter with a brief discussion of all six methods, highlighting what they (dis)agree on. But before diving into all that, I will open the chapter with a few organizational remarks that pertain to this and the next three chapters.}
\end{svgraybox}

To repeat once again, all methods reviewed in this and the next three chapters share the same formulation of the problem: \textit{given a target network, find which of the candidate network models fits (explains, captures) it best} (see section 2.4 for a precise formulation). All 17 reviewed papers present methods that differ from each other in one aspect or another, but share this basic formulation. Studying these methods closely, I found that they can (roughly) be arranged into four categories depending on how they approach the problem. I found that methods in each category have fundamental operational principles in common. This is not to say that the borders between categories are perfectly clear, but this categorization allowed me to better organize the book. Hence I decided to arrange the reviews in four chapters, each chapter devoted to one category of methods for network model selection. This chapter is about six methods that encode networks into vectors using (averaged) values of certain topological (structural) measures (indices), and build network model selection via this framework. Methods in the other three chapter start from a different foundation, which I will state in the abstract of each chapter (similarly as I did in this chapter).

Before I start reviewing the methods, I will make a few remarks and some clarifications that pertain to all four chapters, starting with this one. First remark is to establish a standard way to do reviews, i.e., define my review template: I will present each method in three short paragraphs, each paragraph with a specific content. Those three paragraphs are:

\runinhead{The basic principle} I start each review by describing the fundamental logic (principle, framework) of the method: How exactly it works, does it use ML, physics, or something else, how it chooses the best network model, etc. This is the core of each method, the `trick' it uses to perform network model selection. I will seek to explain it as best I understand it, using a consistent terminology across all 17 methods. I will frame this `trick' as a series of algorithmic steps and briefly explain each step. This basic operational principle is also the main direction along which the methods (chapters) differ. Some principles are similar but none are exactly the same, especially between methods that belong to different chapters. Each reviewed paper proposes a new method or at least a major upgrade of an existing one, so there is a good one-to-one correspondence between papers and methods. Nevertheless, I will review \textit{only the method}, and leave (most of) other details aside.

\runinhead{Other details and comments} Here I mention what else I consider relevant for appreciating each method, beyond its main operational principle. Some of these other details are relevant for the reader of this book: For example, if the authors applied their method to real networks or if they compared their method with other methods for network model selection. Also, I sometimes use this space to insert my own comments and remarks. These comments are not to be understood as a critical and objective assessment of the paper: I am not writing them as a reviewer looking at one isolated paper, but after I have read many similar papers. These comments are my own observations about the method and its relationship with other methods.

\runinhead{Available software} Each method is really a programming code (or some kind of software) that performs the network model selection given the input. Diligent authors publish these programming codes alongside the paper. This allows the rest of us to implement the method and play with it or even use it in practice. This should facilitate further development of that method and comparison with other methods. In this paragraph I mention what software/codes I found available, whether they can be employed immediately, or if the authors instead instructed us to contact them about the codes. Sadly, I must say that many authors have not been particularly diligent about this. Some methods can be (relatively) easily coded, but most of them can not. \\

The above template is how I will review all the methods in all four pertaining chapters. I will consistently stick to this form of presentation, since I found it simple and accessible, regardless of what category a method belongs to. This also enables an easier discussion of the methods. Reviews in each chapter start from the simplest methods and move towards more intricate ones. This does not always align with the chronology of the papers. Each of four chapters including this one starts (in the chapter's abstract) with explanation of the basic concept behind that category of methods. Each chapter closes with a discussion of all methods reviewed in that chapter. Overall joint discussion of all 17 methods is a part of Chapter 7. I will assume that the reader is familiar with the matter described in section 1.7.

\begin{warning}{An important remark:}
Most of these 17 papers include a part where the performance of the proposed method is tested and/or evaluated. This part sometimes includes comparisons with other methods or at least with similar methods, which I mention too. However, I wish to clarify that for the purposes of this book I did not run any methods myself (except for the paper \cite{Yaveroglu2014} which I co-authored). Therefore, I did not check that the methods actually work, let alone compared their performances. So, I am taking the claims in the papers for granted and I am only reporting them. I realize that blindly trusting the papers (even peer-reviewed ones) may not be the best practice. But as I said earlier, a systematic comparison of methods' performances is the central objective of my future work. This comparison should involve a concise benchmarking and implementation of methods' programming codes, for which I currently do not have either time or resources. 
\end{warning}

Having said all this, I will finally start with with this chapter's six reviews. For simplicity, I put each review in its own section. Each section's title cites the paper being reviewed. All remarks above hold for the next three chapters as well, so I will not repeat them in other three chapters.


\section{Generative Model Selection for Complex Networks by Motallebi et al., 2013 \cite{Motallebi2013}}

\runinhead{The basic principle} The paper opens with its own motivation behind the problem of network model selection: Once a synthetic network closely resembling the real target is found, we can replace the target network with the synthetic one. This leaves more maneuvering space for diverse downstream tasks related to network analysis, dynamics, etc. Authors define their main question as: Among the candidate network models, which one is the most suitable for generating network instances similar to the target network? Authors propose to solve the challenge via method called Generative Model Selection for Complex Networks (GMSCN). Its main steps are as follows. (1) Many artificial network instances are synthesized using the candidate network models. This is done in a way to make their density similar to the density of the target network. (2) A set of topological measures (features) in extracted from each of these synthetic network instances. Obtained values are arranged into a labeled dataset. (3) This dataset is used to train the classifier via chosen supervised learning algorithm (the choice of which is to be done carefully). (4) Same features are extracted from the target network and fed to the classifier. (5) The classifier classifies the target network and return the most compatible network model from the list of candidates. This model is the selected one.

\runinhead{Other details and comments} The method can be customized along three dimensions: the set of topological measures to be extracted (authors here utilize ten of them, which are chosen for efficient computation and size-independence), supervised learning algorithm (they examine several of them and eventually settle on LADTree), and the set of candidate network models (authors consider seven prominent ones in 2013). As a key advantage of their method authors highlight that it is not dependent on the size of the target network. GMSCN is then evaluated and tested on real data, including aspects such as robustness to noise and scalability. I also note that this paper includes a good literature review for model selection up to 2013. I found the method very intuitive, but there is a lot to discuss, especially about the choice of topological measures and the classification algorithm. For example: can we be sure that a different choice of e.g. topological measures would lead to the same model being selected?

\runinhead{Available software} In the Appendix B paper mentions diverse software and libraries used for implementation. My guess is that one could reproduce the method using these. However, I could not find the method packaged into a finished software that could be immediately used.


\section{RF classifier and a choice of topological measures by Caceres et al., 2016 \cite{Caceres2016}}

\runinhead{The basic principle}  This paper proposes a simple method to find which of the considered two network models (classic ER and stochastic block) is closer to the target network. The methods uses random forest (RF) classifier and relies on a long list of topological measures as features. The method starts by (1) building a classifier. Here, one should generate many instances of synthetic networks using both models for a certain range of densities (authors consider sparse regime). Next (2), each instance of a synthetic network is embedded into the topological feature space (topological measures are extracted from all considered networks). (3) We train the RF classifier using this data, i.e., we teach it to differentiate between synthetic networks generated by one or the other candidate model. (4) Finally, our method is ready to take on a new network (it has not seen before) and perform model selection, i.e., decide with of the two models fits it better. This is done by computing the values of target's topological measures and checking them against the typical values for two models. The model whose typical values are (on average) closer to target's is selected. 

\runinhead{Other details and comments} As an additional task, authors set out to find which topological measures play more and which less important roles for their network model selection. Eventually, they find that out of 26 initially considered measures about 15 of them carry most of the discriminative power. This I think is a very important point. Authors then make additional effort to examine the performance of their method in presence of noise. To this end, they generate noisy instances of networks by random node rewiring at different levels. Also, they look at method's sensitivity to network sizes, showing how the performance changes as network's size increases. Finally, authors conclude that the combination of RF classification algorithm and a suitable choice of topological measures discerns real networks very well, at least in case of two considered candidate models. This is a great way to proceed, but I feel this method must be expanded to more network models. This will unavoidably complicate the choice of topological measures. Actually, finding a parsimonious list of topological measures to consider for network model selection is definitely an interesting problem.

\runinhead{Available software} I could not find any software package published alongside the paper, but instructions on how to implement the algorithms are fairly detailed.


\section{Generating synthetic graphs similar to a real target by Nagy \& Molontay, 2022 \cite{Nagy2022}}

\runinhead{The basic principle} Rather than proposing a new method, this paper studies to what extent can real networks be distinguished from the synthetic (model generated) ones, i.e. it asks what is the descriptive ability of network models in general. Authors acknowledge a great research interest in examining how well real networks can be captured by a small number of topological measures and whether network models can help us understand variations of these measures observed in the real world. Authors consider 500 real networks from diverse domains (social, neuro, etc.), in addition to 2000 synthetic networks generated by four popular network models. To make synthetic networks as similar to the real ones as possible, authors follow a simple procedure: (1) they first choose 17 topological measures (features), including size, density, assortativity, clustering coefficient etc., and extract their values for each real network, (2) use thus obtained values as input parameters for candidate network models. These models are calibrated using 8 best topological measures. Each model relies on a few parameters, so the calibration is done separately for each model (e.g., for the classic Watts-Strogatz model authors take size, density, and rewiring probability of real networks as inputs). Four synthetic networks are then generated in correspondence to each real network, (3) to quantify the similarity between real and synthetic networks, authors use the mean Canberra distance between feature vectors (network encoding) containing the values of topological measures. Among other insights, this procedure allows to judge which model best fits the target network. For me this is equivalent to model selection.

\runinhead{Other details and comments} Authors find a number of interesting results, e.g. that the correlation profiles significantly differ across the domains of real network, and that the domain a real network belongs to can be efficiently determined using a small number of graph metrics. Further, authors look at the distinguishability of generated from real networks and search for the graph properties that make such distinction possible. Paper finishes by discussing why and how domains of real networks differ. I could not understand very clearly how did the authors determine that 8 of the initially considered 17 topological measures are the best non-redundant ones. Searching for this non-redundancy is anyhow a common theme to many methods in this chapter. Note that this impacts the downstream question of how much we can rely on mean Canberra distance between feature vectors to quantify network similarity. 

\runinhead{Available software} Authors publish extensive supplementary material, including detailed descriptions of the considered network models, fitting/calibration procedures, and used network data. I could not find any user-ready software or programming codes, but my guess is that the available descriptions should suffice to implement the method.


\section{ModelFit classifier with parameter estimation by Aliakbary et al., 2015 \cite{Aliakbary2015}}

\runinhead{The basic principle}  This is the improved version of GMSCN proposed in \cite{Motallebi2013}. Here, authors propose a better method not just to select the best fitting network model, but also to calibrate it by estimating the model parameters which generate networks most similar to the target. Authors call this new two-step method ModelFit, and claim it reduces the sensitivity of model selection process to noise or inaccuracies in the data. So, ModelFit consists of a model selection component and a parameter estimation component. I will start from the the former, which again relies on topological measures as features, but now also involves a network distance measure. Authors consider a wide range of local and global topological measures. The procedure starts by (1) generating a set of synthetic network instances using candidate models and labeling them according to the corresponding models. Each instance is represented by its feature vector of topological measures. Next, (2) one trains a distance measure called NetDistance using a genetic algorithm. This specially designed network distance separates networks based on what model are they coming from. (3) One now uses $k$-nearest-neighbor (KNN) algorithm to develop a classifier that can classify any input network (including the target) into one of the candidate models. (4) Finally, topological measures (features) are extracted from the target network and the classifier is employed to select the winner model. To be more fair, authors separate synthetic network instances into two disjoint sets, training set and neighbors set. Two sets  are used, respectively, to train NetDistance and to make the network model selection decision.

\runinhead{Other details and comments}  Authors evaluate their noise-tolerant ModelFit and confront its performance with the performance of state-of-the-art baseline methods, including the original GMSCN. They find that ModelFit outperforms them, especially in terms of accuracy and noise tolerance. Next they train an artificial neural network to enrich their method with a model-independent parameter estimation procedure. This way, ModelFit can also tailor the selected model exactly to the target network. Paper closes with a case study using empirical data from CiteSeerX digital library. I really appreciate that authors made an effort to compare their method with some of the contenders (existing in 2015), even if not all contenders were included.

\runinhead{Available software} I could not find any. This is a problem, since implementing this entire procedure from scratch is far from simple.


\section{Combining neural networks and $k$NN by Attar \& Aliakbary, 2017 \cite{Attar2017}}

\runinhead{The basic principle}  Authors extend their earlier results, specifically GMSCN and ModelFit (see above). They note that the problem of model selection is equivalent to the problem of network classification, whereby a given target network is classified into one of the pre-established classes, which correspond to the candidate models. To formulate their new method, authors define the selected model as the one that generates synthetic networks most topologically similar to the target, albeit of different size. More precisely, if the target network is $g$ and the set of candidate models is $M$, the selected model $C_g$ is given by $C_g = \arg \min_{m \in M} D(g,m)$. Here, $D(g,m)$ is the minimum distance of the target $g$ to the synthetic networks generated by the model $m$. These synthetic instances are made by varying the parameters that describe $m$ (e.g. the rewiring probability $p$ in the classic small-world model). With this in mind, authors construct their new method using neural networks and $k$NN classification. The method again relies on the chosen set of topological measures/features, including eight features characterizing the degree distribution. The method is: (1) Each candidate model $m$ is defined by one or more parameters, called generative parameters (e.g. the rewiring probability). For each candidate model, we find the optimal values of generative parameters. At such values, the model synthesizes networks that are as similar as possible to the target $g$. This is done by first synthesizing many network instances by varying the generative parameters, extracting the topological measures from them, and then feeding this entire dataset to the neural network. Upon such training, the neural network computes -- for any target -- the optimal generative parameters for any candidate model $m$. (2) Equipped with optimal parameters for all model $m$, we generate a few network instances from each $m$. (3) Finally, we compare these synthetic network instances with the target using NetDistance metric $D$, developed earlier by the authors. We then use $k$NN to classify the target into one of the candidate classes, i.e., we select the model. 

\runinhead{Other details and comments}  Authors test the performance of their method using six popular network models. Simulations showed that their method outperforms the state-of-the-art methods according to noise tolerance, computational efficiency, and classification accuracy. Finally, authors apply the method to two real networks: email-Eu-core-temporal and p2p-Gnutella, and obtain interesting results. What I like about this paper is that it looks closely into the problem of synthesizing a network via whatever model that is as similar as possible to the target network. This is complicated by the fact that each network model comes with parameters of its own. It is one of the key issues in the network model selection story, and it has been somewhat overlooked in other papers. 
 
\runinhead{Available software}  I could not find any, but I am quite sure we should be able to obtain them from the corresponding author, especially since this group of authors contributed several papers along these lines.


\section{TripletFit: Learning the network distance by Singh et al., 2021 \cite{Singh2021}}

\runinhead{The basic principle}  Authors establish the problem of network model selection as searching for the candidate model which generates synthetic networks most similar to a given target network. They claim that the existing methods for this task suffer from sensitivity to perturbations, dependency on the size, and low accuracy. To remedy this, they propose the method called TripletFit, which relies on deep learning and a vast array of topological measures. Specifically, they train a neural network that learns to differentiate pairs of similar networks from pairs of dissimilar networks. This training yields a new network distance (similarity), according to which, the distance between two similar networks is always smaller than the distance between two dissimilar networks. The method is: (1) Consider candidate models and for each of them create many synthetic network instances of varying size and density. Authors employ 6 popular network models. (2) Calculate the topological measures for each instance. Synthetic network instances in the resulting dataset are represented as topological feature vectors. Each is labeled by its model. (3) Put instances in triplets in a way that each triplet comprises a positive, a negative, and an anchor instance. Positive and anchor instances come from the same candidate model, while the negative instance comes from a different model. Train a neural network to assign distances (similarity scores) to pairs of instances. The training process encourages the neural network to converge towards a metric, whereby the distance between positive and anchor instance is always smaller than the distance between negative and positive or negative and anchor instance. This neural network eventually learns to assign the distances in a way that networks synthesized by the same model are always closer to each other than the networks synthesized by different models. (4) Thus trained neural network lends itself to a network classifier that can classify any input network and hence perform model selection. The target's topological measures are now computed and used to select the model.

\runinhead{Other details and comments}  Authors carry out a systematic analysis of their method's performance. They compare the accuracy of their method with the accuracy of six other methods. Most of them are featured in this book. They also test the performance of all competing methods to noise in the network structure (random rewiring of edges). They find that their method outperforms all contenders in both accuracy and noise-tolerance, while also being independent of size. Finally, authors examine 7 case studies of real-world networks to showcase the practical effectiveness of TripletFit. This is indeed an extra effort compared to many other papers. TripleFit could be the best method in this chapter (or close to). Still, I missed a better discussion on how synthetic network instances are created in step (1), some of which has been done in other papers. I am also somewhat skeptical about training the network distance this way. Coming (or not coming) from the same network model does not guarantee topological similarity (and difference).

\runinhead{Available software} I could not find any, but I am guessing we should be able to obtain it from the authors. I feel it would be quite hard to recreate the entire TripleFit. \\


This ends the reviews of six methods belonging to this chapter. Clearly, their central theme is relying on topological measures and using their values to encode networks via feature vectors. I will conclude this chapter discussing a few points of no consensus among these methods. These points are key to improving them, or even unifying them in relation to methods from the next three chapters, which depart from different frameworks. 

\begin{svgraybox}
\begin{itemize}
    \item  Which topological measures should we consider? Clearly, there are important redundancies among them: some measures are correlated between each other, or at least not perfectly independent. This calls for finding the optimal set of topological measures, characterized by having minimum redundancy possible. In this optimal set, each measure contributes independent information that can not be inferred by knowing the values of other measures. Only then can we consider feature vectors (whose components are the values of topological measures) as good network encoding. My feeling is that there is probably not more than 10 such measures. Such encoding can then be used to e.g. establish a distance (similarity) measure between networks. Naturally, finding this optimal set would have a huge impact on many other problems in network science. Among the above methods, I actually found conflicting results about this (see e.g. \cite{Caceres2016}, \cite{Nagy2022} and \cite{Singh2021}). I am not aware of status of this problem in all of network science, but to my best knowledge it is still open.
    \item  Given a target network, exactly how should we make synthetic networks from some candidate model so that they best resemble the target? Synthetic networks should be as similar to the target as possible, except for being generated by the candidate model. Is is enough to require the same size and density \cite{Motallebi2013}, or should we fix some other topological measures as well? If so, which ones? For example, say that in addition to size and density, we also require that synthetic networks have the same number of triangles as the target. How will this impact the model selection? My intuition is that the optimal set of topological measures should be used to adjust the synthetic networks to the target. 
    \item   Another issue revolves around model parameters. Some models usually have more of them, others less. For what values of model parameters should we look for this resemblance between synthetic networks and the target network? Some authors went as far as designing an artificial neural network which tunes the model parameters in a way to make synthetic networks maximally similar to the target \cite{Attar2017}. It might be useful to fix the model parameters so that, for instance, each candidate model has only one parameter.   
    \item  Lastly, once we have the optimal set of topological measures, and once we extract them from all synthetic networks and the target, how do we use all this to select the model? Should we employ some classification algorithm such as decisions tree \cite{Motallebi2013}, develop a suitable distance metric \cite{Aliakbary2015}, or both? This an important methodological step that should be better researched. The question of consistent network distance (similarity/difference) measure is an open problem in its own right. I know that there is a considerable progress in this direction, but I will not be reviewing it here.
    \item  I am surprised that noone tried to do model selection using unsupervised learning: you simply add the feature vector of the target network to the dataset of feature vectors for all synthetic networks (coming from the candidate models). Then you employ a clustering algorithm on the entire dataset, and you look at the synthetic vectors that the target vector best clusters with. It might be that most of them (or even all of them) come from the same candidate model.   
\end{itemize}
\end{svgraybox}

\chapter{Methods based on encoding networks via graphlets}
\label{encoding-graphlets}

\begin{svgraybox}
\abstract{This is the second of four chapters in which I review the methods for network model selection. The distinctive trait of methods in this chapter is that they rely on examining \textit{graphlets}, and are looking at various statistics describing their presence in the examined networks. Graphlet statistics can be (relatively) easily computed. It can then be used to create feature vectors that encode networks in a way similar to how topological measures are used for this encoding by the methods from the previous chapter. Graphlet feature vectors can now serve as input for various network comparison (similarity) measures and/or diverse ML classifiers to perform network model selection. Five methods (papers) that I will review in this chapter share this framework. I will finish the chapter discussing several open questions regarding this category of methods. The organization of this chapter is the same as of Chapter 3.}
\end{svgraybox}


Many heuristics for network analysis rely on counting small subgraphs (subnetworks) of the studied network and extracting various information for these counts. Given the ubiquity of this approach, researchers started using special names for small subgraphs, one of which is graphlets (another one being \textit{motifs}). Let me define graphlets as small, connected, non-isomorphic subgraphs of the examined network, usually with up to 4 or 5 nodes. Recall that two graphs are called non-isomorphic if there is no way to relabel the nodes of one graph to make it identical to the other graph. A simple exercise reveals that there are 6 such graphlets with up to 4 nodes, 21 graphlets with up to 5 nodes, etc. (of course, all in the context of non-directed and non-weighted networks).

Once we establish the concept of graphlet, we can count graphlets of various sizes and types in the entire network, capture the local neighborhood of the studied node by looking at how many graphlets of different types are touching it, and many other things. One can be even more precise than this, and note that some graphlets -- those whose nodes are not all topologically equivalent -- can `touch' the studied node in different ways \cite{Yaveroglu2014}. By virtue of all this, graphlets can be used for all kinds of ML algorithms, for sophisticated distance measures, and for much more. Below I present five new methods for network model selection that rely on this paradigm. Each method employs the graphlet paradigm in its own interesting way. 

Otherwise, this chapter is organized in the same way the previous one. My review template and all remarks made at the beginning of the previous chapter carry on to this chapter. One could argue that the difference between the methods from this and from the previous chapter is somewhat unclear because graphlet counts can be considered as a special type of topological measure. More about this in the discussion at the end of the chapter.


\section{Model selection via raw graphlet counts by Middendorf et al., 2005 \cite{Middendorf2005}}

\runinhead{The basic principle}  Authors start by claiming that many network models can easily reproduce the structural characteristics of many real networks, such as degree distribution, etc. However, when only one such characteristic is considered and the model parameters are flexible, we get that completely different network models may reproduce any of the considered characteristics. Based on this motivation authors propose a network model selection method that requires no prior choosing of which structural characteristic will be considered. They establish seven candidate network models. Since the authors approach the problem from the angle of biological networks, these seven model are chosen from that context (in 2005). The method begins by defining the target network and proceeds as follows: (1) Generate 1,000 synthetic networks for each candidate model, meaning 7,000 network in total. Each synthetic network has the same size and density as the target network. All other relevant network parameters are sampled uniformly, (2) The topologies of all synthetic networks are quantified via exhaustive counting of graphlets up to a prescribed cut-off. Such cut-off is to be specified via number of vertices, number of edges, or via length of a walk. Authors apply two cut-offs: a walk of the maximal length of eight steps, and, graphlets with at most seven edges. No tests for statistical significance are done. Instead, authors rely on raw graphlet counts. (3) These counts are next fed to a classifier as input features. Authors employ Alternating Decision Tree (ADT) implemented via ensemble ML Adaboost algorithm. Graphlet counts of all synthetic networks serve as the training data to build the classifier. (4) Once the classifier is completed, graphlet counts of any input network determine paths dictated by the inequalities in ADT. Such classifier is then applied to the target network and the winning network model is selected. 

\runinhead{Other details and comments}  Authors apply their method to select the model for the protein–protein interaction network for yeast (\textit{Drosophila melanogaster}) using the best available data at the time. Applying their method they find that duplication–mutation–complementation model fits best. They claim that strong sides of their method include robustness against both noise and data sub-sampling, in addition to absence of any prior assumptions about important network features. I found the method interesting, also since it was one of the first methods for network model selection. I am somewhat surprised that authors did not consider the usual topological measures as features in addition to graphlet counts. One could also discuss non-redundancies in graphlet counts.

\runinhead{Available software}  Their is no software or codes associated with the paper, but from what I understand, the source code should be available from the last author upon request.


\section{Model selection via comparing graphlet degree distributions by Pr\v{z}ulj, 2007 \cite{Przulj2007}}

\runinhead{The basic principle}  The core result of this paper actually is a new heuristics for comparing networks, i.e., a new measure of agreement (or similarity or distance) between two networks, be it empirical or synthetic. Author starts by acknowledging that successful heuristics in network analysis often come from looking at the statistics/counts of graphlets. Expanding on this, author considers graphlets with up to 5 nodes. Local neighborhood of any node can be represented by counting how many different graphlets `touch' this node. In particular, it turns out that we can assign 73 values to a given node in any network by carefully counting the graphlets that this node touches. Now, if we put together such 73-dimensional vectors for all nodes in the examined network, we get 73 graphlet degree distributions (GDD). Each GDD pertains to one graphlet and describes its presence over the entire network. This spectrum of 73 GDD lends itself to a network comparison measure that relies on local network characteristics and it is called GDD agreement. Equipped with GDD agreement, author proposes the following method of network model selection: (1) Consider the target network and compute its full GDD.(2) For each candidate model generate some number of synthetic networks of the same size. The number of links in these synthetic network should at most 1\% off compared to the number of links of the target. Four network models (popular for biological networks) are considered. (3) Compute the GDD agreements between the target and all synthetic networks. Compute the average and the standard deviation of all these GDD agreements. (4) Use this information to select the model whose synthetic networks are on average closest to the target. In other words, leverage the distance (agreement) measure to determine which model generates networks most similar to the target. 

\runinhead{Other details and comments}  The paper is written in the context of analysis of biological networks, specifically protein-protein interaction (PPI) networks. However, from what I can judge, the method is suitable for any network. Author further considers 14 empirical PPI networks, including human, resulting from various high-throughput experimental techniques (remember that this was in 2006 -- newer data has emerged since). The key result is that the model called 3-dimensional Geometric Random Graph has exceptionally high GDD agreement with all of the 14 empirical networks, much better than the other three candidates. I find the method very elegant, although it relies a lot on the proposed distance measure (GDD agreement). There are newer distance measures proposed since, so it would be interesting to see how the selected models depend on the choice of distance measure. A side observation: it seems that graphlet framework took root in the bio-networks community as opposed to physics/computer science communities, which prefer the framework of topological measures. In other words, the culture of the field plays a role in authors' decision on which framework to rely on.

\runinhead{Available software}  The paper itself is not accompanied by any software that I could find. However, author's later publications (some reported here) contain extensive upgrades and improvements of this method, including ready-to-use software packages.


\section{Model selection via topological measures and graphlets by Janssen et al., 2012 \cite{Janssen2012}}

\runinhead{The basic principle}  Authors first state their interpretation of the network model selection problem: Looking for the model that generates graphs most similar to the target. Then they acknowledge that there is no consensus on how to measure `similarity' between networks. Their method extends \cite{Middendorf2005}, albeit it is placed in the context of social (rather than biological) networks. There are three novelties compared to the earlier result: considered networks are larger and denser, another type of decision tree is used, and several new candidate models are considered. The method is the following: (1) We first generate the training data, consisting of 1000 synthetic networks from each of the candidate model. There are six candidate models chosen from social networks literature, which lead to 6000 synthetic networks. The model parameters are chosen at random. Size and density of synthetic networks are kept approximately equal to those of the target. Since the method is based on counting graphlets, which depends heavily on the network's size and density, authors stress that each target requires its own training data. (2) Next, the synthetic network data is represented via features. For features we can choose either of the three: (i) global features, which are values of the topological measures as considered in the previous chapter, (ii) graphlet counts, which authors consider as local features, or (iii) both topological measures and graphlet counts (please check the paper for details on which topological measures are considered and how are the graphlets counted). (3) Regardless what features are chosen, this entire data is now used to build a multiclass alternating decision tree (ADT). Training of ADT leads to a classifier. (4) This classifier is now used to select the model. One extracts the same features from the target and runs them through the classifier. This yields a score for each candidate model corresponding to how well that model fits the target. The model with the highest score is selected. 

\runinhead{Other details and comments}  Authors showcase their method using four social networks (as targets) obtained from Facebook. They find that models based on preferential attachment are the best match for all four targets. Moreover, they test the robustness of their method on perturbations of network structure (changes or rewiring of edges), and find that the method is robust when perturbing up to 5\% – 10\% of edges. Authors develop three versions of the method depending on which features are considered and test their performances. They find that the version relying on graphlets only performs as well as the version based on both feature sets. This is an interesting conclusion, indicating that topological measures are less important for network model selection, at least where ADT classifier is used. I would expect this to (also) depend on which topological measures were (or were not) considered, not to mention redundancies in both topological measures and graphlets. Note that in contrast to the previous paper this method does not rely on network distance (agreement).

\runinhead{Available software}  Authors use Weka to train ADT, but I was unable to find any programming codes or software that one could immediately use.


\section{Network distance from non-redundant graphlets by Yavero\u{g}lu et al., 2014 \cite{Yaveroglu2014}}

\runinhead{The basic principle}  This paper expands the ideas from \cite{Przulj2007} and proposes an improved measure of network similarity (distance). The improvement lies in removing the unwanted redundancies in graphlet counts. These redundancies is easy to observe: consider a node in some network and consider its neighbors. Their number defines the node's degree $k$. Some of these neighboring nodes are connected between them, others are not. Now, say we wish to calculate (i) how many triangles is this node part of (graphlet $C_2$), and (ii) how many paths of length 3 have this node as the middle node (graphlet $C_3$, please refer to \cite{Yaveroglu2014} for graphlet labels). It is easy to spot that knowing the value of degree $k$, $C_2$ and $C_3$ are not completely independent. In fact, $C_3$ can be easily calculated from information about $k$ and $C_2$, meaning that one of them is redundant and should be excluded from graphlet statistics. Proceeding this way further authors derive a system of 17 linear equations describing all such redundancies. After eliminating them, authors show that for graphlets with up to 4 nodes there are 11 non-redundant counts that should be kept. All the rest is redundant. This yields a new Graphlet Degree Vector, now consisting of 11 coordinates (compared to 73 from \cite{Przulj2007} but for graphlets up to 5 nodes). This allows us to encode the structure of any network into a symmetric $11 \times 11$ matrix, called Graphlet Correlation Matrix, which expresses the correlations between the counts of all 11 non-redundant graphlets. So, the improved network distance is the difference between the matrices corresponding to the two compared networks, and is called Graphlet Correlation Distance (GCD-11). Equipped with GCD-11 -- which is clean of redundancies -- authors propose a new method for network model selection: (1) given the target network, generate many synthetic networks from each candidate model. Authors consider 7 popular network models and generate 30 synthetic networks per model. Each synthetic network is of the same size and density as the target. (2) Compute GCD-11 between the target and all synthetic networks. Also, compute all values of GCD-11 between all pairs of synthetic networks, but for each model separately. Store all these values. (3) This leads to two distributions. The first is the distribution of data-vs-model distances (target vs synthetic). The second is the distribution of model-vs-model distances (synthetic vs synthetic). The fit between the target and a candidate model can be quantified by looking at the intersection between the data-vs-model distribution and the model-vs-model distribution for that model. Perfect fit means perfect overlap, no fit means zero overlap. Authors employ a particular non-parametric test to compute the value of this intersection. This value is formalized as goodness-of-fit between the target and a candidate model. (4) The winner model is the one with maximal goodness-of-fit.

\runinhead{Other details and comments}  Authors find that many networks from very different domains are best captured by only three of the candidate models. Paper further uses a huge sample of real networks to test and showcase the method's performance, also for purposes other than network model selection. The method shows excellent robustness to noise and missing data. Method can also pinpoint the structural role of a node in its local neighborhood, as showcased by authors via examining the world trade network. I am very happy to have contributed to this paper. But coming from the context of all the papers reviewed here, I wonder if there are any topological measures that are independent from all graphlets and should have been included in non-redundant encoding of networks? Also, what if non-redundant graphlet statistics were fed to a classifier rather then used for designing a distance measure? What is the best way to generate synthetic networks for some model which are similar to the target?

\runinhead{Available software}  Algorithms and programming implementation of this method (and its earlier and further developments) are available at last author's webpage, e.g. via GraphCrunch software.


\section{Model selection from various distance measures by Ospina-Forero et al., 2018 \cite{Ospina-Forero2018}}

\runinhead{The basic principle}  This paper opens by recognizing that there are many measures of network distance (comparison) in circulation, but it is unclear which of them is best suited for network model fit and network model selection. So, authors propose a method where we fit each candidate network model onto the target relying on a sample of diverse distance measures. In particular, authors consider three distance measures: Graphlet Correlation Distance, Graphlet Degree Distribution Agreement, and Netdis. In addition, they include one measure of network alignment called Netal, which is this case operates as a measure of network distance (see literature for precise difference between network comparison and network alignment). All these measures are based on graphlet statistics. Authors next use a framework based on statistical analysis to evaluate if the target network can be considered as coming from a specific candidate network model. The method is: (1) Take the first candidate model and use it to generate $M$ synthetic networks. For each of these $M$ networks, generate additional $N$ synthetic networks using the same candidate model. Take the first distance measure, and use it to calculate the average distance between the first among $M$ synthetic networks and the $N$ additional networks corresponding to it. Redo this for all $M$ initial networks. This calculation leaves us with $M$ averages called $\bar{S}_1, \bar{S}_2, \hdots \bar{S}_M$. (2) Now generate $N$ brand new synthetic networks from the same candidate model, and for each of them calculate the distance $S$ to the target network, still using the same distance measure. Find the average value of these distances and call it $\bar{S}_0$. (3) Now we compute how well this candidate model fits this target under this distance measure. If $\bar{S}_0$ is the $k$-th value on the ordered sample (with ties broken randomly), then the statistical $p$-value of the test is $p=\frac{k}{M+1}$ (please see the paper for exact details). As usual, we reject the null hypothesis when the $p$-value is small (typically $p \le 0.05$). This will tell us how well (and if at all) the considered candidate model fits the target. (4) The procedure starting from (1) can now be repeated for another candidate model and/or for another network distance measure. We compare the obtained $p$-value to the $p$-value obtained for the first candidate model. This enables us to decide between these two candidates. Repeating this for all candidate model leads to the model with minimal $p$. This model is selected.

\runinhead{Other details and comments} Note that this is not really the problem of model selection (model-vs-model comparison), but rather a problem of model fit (data-vs-model comparison). Also, authors aim primarily to compare the behavior of different network distances under deviations from the null hypothesis (‘$H_0$: Network $G_0$ is a realization the fully specified model $B$’, based on a network comparison $S$). Hence, their main goal is not model selection. Still, I decided to include this paper because it considers three network models (classic ER, Chung-Lu, and a duplication–divergence) and it can be easily adjusted for the purpose of network model selection. The data considered are protein–protein interaction (PPI) networks of Yeast, Fly, Worm, Human, Escherichia Coli (data from October 2015), plus five herpes virus networks and five social networks. Still, I wonder how the performance of this method compares to the traditional model fit approaches from the literature (Stochastic block models, etc.). I also noticed that authors considered only the network distance measures all based on graphlets, despite a huge literature on various network distance measures (based on graphlets or otherwise). I did not catch if and how are synthetic networks related to the target, but I find it interesting that this method does not encode networks via any features. It relies \textit{only} on the distance measure. 

\runinhead{Available software}  I could not find any programming codes published alongside the paper. However, this work largely relies on algorithms published elsewhere (network comparison and alignment methods), so it should not be hard to find them. The rest should be pretty straightforward to replicate. \\


This completes my reviews of five methods for network model selection that are based on graphlets and their statistics. Interestingly, some of them employ distance measures quite heavily, other not at all. I will close the chapter with a few overall remarks about these methods. They are similar to my remarks about the methods in the previous chapter. This is not surprising, since graphlet counts can be seen a specific case of topological measure.

\begin{svgraybox}
\begin{itemize}
    \item The paper that I contributed to \cite{Yaveroglu2014} deals with the problem of non-redundant graphlets. That is to say, a set of graphlets of all which carry independent information, and whose counts can not be predicted from one another. As far as I can tell, other papers in this chapter did not consider the problem of graphlet redundancy. As I reported in the previous chapter, various attempts were made to construct a set of non-redundant topological measures, but (as far as I can see) with inconclusive results. My sense is that for the overall approach of encoding networks via feature vectors, we must finalize this process and construct the optimal set of non-redundant features. This optimal set of features might end up being composed of graphlets, topological measures, or some combination of both. My intuition is definitely a combination of both. The way I would proceed about this is to start from 11 non-redundant graphlets identified in our paper \cite{Yaveroglu2014}, and then check what topological measures we can add to this set while preserving non-redundancy. In other words, we need a list of topological measures (considered in papers from the previous chapter) that are independent from above mentioned 11 non-redundant graphlets. On the other hand, there is a volume of literature on node embedding (encoding) using different approaches such as ML. To my best knowledge none of them have been used for network model selection.     
    \item Both graphlet counts and topological measures allow to build various measures of distance or similarity between networks. The aim of these is to compare networks: two structurally similar networks are less apart than two structurally different networks. Most papers in this chapter employ some form of network distance (e.g. \cite{Przulj2007,Yaveroglu2014,Ospina-Forero2018}). In fact, \cite{Ospina-Forero2018} relies primarily on various comparison measures. However, the authors of this chapter did not consider the entire market of comparison measures that exists out of there. Many of them are not based on network encoding (embedding) via either graphlets or topological measures. As I will show in the next chapter, some distance measures rely on concepts from statistical physics of non-equilibrium systems, and are not coming from computer science literature. To my best knowledge, there is no consistent comparison of the performance of all these distance measures. This I think should be a prime research direction: having a validated and robust distance measure would be a game-changer for many network science problems. I wonder if having such as a distance measure means we no longer need a non-redundant set of features? 
    \item  Rather than developing and/or employing network distance measures, some papers rely on decision trees and other classification algorithms \cite{Middendorf2005,Janssen2012}, even if classification involves graphlet counts as features. Same is found for papers that encode networks via topological features (previous chapter). So, which is better: distance measures or classification algorithms? This I think should be answered by testing and benchmarking the performance of these methods. I wonder if it will make any difference at all. Maybe one can show that in some statistical sense it boils down to the same?  
    \item How to generate synthetic networks from candidate models that are similar to the target? I like to call them surrogate networks (despite slight abuse of the terminology): they should be as similar to the target as possible, except for being synthetic. Just like in the previous chapter, many methods resort to synthetic networks, which opens the question of how to generate them. From what I can tell, papers in this chapter resort to keeping only the size and the density the same as the target. But I feel there is much more one could keep fixed. How about keeping the the graphlet counts fixed, for example, the counts of the 11 non-redundant graphlets? How exactly we can generate a synthetic network from whatever model so that it has a pre-determined count of certain graphlets? I must say I find this task extremely daunting.
    \item All (or most) of the papers from this and the previous chapter come from computer science authors and literature. With that in mind, I find it surprising that very little attention is devoted to computational cost of these methods. Is counting graphlets -- especially those with 4 nodes or more -- more or less computationally expensive than computing the values of (many) topological measures? Which is cheaper: training decision trees on synthetic network data or calculating network distances among the pairs of these synthetic networks? How about the computational costs of generating a sufficient number of synthetic networks? My guess is that improving the efficiency of network model selection should consider this dimension as well.  
\end{itemize}
\end{svgraybox}

\chapter{Methods based on spectral properties of network matrices}
\label{spectral-properties}

\begin{svgraybox}
\abstract{This is the third chapter in which I review the methods for network model selection. As opposed to the methods from two previous chapters, methods from this chapter do not rely on encoding network via feature-vectors based on topological measures or graphlets. Instead, they rely on theoretical calculations involving adjacency matrix and other relevant network matrices. This still leads to a vector-encoding of networks, but now this vector (or even distribution) is the \textit{spectral density or entropy}. The key advantage of this framework is that now we do not need to choose a set of network descriptors (topological measures/graphlets) for encoding, which eliminates the problems associated with (non)redundancies. These network measures are more holistic in the sense that they come from considering the network topology in its entirety. After reviewing four methods (papers) that share this operational framework, I will finish the chapter with a brief discussion of issues pertaining to this category of methods.}
\end{svgraybox}


Some years ago while I still was a physicist at heart I had an argument with a colleague who was a computer scientist at heart. The argument was about which of us -- physicists or computer scientists -- know more mathematics. She argued it was them, because they deal with all those hard algorithmic problems as part of theoretical computer science. I argued it was us, because we deal with all those hard differential equations as part of quantum and statistical mechanics. The dispute was settled in a peaceful way by the intervention of a third colleague, who was a mathematician. He proposed a compromise: computer scientists know more \textit{discrete} mathematics, while physicists are better at \textit{continuous} mathematics. This made sense to me, since indeed each discipline needs its own part of mathematics. 

I have seen this tendency multiple times, including in this book. While the papers in the previous two chapters came from computer science mindset, the four papers reviewed in this chapter largely come from physics approach to network model selection. Rather than utilizing a discrete set of network descriptors as features, papers in this chapter encode networks using distributions. These distributions include spectral density and entropy. They are calculated via operations with adjacency matrix and other relevant network matrices. While spectral density and entropy are technically still realized as vectors, this approach is more holistic in the sense that these distributions are obtained by considering the whole network structure at once. This means we no longer worry about which set of features to choose. Nor we worry about (non)redundancies among them. Of course, this calls for an entirely different mathematical apparatus, one founded on statistical mechanics and information theory. Various versions of this framework is the common thread of all four papers in this chapter.  

On the other hand, an important similarity with the methods from the previous two chapters is in the need to generate instances of synthetic networks using candidate models. This again means we are in the need of a consistent measure of network distance or similarity, which in this chapter comes from Kullback-Leibler and/or Jensen-Shannon divergence. Unfortunately, from what I can tell, no paper in either of these three chapters looked into comparing the performance of feature-based vs. distributions-based methods. This I feel is a crucial direction of future work. Otherwise, this chapter is organized in the same way as the previous two.


\section{Model selection via spectral density by Takahashi et al., 2012  \cite{Takahashi2012}}

\runinhead{The basic principle}  Although this paper is situated in the specific context of biology and neuroscience networks, the proposed method is applicable in any setting. It relies on the concept of spectral density $\rho$ -- the distribution of eigenvalues of network's adjacency matrix. If network is of finite size, this is distribution is of course a vector, but we can still think of it as a distribution (density). This makes $\rho$ into a network's ``fingerprint''. In other words, we can use $\rho$ for encoding (embedding) a network just as we did in the previous two chapters with topological measures and graphlet counts. Authors open the paper by stating that networks generated by the same random process have the same spectral densities $\rho$. This means that we can use it to identify the generating mechanism behind any given network, including a target network considered in the context of network model selection. Authors define Kullback-Leibler (KL) divergence between two spectral densities $\rho_1$ and $\rho_2$ and call it $KL(\rho_1 | \rho_2$). They interpret KL divergence as a distance/comparison measure between two networks. Using three candidate network models, authors propose the following method. (1) compute $\rho$ for the target network. (2) consider now the first candidate model and use it to generate many synthetic networks. They must match the target in parameters that define that model (such as rewiring probability, etc). Let this family of synthetic networks be parameterized by index $\theta$. Now, compute the spectral densities of all these synthetic networks. It is a family of densities called ${\rho_\theta}$. (3) find the specific value of $\theta$ for which $KL(\rho | \rho_\theta)$ is minimal. This value indicates how closely the first candidate model can (at best) explain the target network. (4) repeat steps (2) and (3) for all remaining candidate models. This will give us one minimal value of $KL(\rho | \rho_\theta)$ for each candidate model. The model whose $KL(\rho | \rho_\theta)$ is minimal across all candidates is selected.

\runinhead{Other details and comments} Authors test the performance of their approach as a function of number of nodes. They illustrate its practical use via protein-protein interaction networks of eight different species (Scale-free model wins in all cases). Authors then introduce Jensen-Shannon divergence between $\rho_1$ and $\rho_2$ and use it to identify whether two given network are generated by the same process or not. My first question is: are the spectral density and the KL divergence better framework than topological measures and graphlets? Which network fingerprint is more unique: spectral density or a combination of descriptors considered in the previous two chapter? This is the key to improving network model selection. Furthermore, I am skeptical about the idea that networks generated by the same process always have the same $\rho$. What about small(er) networks, where statistical patterns of the generating process are not (yet) clearly visible? 

\runinhead{Available software}  I was unable to find any software published with the paper. This is bad, since complicated procedure such as above is hard to code and implement from scratch. Still, given good Materials and Methods section, this is not impossible.


\section{Improving the spectral density method by Santos et al., 2021 \cite{Santos2021}}

\runinhead{The basic principle} This is an extension the method above proposed by Takahashi et al. \cite{Takahashi2012}. Authors again identify the spectral density $\rho$ as a unique fingerprint of any network. This means that a suitably defined difference between two spectral densities should be a good proxy for the structural difference between the corresponding two networks, i.e., a comparison (similarity) measure. The key advance of this paper compared to \cite{Takahashi2012} is that now authors use $l_1$ distance instead of KL divergence to compare the distributions. Authors show that with this improvement, empirical spectral density (for many real networks) converges to some limiting distribution. This allows for a more consistent network model selection. The improved method is the following. (1) we start from a set of candidate models $\{ P^i_\theta ; \theta \in \Theta^i \}$. They are parameterized by the index $i=1,\hdots, N$ so that for each $i$ we have a different model ($N$ models in total). For some model $i$, $P^i_\theta$ is the probability distribution of instances of synthetic networks generated by that model. These network instances are parameterized by $\theta$, which summarizes the variable(s) within that model (such as link probability in ER random graph). (2) Now, let $\mathcal{G}$ be the target network described by the distribution $P$ (that may or may not belong to $P^i_\theta$). The goal is to find the model $i$ and the corresponding parameters $\theta$ such that $P^i_\theta$ is as close as possible to $P$. This would select $i$ as the best model for the target $\mathcal{G}$. (3) Next, for each candidate $i$ and each (discretized) $\theta$, we sample $M$ synthetic network instances. We then use kernel density estimators and $l_1$ distance to find the best fit to the target $\mathcal{G}$, taking into account all synthetic network instances. (4) Eventually, this procedure yields not just the selected network model, but also the best fitting synthetic network instance.

\runinhead{Other details and comments} Authors test the performance of their model selection method using 5 standard network models, each defined by one or more parameters ($\theta$). They also compare their new method with \cite{Takahashi2012} and show its advantages. To its credit, this is the only paper that offers precise mathematical arguments, especially regarding the convergence properties of spectral distributions. However, from the practical side I have a few doubts. What makes $l_1$ distance (and KL divergence) better comparison measures than those from the previous two chapters? How are the properties of $P$ influenced by the fact that the target $\mathcal{G}$ is a network of finite size, not necessarily large? More broadly, how can we claim that the spectral density of a specific real network converges to some limiting distribution if that real network is finite? On the other hand, I am not sure I understand how should we generate synthetic (surrogate) networks in this case as opposed to how it was done in other methods. There is no consensus on this. Still, the key advance of this paper is a strong argument that $\rho$ is an excellent fingerprint to encode any network. I really feel that we need to compare it to fingerprinting via topological descriptors from the previous two chapters.

\runinhead{Available software} The paper is accompanied by R codes used for simulations and websites where these codes can be found. I feel implementing this method should not be difficult.


\section{Model selection inspired by quantum mechanics by De Domenico \& Biamonte, 2016 \cite{Domenico2016}}

\runinhead{The basic principle}  Inspired by how entropy is calculated in quantum systems, authors first introduce a matrix $\mathbf{\rho}$ associated with a given network (this $\mathbf{\rho}$ has nothing to do with $\rho$ from the previous two papers). The matrix $\mathbf{\rho}$ is based on network connectivity and it captures its topology. In addition, this matrix is somewhat analogous to the density matrix that is utilized to represents mixed states in quantum mechanics (hence the name). Specifically, it satisfies the mathematical properties of such a matrix: it admits a spectral decomposition and it has trace equal to one. This enables us to define the entropy $S$ for any network as $S(\mathbf{\rho}) = - \mbox{Tr} (\mathbf{\rho} ln \mathbf{\rho})$. To clarify, the matrix $\mathbf{\rho}$ has nothing to do with quantum mechanics, but the logic of constructing and employing it is (almost) identical. Thus defined entropy $S$ allows us to introduce information-theoretic tools such as R\'enyi $q$ entropy, Kullback-Leibler and Jensen-Shannon divergence. Still, we lack the info on how to construct the matrix $\mathbf{\rho}$ itself. Drawing on more analogies, authors define $\mathbf{\rho} =  e^{- \tau \mathbf{L}} / \mbox{Tr} e^{- \tau \mathbf{L}}$. Here, $\mathbf{L} = \mathbf{D} - \mathbf{A}$, where $\mathbf{A}$ is the adjacency matrix and $\mathbf{D}$ is the diagonal matrix of nodes' degrees. It should be stressed that in network science $e^{- \tau \mathbf{L}}$ is the diffusion propagator, operating as function of time $\tau$, and used for studying random walks on networks. I will not go into more mathematics here, but authors eventually formulate their model selection method as follows. (1) Consider the target network described via above matrices. Consider also a set of candidate models $M_1, M_2, \hdots M_n$, whereby each model $M_i$ is described by a certain number of parameters $k_i$. (2) For each model $M_i$ we calculate the likelihood $\mathcal{L}_i$ using the information-theoretic machinery introduced above (more detail in the paper). This likelihood is the probability of observing the target network given a candidate model and its parameters. (3) once we have a likelihood for each candidate model, we use either Akaike information criterion (AIC) or Bayesian information criterion (BIC) to identify the most suitable candidate. This model is selected. It has the best trade-off between divergence from the target and the number of model parameters. 

\runinhead{Other details and comments}  There are other two important by-products of the proposed framework. First is its own network comparison (similarity) measure. The other is a method to fit a proposed network model to a given target (i.e., estimate optimal model parameters for that target). Finally, authors highlight what they see as the main advantage of their approach: it takes as input the network as a whole, not just a set of topological measures or graphlets. I totally agree, although that is true for the previous two papers as well. As another important advantage, I would add that (from what I understand) this method requires no synthetic network instances to be generated from the candidate models. However, it still relies on encoding networks with a quantity of its own, in this case the entropy $S$. This brings us back to the debate on how to best describe a network: via some sort of holistic measure such as in the last three papers or via chosen set of descriptors as done in the previous two chapters? I would very much like to see this comparison, not to mention comparison of the performances of all these methods. Otherwise, this method stunned me with its mathematical complexity.

\runinhead{Available software} I could not find any. This could be a problem, since recreating a procedure this complicated from scratch is (probably) not easy. Moreover, these physics-based algorithms are not easy to find online.


\section{Model selection via ``word space'' by Middendorf et al., 2004 \cite{Middendorf2004}}

\runinhead{The basic principle}  This paper also relies on operations with adjacency matrix, but not in the same way as the three papers above. Still, it best fits in this chapter. Authors stage this work in the context of biological networks. They develop a classifier based on encoding networks as vectors, which is done by expressing their topology via several (or even infinite) coordinates. These coordinates later become features for the classifier. Specifically, authors propose to map networks into a high-dimensional ``word space'', in which the coordinates are ``words''. These words have to do with looking at open and closed walks (of given length) between pairs of nodes. Given any network, we can encode it into a set of words. This encoding is not unlike the encoding via topological measures or graphlets. This encoding leads to a method for network model selection as follows. (1) Start with a primitive alphabet composed of `letters' $\{A; T, U, D \}$. Here, $A$ is just the usual adjacency matrix, while $T$, $U$, and $D$ are various matrix operators (note the similarity with DNA alphabet). This alphabet allows to construct a series of operations of arbitrary length for a given network described by its $A$. Such a series of operations is a word: it is nothing but a series of transformations of network's adjacency matrix. Each word is hence a matrix, and this matrix determines two relevant network characteristics: the number of distinct walks and the number of distinct pairs of nodes (see paper for details). This is how words enable the encoding of networks into vectors. Obviously, such encoding can be done for any network, target or otherwise. (2) Author then use no less than 17 candidate models: Some of them are found in the literature and others are introduced by the authors in this paper (with biology motivation). Each model is used to generate some number of instances of synthetic networks (much in the same way as in other similar papers). (3) Support Vector Machines (SVMs) is employed to build a classifier that discriminates between network generated by different candidate models. SVM is trained on all instances of synthetic networks, which are labeled by a proper class (network model). (4) This classifier can now be used to make the model selection for any target network. 

\runinhead{Other details and comments}  Authors test their method on \textit{E. coli} genetic network, \textit{S. cerevisiae} protein interaction network, and \textit{C. elegans} neuronal network (using then available data). This encoding approach offers other opportunities for network analysis. One of them is a systematic enumeration of network features, which can (at least in principle) be infinite. This is the oldest of all methods I found for network model selection. Interestingly, I could not find any further developments of it. I wonder how this encoding compares to other encoding ideas, both from this chapter and the previous ones. My gut feeling is that it is equivalent to some combination of topological measures and graphlets. Also, I am curious as to why authors use SVM classifier instead of designing a distance measure based on words? To what extent is the method applicable to non-biological networks?

\runinhead{Available software}  Authors say that their source code is written in MatLab and is available at the link provided in the paper. However, clicking on it (at the time of writing this book) I got ``Access forbidden''. I guess the method could be reconstructed from scratch, but not too easily. \\


These were the four methods based on spectral and other properties and operations with adjacency (and other relevant) network matrices. They offer a perspective different, albeit not entirely orthogonal, to the methods from the previous two chapters. This chapter will again end with some general remarks and observations about these methods, confronting them with the methods from previous chapters.

\begin{svgraybox}
\begin{itemize}
    \item  These four methods use completely different ways to encode networks. Instead of choosing which topological measures and/or graphlets to rely on, they employ various matrices (e.g. adjacency) that capture network structure, and use them to derive other quantities of interest. Therefore, one does not have to choose which topological descriptors to include or not to include, since the adjacency matrix by definition represents the entire network structure. There is no discussion about (non)redundancy to be had here. Network encoding is now based on the entire topological information. Such an encoding is indeed more holistic, be it in a form of  distribution/density as in \cite{Takahashi2012,Santos2021,Domenico2016}, or a vector as in \cite{Middendorf2004}. 
    \item Nevertheless, it is not immediately obvious (at least to me) that such encoding necessarily works better than encoding via topological measures and graphlets. What spectral density discussed in \cite{Takahashi2012,Santos2021,Domenico2016} really consists of is $N$ values (e.g. adjacency matrix eigenvalues), where $N$ is the network size. But are all these $N$ values independent, or is there a (non)redundancy discussion to be had here after all? With a bit of imagination, I could ramp up the number of graphlets to that very same $N$ (at least for small to moderate $N$). Naturally, not all of them will ever be non-redundant. But my guess is neither all elements in a spectral density will be non-redundant. So, which encoding is better: the one relying on $N$ graphlets or the one relying on $N$ values that compose the network's spectral density? I would argue that there is a discussion to bed had here. Moreover, to make this discussion complete, one should also consider the encoding ideas from the rest of the complex networks literature, including concepts not related to network model selection.        
    \item  Different ways to encode networks lead to different network comparison (similarity) measures. By the same token, different encoding approaches will play out differently when used to build classifiers. So, should we employ network comparison or classification algorithms for network model selection? How this choice depends on the choice of network encoding approach (spectral distributions vs. topological descriptors)? Papers in this chapter (with exception of \cite{Middendorf2004}) do not employ classification algorithms. My guess is that this is because they are physics-inspired (\cite{Domenico2016} is a good example). However, would considering classifiers (instead of or alongside distance measures) offer any benefits?     
    \item  The method \cite{Domenico2016} does not rely on simultaneously considering all candidate models. Instead, it employs a frameworks similar to hypothesis testing, whereby one tests a single model at the time. The winning model is selected based on some kind of suitability score, computed separately for each candidate model (actually, we have seen similar methods in the previous chapters). This begs an important question: does this framework of one-at-the-time network model selection has benefits over the all-at-once network model selection (such as in e.g. \cite{Middendorf2004})? My guess is that such ordering does not ultimately matter, but I can not be certain. Actually, I will show more examples of methods like these in the next chapter.
\end{itemize}
\end{svgraybox}

\chapter{Fitting methods for mechanistic network models}  \label{fitting-mechanistic}

\begin{svgraybox}
\abstract{This is the last of four chapters devoted to reviewing the methods for network model selection. Two methods reviewed in this chapter are suited specifically for mechanistic network models. These methods do not employ the concept of all-at-once network model selection, whereby one looks for the most suitable model from a list of candidates. Instead, these methods do one-at-the-time network model selection, similar to model inference methods available for static network models such as Stochastic Block Models (SBM). This is an attempt to level the playing field between static and mechanistic models in terms of network model selection. As such, this approach is a conceptual alternative to all methods reviewed in the previous three chapters. I will first review the methods and then close the chapter with a brief discussion.}
\end{svgraybox}


As I discussed at length in chapter 2, basically all network models can be thought of as either mechanistic (formation, growth) models or as static (statistical, probabilistic) models. This distinction shows very clearly from the point of view of network model selection: Static models come with very elaborate inference/fitting methods. Such methods allow us to find, in a very concise way, the best fitting version of a given static model for a given target network (one-at-the-time model selection). The most beautiful example of this are the SBM, but a similar inference scheme can be developed for pretty much any static network model. In contrast, mechanistic network models do not come with an inference framework of the kind, or at least not an immediately obvious one. 

But what if we could develop such an inference technique for at least some mechanistic models? Or maybe find another way of fitting a mechanistic model onto a given target network? This would level the playing field between static and at least some mechanistic models and offer an entirely new approach to network model selection. This would be a one-at-the-time network model selection, much like in the case of SBM, but for mechanistic models. 

In this chapter I review two methods that are based on this logic, which is contrary to most methods from the previous three chapters. One of them is not even a method of network model selection, but a method of converting a mechanistic model into a static model. These methods are, of course, not as developed as those from the previous chapters. Nevertheless, they offer an interesting alternative worth investigating.


\section{Selecting mechanistic network models by Chen et al., 2019 \cite{Chen2019}} 

\runinhead{The basic principle}  Authors start by clarifying that their method is suited specifically for mechanistic network models. Instead of independent candidate models, authors consider `nested' network models. There is an overarching full mechanistic model as the candidate zero, plus a number of its submodels. Full model is specified by a (large) set of microscopic rules that dictate how the network can grow. Each submodel is established by `switching off' one (or more) of such microscopic rules. Switching more microscopic rules off makes the resulting submodel simpler. Now, this entire family of submodels (together with the full model) compete as equal candidates for network model selection. Authors propose a sophisticated procedure based on Machine Learning. In particular, they deploy an ensemble learning approach called Super Learner (SL), which I will not describe here in detail (reader is invited to consults the Figure 2 in the original paper). For now, it is enough to think of it as an algorithm that is trained on network (sub)models. The method itself goes as follows: (1) Choose relevant statistics (features) that highlight the differences between the candidate models. Author do not specify a unique choice that is to be used for all situations. However, in their simulations, authors consider a combination of usual topological measures and measures related to $k$-hop reachability (number of distinct nodes that can be reached in $k$ hops). (2) Once the set of features is established, we generate the training data. Specifically, we generate a certain number of instances of synthetic networks from each candidate model. These synthetic networks are grown until the desired network size. (3) Then we split this training data into cross-validation sets. We train and evaluate each candidate on the corresponding training/validation pair. (4) Next, train the SL on the results from each candidate. (5) Finally, re-train each candidate model on the entire training data. Use thus prepared candidate models to select the winner network (sub)model for the given target network.

\runinhead{Other details and comments}  This method is too intricate to be fully described here, so I again refer the reader to the paper. Authors demonstrate method's usefulness via several versions of ER random graphs and showcase its performance via PPI network of \textit{S. cerevisiae}. I found this `nesting' of mechanistic network models very interesting. I wonder what part of growth models' diversity is covered by this nesting? Are there important growth models left out? This is the only time I have seen SL, so I wonder what benefits it offers over classical ML algorithms. My guess is that this method would not be hugely different with employing the usual ML classifiers instead of SL. At the same time, my understanding is that this method involves no distance/similarity measure. Is this a benefit? The choice of features for training excludes graphlets and spectral distributions from the previous chapter. But it contains $k$-hop reachability, which makes me wonder if $k$-hop reachability can be traced back to graphlets or it offers an alternative to them?

\runinhead{Available software} I could not find any. This could be a problem since recreating a procedure this complicated from scratch is (probably) not easy. I guess one should try contacting the corresponding author.


\section{Converting mechanistic to static network models by Goyal et al., 2023 \cite{Goyal2023}} 

\runinhead{The basic principle}  This paper does not present a new method for network model selection, but I felt important reviewing it anyway. Namely, authors present an interesting framework for converting a mechanistic/growth network model into a static/probabilistic network model. They start by continuing their discussion from \cite{Chen2019} (some authors overlap) about nested mechanistic models, which capture various combinations of microscopic rules of network growth. They observe that for any given set of growth rules $\gamma$, some synthetic network instances are more likely to occur than others. However, it is hard to compute the probability (likelihood) $P_{\mathcal G} (G = g | \gamma)$ of observing a specific network instance $g$ from a set of possible instances $G$ under rules $\gamma$. This is what precludes easy calculation of statistical inferences and model fitting for all mechanistic models. In contrast, static/probabilistic network models are actually defined exactly via such a likelihood (for example SBM), which automatically enable us to make statistical inferences. This justifies the need for this conversion framework: being able to perform statistical inferences and model fitting for mechanistic network models. Authors develop their framework as follows: (1) We learn the joint distribution of the network properties that characterize the mechanistic model in question (please see paper for all details). (2) Next, we use this joint distribution to generate a collection of synthetic networks. We do this in a probabilistic fashion, i.e., using the distribution only. Now we check if this collection of synthetic network is indistinguishable from the synthetic networks generated (in the usual way) by the original mechanistic model. (3) If they are indistinguishable, then this joint distribution can be considered the static/probabilistic equivalent of the original mechanistic model. Once the model conversion is done, we can use the probabilistic equivalent to make statistical inferences and interpret the results as true for the mechanistic original. For example, for a given network we can check if some property is over- or under- represented as compared to the reference model. 

\runinhead{Other details and comments}  Authors test their conversion framework on mechanistic Kretzschmar-Morris model and show that the obtained static model indeed lends itself to statistical inferences. The paper does not really go into the problem of network model selection, but my guess is that with this framework we should be able to do model fitting for mechanistic network models. To repeat again, for this situation I use the name model fitting or one-at-the-time model selection. In opposition to this, I use the name all-at-once model selection for the usual situation involving multiple competing model candidates. That said, I wonder how general can this framework be and what range of mechanistic models can be included in it? The prospect of being able to convert any mechanistic model into an equivalent probabilistic model is fascinating. However, reading this paper I could not grasp whether this approach is readily available for any mechanistic model (e.g. preferential attachment, etc) or only for Kretzschmar-Morris model? Also, how many instances of synthetic networks I need to generate via either probabilistic or mechanistic rules to verify that they are indeed indistinguishable? Static models are supposed to describe network structure as we observe it, while mechanistic models are supposed to capture the network growth over time. Does this still hold after we make the conversion? I feel this a promising research avenue but it needs more work. 

\runinhead{Available software} Again, I could not find any, at least not immediately. This could be a problem since the procedure is not simple. I guess one should try contacting the corresponding author. \\


Above two reviews were the last two paper reviews of this book. As opposed to all methods from the previous three chapters, these two methods (proposed by overlapping groups of authors) are founded on somewhat different premises. First, they apply exclusively to mechanistic network models, under presumption that for static models we can always use well-developed inference techniques. Second, they rely on on-at-the-time network model selection framework without competing candidate models (well, the method \cite{Chen2019} has a competition, but it is among the submodels of the same overarching model). Therefore, these methods make quite an alternative contribution to the problem treated in this book, complementing the methods from the previous three chapters. As before, I will finish the chapter with a few overall remarks.

\begin{svgraybox}
\begin{itemize}
    \item First, the are obvious questions associated with technicalities of each method: which topological measures or features, which distance measure, which ML algorithm(s), how exactly to create synthetic network instances, etc. These are more or less the same as in the previous three chapters, so I will not dive further in them. These are the issues that relate, to one extent or the other, to all methods considered in this book.
    \item More interestingly, we could ask exactly which mechanistic models can be included in this framework? There is a huge variety of those models in the literature with diverse thinking behind them. This begs the question: How easy would be to generalize above methods to other mechanistic models? Is there a way to develop a general framework that would extend \cite{Chen2019} and \cite{Goyal2023} to all mechanistic models? This would mean we have an inference/fitting framework suited for mechanistic models that complements the existing inference framework for static models (such as SBM). These two framework operating in concert would be an interesting counterpart to all methods from the previous three chapters.
    \item Now, say we have a mechanistic model that we wish to fit to a given target network. Is it better to use one of the all-at-once model selection methods from the previous chapters or the framework proposed in \cite{Chen2019} and/or \cite{Goyal2023}? Which works better in practice? Should we use the mechanistic model as it is, or should we first convert it to a probabilistic model and then fit? Is the performance of model fitting via \cite{Goyal2023} of the same precision level as the traditional model inference developed for SBM? These (and many other) questions pertain to comparing the performances. 
    \item  There is an important distinction between mechanistic and static network models from the point of view of network model selection. This is neatly emphasized by these two papers. But what about interpretation of results? What does it mean that we find some mechanistic model to be a good fit for some target network? Is this an insight about target network's current structure or about its growth process up to now? Does this depend on whether we apply the methods from the previous chapters or from this chapter? Fitting any target network to a static model means explaining a snapshot taken at some point in time. In contrast, fitting a target to a mechanistic model means trying to explain its growth/evolution process over time. In other words, how do we interpret the outcome of network model selection? More about this in the next chapter. 
\end{itemize}
\end{svgraybox}

With this summary I conclude the four chapters devoted to reviewing the methods for network model selection. The next chapter is the last one in this book. In it I summarize all the insights (from all four chapters) and discuss their implications.

\chapter{Overview and discussion} \label{Overview} 

\begin{svgraybox}
\abstract{This is the final chapter of this book. It starts with a comprehensive overview of all methods covered in the previous four chapters and a general discussion of their similarities and differences. What I emphasize in particular is the need to test and compare them in a practical applicative setting. Along these lines I will also discuss possible directions of future research. Further in this chapter I briefly review a few alternative approaches to identifying the network models underlying real networks, which are not part of network model selection framework. I will also open the question of how we interpret the outcome of any network model selection. The chapter (and the book) ends with some concluding remarks about things I learned writing this book.}
\end{svgraybox}

This is the final chapter. Let me open it by repeating the main reason I decided to write this book: To trace the research steps towards the \textit{best method} for network model selection. Applying this method to real-work networks should, at least in principle, tell us how they grew over time. The content of this book is the first step in this journey -- a comprehensive review of the state-of-the-art, i.e., a review of existing methods for network model selection. I am assuming that the best method will come from systematically examining the existing methods for the same purpose. In the previous four chapters I reviewed and summarized 17 of them. These are \textit{all} the methods for network model selection that I was able to find in the literature. As I described in the chapter 1, I conducted a very extensive literature search involving many papers available at the time of preparing this book. 

The goal of this last chapter is an overarching discussion of all 17 methods. This discussion is aimed at identifying where exactly should we focus research efforts if we want to find which of the methods work best, or, develop a new method by combining and improving these 17 methods. As I will show, it might be interesting to integrate certain algorithmic steps from one method (or from one category of methods) with the algorithmic steps from another method of another category of methods. This is the added value of having divided the reviews in four categories (corresponding to last four chapters).

To clarify, at the end of each of the previous four chapters I included a short discussion pertaining only to methods from that chapter (i.e., that category). Now, I wish to confront one category against another, extract the similarities and differences among them, and see what we can learn from all that. This will also include ideas about benchmarking the methods across categories and compare their performance and efficiency. Hopefully, this should clarify the steps we must take towards building/designing a single best method.


\section{Overarching discussion}

In this section I will overview, compare, and discuss all four categories of methods for network model selection. For clarity, I will first reiterate the core concepts related to each category, especially the concepts that I will discuss across all categories. 
\begin{enumerate}
    \item  The first category of methods relies on a simple way to encode (embed) a network: Choose a set of topological measures that are easy to compute for any network. These typically include all kinds of centrality measures, versions of assortativity or transitivity, global characteristics such as density, various statistics of shortest paths, etc. Computing the values of all these measures for a given network amounts to encoding that network as a feature vector. This feature vector is nothing but a representation of the network as a point in a metric space. Once we have that point, we can use it to quantify the similarity between a pair of networks, plug it into an ML classifier, into a neural network, etc. Along the way, we usually have to generate some instances of synthetic networks from the candidate network models, keeping some characteristics of the target network fixed. Eventually, we select the most suitable candidate model for the target network.
    \item  The second category is really the same as the first, except that now we rely on a very special form of topological measures called graphlets. Graphlets are small, connected, non-isomorphic graphs. Counting their presence in the studied network (in form of subnetworks) leads to various statistics. We can use these statistics to build another type of encoding into feature vectors: instead of values of topological measures, feature vectors now contain statistics of graphlet counts. And once we have feature vectors we are back in business -- network comparison, ML classification and clustering, neural networks -- it is all at our disposal. As before, we may have to generate samples of synthetic networks from the candidate models. But at any rate, we end up with the best candidate model selected for the target network. I do realize that the first two categories are very similar. However, I prefer to keep them apart, since classical topological measures and graphlets are covered somewhat differently in the literature, particularly the literature reviewed here.  
    \item  Methods in the third category depart from somewhat different foundation. They still employ the idea of network encoding, but not via feature vectors based on topological measures or graphlets. Instead, these methods use network's adjacency matrix to calculate distributions such as spectral density and entropy. Converting a network into a distribution means we are back in business of network model selection. The key novelty is that we no longer need to choose the set of features and worry about the redundancy. Mentioned distributions are computed from the entire network information. So in theory, all structural details of a given network is hidden somewhere in the corresponding distribution, which makes it a more holistic network encoding than the previous ones. This opens the door to network distance/comparison and various classification algorithms via operations such as Kullback-Leibler or Jensen-Shannon divergence. The rest of the overall procedure in this category is fairly similar to the two categories above, so that eventually, the best network model is selected.
    \item  Lastly, the fourth category of methods relies on a premise different altogether. These methods are not really aimed at network model selection, but at model fitting/inference tailored specifically for mechanistic/growth models. Recall that current model inference approaches work for static/probabilistic network models only (e.g. SBM). They are meant to fit a single network model to the target network, i.e., find the most suitable combination of parameters. The methods in the fourth category (previous chapter) seek to extend this framework to mechanistic models, which is not simple. Naturally, they are not as developed as all-at-once model selection approaches from other three categories. Moreover, as best I could tell, such approaches are not yet available for all mechanistic models. Nevertheless, they offer an interesting alternative worth further development.
\end{enumerate}


\begin{svgraybox}
This was a brief overview of all four categories of methods. Let me call it vertical overview. What I will do in the rest of this section is a \textit{horizontal overview}, meaning a discussion that relies a comparison across categories. Each subtitle is devoted to one particular aspect that jointly concerns all categories. I will start with very concrete points and finish with discussing several broader issues. I wish to highlight that there are clear directions of future work where advancements are within reach, keeping in mind that our goal should be to design a single best method and apply it to real-world networks. 
\end{svgraybox}


\runinhead{Network encoding into vectors and distributions}  If you ask me, the most important directions of future work is finding the best way to encode networks as (feature) vectors. The methods so far reviewed involved several ways of doing this: various combinations of topological measures and/or graphlets and various `continuous' distributions such spectral density. But which encoding is the most optimal/parsimonious one? The first issue associated with this is redundancy. This means that taking additional network descriptors (increasing dimensionality of feature vectors) may or may not improve the encoding, since some vector components will (very) likely be correlated. On the other hand, searching for the combination of network descriptors with no redundancy is far from simple. Moreover, to best of my knowledge, no one has looked for redundancies within spectral densities and similar encoding approaches. I might be wrong, but my guess is that these distributions, although `holistic', may not be parsimonious. My intuition is that the best encoding should be a combination of all this: specific topological measures and graphlets together with specific network matrix eigenvalues, i.e., parts of spectral densities. That said, there is a vibrant research on network encoding/embedding \cite{zhang2021,gu2021,rozemberczki2021,dehghan2022,baptista2023,dong2024,milocco2024}, but I am not aware that any of these novel approaches were used for the purposes of network model selection. At any rate, a review paper summarizing all current approaches of network encoding would be highly welcome. Another question is how do we search for the best way of encoding networks? What criteria should an encoding principle fulfill to be the optimal one? My guess is as follows: we should look for a simple encoding with up to 10 components (dimensions of feature vectors). Ideally, those 10 network descriptors should be easy to calculate even for large networks. They should not (necessarily) have a particular interpretation in the context of network analysis (such as node degree, etc). In my view the key criterion should be that this encoding represents a given network as a point in 10-dimensional space, so that this point identifies that network as uniquely as possible. This will still be heuristic and not perfectly parsimonious/non-redundant, but it will simplify any method of network model selection that employs it. Needless to mention, such user-friendly and cheap encoding may also benefit other areas of network science.

\runinhead{Quantifying distance (similarity) between networks}  The next key direction of further work is looking for the most suitable way to compare networks, i.e., express/quantify similarity between two networks, for example, in a form of a distance. Naturally, the distance between two networks should be zero if and only if the two networks are topologically identical. I already introduced several ways to measure network distance, since many reviewed methods rely on them. They typically involve some sort of difference between the two feature vectors/spectral densities corresponding to the two networks (one or both of which are often synthetic). Note that the question of redundancy in network encoding also plays a role when looking for the best network distance. The distance measures from the papers I reviewed come across as rudimentary when confronted with the recent literature, which contains a wide range of proposals, some far more sophisticated \cite{schieber2017,bento2019,bagrow2019,hartle2020,wills2020,felippe2024}. There are even papers comparing their performance, whereby one must clearly specify the criteria used for such a comparison. A particular issue in this regard is whether the compared networks are of the same size or not. In the context of network model selection this problem can be easily avoided, since the concrete procedures can be reduced to comparing only networks of the same size, which is the size of the target. Coming back to feature vectors, once we encode a network as a vector, it is straightforward to use something like Euclidean distance. What I am not sure about is whether we can have meaningful network comparison without network encoding: Can we express distance/similarity between two networks without transforming them any kind of vectors? On the other hand, some of reviewed methods avoid using network distance altogether. They instead resort to Machine Learning and various classification algorithms. Once we have feature vectors (of synthetic networks), each labeled by the candidate model that generated it, it is not hard to train an ML classifier. In my view this begs an interesting question: Is it better to use a distance measure (perhaps in combination with a clustering algorithm) or a classification algorithm? To me it looks more natural to use distance, but this should be decided by examining and the performance of the corresponding methods for network model selection.

\runinhead{Synthetic networks for selecting the model}  Most methods that I reviewed rely on generating instances of synthetic networks from the candidate models. Together with an adequate distance measure, these synthetic instances allow us to select the winner model. This intermediate step, however, opens several dilemmas. First, which characteristics of the target network we should keep when generating these synthetic networks? How the overall ensemble of synthetic network instances depends on what is kept constant? Most methods rely on the simplest choice: keep the network size fixed, i.e., same as the target's. Other methods go a step further and fix the network's density and perhaps other descriptors too. What I wonder about is the following: how close to the target we can get by generating more and more networks from a given model? Ideally, if I can make a synthetic network which is -- under some distance measure -- (almost) identical to the target, this basically means that I found the winner model. Is this possible by keeping many characteristics of the target fixed? Can the literature on the Configuration model and Exponential random graphs help? Let me put it a bit differently: Can I generate a surrogate network, which is identical to the target except that it was generated by a given network model? I am not sure that I am using the term `surrogate' correctly here, but I guess the reader understand what I mean. Actually, there is a volume of research in statistical physics and elsewhere about generating surrogate data (e.g. time series) for all kinds of purposes \cite{schreiber2000,luo2005,lucio2012,lancaster2018,moore2022}. I am not sure to what extent this literature involves networks, except perhaps for the case of null models, which is somewhat similar \cite{hobson2021}. Should synthetic networks always be of the same size as the target? My first guess is that they should, but it might be interesting to see what happens if we make them larger than the target. Obviously, larger networks contain more statistics and particularities of the underlying model and are hence more robust. So, equipped with a distance measure that does not care about network size, I guess we could take this on.

\runinhead{Completing the list of candidate models}  All reviewed methods include candidate network models, which authors use to demonstrate the performance of their method. However, it is not hard to spot that all considered candidates are just the classic textbook models: (preferential) attachment model and many of its variations, forest fire model, duplication-divergence models, etc. The purpose of employing these basic models is to provide a proof-of-concept for a new method of network model selection. On the other hand, we can not expect them to capture the actual growth of a real network. The intuitions on which they are based are routinely oversimplified. For example: Small-world model is a beautiful illustration of how real networks can have high average clustering together with small average distances, but this illustration should not be taken literally when studying real networks. Similar is true for most mechanistic models considered in the reviewed methods. However, recent literature abounds with network models of all kinds -- far beyond the simple textbook ones. I presented and summarized some of them in sections \ref{static-models} and \ref{mechanistic-models}. But the volume of network models in the market is much bigger and I can not review it all in this book. For instance, there are models tailored for very specific science domains and/or intended for very specific situations \cite{kudelka2017,bunimovich2019,beckage2019,villegas2020,molinero2020,arora2022,mcmillan2022}. They tend to be much more realistic in describing the formation of real networks in respective domains. For all I can tell, they can be immediately plugged in as candidates in all (or most of) reviewed methods. Actually, I am keen on seeing how would they perform compared to the usual textbook models. My guess is they would easily win, but I could be wrong, which would be very interesting. Furthermore, there are new results about network models themselves, including the limitations of their ability to synthesize realistic networks \cite{toivonen2009,falkenberg2020,valles2018,valdano2019,airoldi2011,alves2020,arora2020,sikdar2023}. Studying these results should answer which models are worth taking as candidates in the first place. So, as far as the list of candidate models goes, I would argue that a good method of network model selection should: (1) implement most recent network models, in particular those that are specialized for the scientific domain of the target network, and (2) be adjustable, in the sense that new models can always be added to the list. There is no reason to believe that researchers will stop proposing new network models. I think the methods should be flexible to this.

\runinhead{Comparing the performance of methods}  As I stressed many times, the goal of all this is to find or design the \textit{best} method for network model selection. The main contribution of this book is the list in which we should look for it. I am confident that I included all relevant methods, and that I did not overlook any important ones. There are methods which I intentionally did not review, but they are reiterations (to one degree or another) of those that I did review. The next logical step in this research is to compare the performance of all these methods and find the best one. However, it is not immediately obvious how to do this comparison. We first need a testing procedure that can be equally applied to all methods and a criterion according to which we can say which method is better or best. As I see it, the simplest approach is to equip all methods with the same candidate models and take as many of them as possible. Then use each candidate model to generate a large number of synthetic networks. Define these synthetic networks to be our targets. Now, take the first method from the list and use it to select the model for each target. The number of correct selections can be easily obtained, since we know the model underling each target. The number of correct selections is a good measure of that method's performance. Repeating this procedure for all method will yield the method with the highest score. This is, in principle, our best method. To implement this approach it might be necessary to adjust (or slightly modify) some methods. I am curious to know what would applying such a method to real data reveal. However, this idea is not free from caveats. One of them is that it relies on synthetic data, whereby the correct model is always known. This does not tell us much about methods' performance in a realistic setting, where the correct model is not just unknown, but it may not even be among the candidates. For a more robust comparison we should find real networks for which the generating network model is known. Such data are not easy to find, but my guess is that it might be enough to at least know something about the real network's formation mechanisms. I am aware, however, of accumulating datasets on temporal networks, which I think should be close to this. Another caveat has to do with measuring `best-ness' of a method. What if the score for each method can not be simply calculated as I described above? This could occur for many reasons, for example, if several models turn out to be equally suitable for some target. Or if the ability of some methods to make correct selections depends on the amount of ML training, on the simulation run-time, etc. There is also the aspect of computational complexity and cost, which I will discuss in the next paragraph. In other words, it may be somewhat complicated to determine what exactly we mean by one method being `better' than another one. Finally, studying all these methods I feel that the best one might not be on the list, but should be developed by combining elements from different methods. For instance, some methods involve network encoding, but seem to miss the newer insights about how to encode networks efficiently. Other methods rely on measures of network distance, but overlook how network distance is measured in other papers, including papers that are not related to network model selection. I think there is room for improvement here. But this is arguably a harder problem than a mere performance comparison that I just outlined.

\runinhead{Computational cost and other practicalities}  All reviewed methods devote very little attention to computational (algorithmic) complexity. This is relevant because it has to do with the computational cost of running a method. In fact, computational efficiency is an important dimension of comparing the performance of methods. Which methods have the optimal balance between computational cost and quality of the model selection? Is it useful to have a method that is very accurate in selecting models, but it requires huge computing resources and is time consuming to run? This dimension must be taken into account when identifying the best method, which should be accessible even for practitioners with limited resources. Another practical consideration has to do with availability of the programming codes and software used for network model selection. As I discussed earlier, not all authors have been diligent in making their codes (easily) available. An effective comparison of methods should start from implementing all the codes in one platform using some standardized benchmark (e.g. same amount of ML training, same number of synthetic network instances, etc.), so that we can simply feed a target after target into this platform and compare the results. But this requires us to first sort out the problem of codes availability. Let me close this section with the role Artificial Intelligence (AI) could play in this. At the time of finishing this book, anyone hardly talks about anything else. Benchmarking of programming codes (e.g. implementing them all on the same platform and/or in the same programming language) could benefit from AI. Ultimately, network model selection can be seen as a classification problem, whereby candidate models are the classes. Since AI `understands' classification algorithms well, it might be of help here. I am not familiar with how reliable AI really is with tasks of this kind. Nevertheless, I am of the opinion that AI should -- eventually -- play an important role in this research.


\section{Interpretation of results of network model selection}

\begin{svgraybox}
In the previous section I discussed general points related to performance, future improvements, and comparison of methods for network model selection. There is an additional very important point to discuss. It merits an entire section. It regards the way we should understand and interpret the results/outcomes of applying network model selection to any target network.
\end{svgraybox}

Let us imagine that we apply network model selection (whichever method) to some target and select the model. Assume now that the selected model is a mechanistic one. Can we safely argue that the target network was \textit{actually formed} via growth process captured by this model? Let us even assume that the winner model was selected with a great confidence, much better than any other candidate. Does it mean that this model reveals the full history of how the studied target network developed? Or, the act of model selection merely means that the observed structure of the target network is \textit{consistent with} formation via selected model? In other words, the studied target looks as if it grew via selected model, but we can not know for sure whether it actually did. These are two possible interpretations of the results of network model selection. As I see it, they have different scientific values and implications.

Let me make an analogy with cosmology. Trying to explain the Universe as it is (e.g. the observed distribution of galaxies) is much simpler than trying to explain how the Universe got to be the way it is. Obviously, the latter problem is much harder and more ambitious, meaning that any claim in that direction should be substantiated with stronger evidence. Coming back to networks, we must remember that we are applying network model selection only to one snapshot of the target network, captured at a specific instant of its (presumably long) formation/evolution. All reviewed methods work this way and require no information about the history of network's changes. They base their conclusions solely on one snapshot. With this in mind, how should we interpret the results of network model selection: as `target network has \textit{actually formed} via this model' or as `target structure is \textit{consistent with} growth via this model'? 

I do not have a conclusive answer to this dilemma, but I will say the following. The latter interpretation (`\textit{consistent with}') is, in my opinion, always applicable and is far from useless. A demonstrable consistency of the studied network with the networks grown via known formation model is an important result. So, even if we can justify only the latter (weaker) interpretation, the endeavor of network model selection still has scientific merit.

If and when the stronger interpretation (`\textit{actually formed}') is applicable should be examined further and supported by evidence, as I will discuss later in this section. Let me first clarify what does this interpretation mean via simple example. Many real networks studied via empirical data -- as it has been shown in countless studies -- exhibit a hierarchical connectivity of one sort or another, usually revealed as a power-law degree distribution. Such network properties are known to arise when we grow synthetic networks via some kind of preferential attachment. Say that we come up with a preferential attachment model that generates synthetic networks that are basically identical to real networks with a power-law degree distribution. Question: were these real networks \textit{actually formed} via preferential attachment or is it just that their characteristics happen to be \textit{consistent with} preferential attachment growth? As far as I know, the evidence about this is inconclusive, although it might exist for specific cases \cite{smolyarenko2013,ubaldi2016,young2019}. Similar is true for other growth models.

I should stress that in contrast to mechanistic models, model selection involving static network model does not suffer from this problem. Inference/fitting methods, such as SBMs, do not attempt to explain the formation history of the studied network, but only its current structure. This is a less ambitious task, which is why its results are not ambiguous. SBMs are using a snapshot of network's structure exclusively to explain the snapshot's structure itself.

The ambiguity of interpreting the results of network model selection do not end here. Say we find a target to somehow be halfway between two models, meaning either of them can be selected with equal confidence. Is this a reason to believe that both models have contributed (equally or in some hybrid form) to target's formation? Or it means that none of the two models is responsible and that we should look for the third one, perhaps by combining the first two? What if a target is equally close to three or more candidate models? What if network model selection yields no result, in a sense that none of the candidates is selected with a reasonable confidence? This suggests that the correct model was not among the candidates, but may never be able to complete the list of candidate models as to account for all imaginable targets. My impression is that we could encounter these outcomes rather frequently in practice. In reality, it is reasonable to assume that the formation process of a real network is governed by various mechanisms acting simultaneously and changes over time in a noisy way. How would this reality show up in outcomes of applying our methods?

In my mind, the only way to settle this is by testing on real empirical networks for which the formation history and/or the growth mechanism is known. While I admit that I am not familiar with such available data, I am very curious to see how would they play out as targets for our methods. With such data available, we could test many interesting things. For one, check what happens depending on what candidates we include, especially when the correct model is not included. Or examine the confidence with which the correct model is selected against other candidates. Perhaps we could design a sequence of network models ever closer to the correct one and then look at how the confidence of model selection changes. We might also check how the method reacts when different snapshots of the same evolving network are taken as targets: a good method should always select the same model regardless of which of its snapshots is considered. We could also use this scenario to compare the performance of methods: which one will be the fastest or cheapest to select the correct model? These steps should be a neat way not just to calibrate the methods, but to build a robust framework of interpreting the results.


\section{Alternative ideas of inverse network modeling}

\begin{svgraybox}
Inverse network modeling, as I explained earlier, is the problem of identifying the mechanisms governing the formation and evolution of the studied network. Network model selection, which relies on looking for the best network model among several candidates, is one way of doing it. However, there are other ways of doing it, in particular those that do not utilize the competition among candidate models. In addition to them, my literature search revealed several adjacent ideas and concepts that can -- with some adjustments -- be applied for the same purpose. In this section I will briefly review some of them.
\end{svgraybox}

Actually, somewhat to my surprise, I found papers that propose methods for identifying the growth models \textit{directly} from a single network snapshot. The first such method \cite{menezes2014} applies symbolic regression to automatically detect growth models. It employs an ML technique inspired by natural selection, treating network models as computer programs (codes). Similarly, the method \cite{medland2016} uses ML to directly infer growth models for directed networks. Another work \cite{tran2019} starts by transforming the target network into a point cloud and then applies topological data analysis to identify the underlying growth model. Finally, the authors of \cite{xiao2021} use fractal analysis to directly reveal the generating rules behind the target network. As I understand it, above four papers seek to solve the same problem as the methods of network model selection reviewed in the previous chapters. Obviously, they rely on very different principles and start building the model `from scratch'. They involve no candidate models that compete over suitability in describing the target. This immediately solves the issue of choosing the list of candidates. I am curious as to how these four methods complement and/or offer a viable alternative to earlier reviewed methods. I will not review these approaches in detail here, but this is definitely an interesting part of future work.

I also came across three papers \cite{staudt2017,arora2017,attar2020} that deal with the following problem: given a target network, create a single synthetic network that is as similar as possible to the target. This similarity can, as always, be quantified in diverse ways, for instance using a distance measure or via some statistics of the values of topological measures. The obtained synthetic network can be called a \textit{replica}. The point, however, is that if such a replica is indeed very similar to the target, then the process of generating the replica can be seen as the growth model for the target, or at least very close to it. So, if we have a good model that we can iterate and make replicas as similar to the target as we want, this for me is nothing but the correct model for the target. This is interesting, but I must note that it again depends on how we measure the network distance or similarity. The idea can be thought of like this: given a target network characterized by some kind of power-law distribution, can we make an attachment model so specific that it generates networks as similar as we want to it?

Along the same lines, I found a paper \cite{fujita2020} that proposes an approach to check whether two (or more) networks are generated by the same network model. But this means that if I know exactly what model underlies one of the tested networks, and if I find that it is the same model as for the other tested network (target), then I identified the correct model for the target. So, this method too can be utilized for the same purpose as the methods for network model selection.  

Lastly, there are three papers that deal with the question of having multiple (instead of just one) snapshots of the studied target network, one of which I mentioned earlier \cite{young2019}. The second one \cite{wang2012} assumes we have two such snapshots, taken at times $t_1$ and $t_2$, and a few candidate models attempting to describe how did the network evolve from the state at $t_1$ to the state at $t_2$. Authors propose to calculate the likelihood for each candidate and define as the winner the model that maximizes it. Similarly, the paper \cite{reeves2020} presents an interpolation scheme to build a sequence of synthetic networks that captures the transition between two end-point networks (belonging to the target). Again, the mechanism that generates these interpolation networks, for me, is just as good as the evolution model for the target. Clearly, these approaches speak to the question of improving inverse network modeling with information coming from having several snapshots of the target. It would be interesting to combine these approaches with above results. It is obvious that having more snapshots of the target helps, but devising a method that exploits this benefit may be less obvious.


\section{Concluding remarks}

\begin{svgraybox}
Let me close the book with a few final remarks. They refer to what I learned about network science while preparing this book. These remarks are my personal observations and opinions and they might be subjective. Naturally, I welcome any disagreement or criticism. 
\end{svgraybox}

In this book I did my best to consistently and comprehensively review all the methods for network model selection that I could find in the literature. The problem of network model selection (as I defined it) can be considered a small sub-field of network science. I studied all sorts of results in this direction, including those reviewed just above. All this reading made me realize something. Not to make this section too long, I will make just one observation, which I feel is crucial. A careful reader might have made it already.

I can not escape the impression that these methods develop in isolation from each other. I found the network model selection to be a \textit{fragmented} field. Each method relies on an idea of its own. These ideas are brilliant, sometimes genius. But each new method seems to start from scratch rather than pick up where the previous method left. There is no continuity of development. There is no tendency to examine the downsides of the existing methods and seek to improve them. There is some effort to compare a new method to the previous ones, but never in a systematic way. Rather than building on each other, the published results are scattered, with no clear direction. This is in stark contrast with (traditional) scientific fields where results may contradict or extend each other, but in the long run accumulate in a well-established direction and create new knowledge.

This situation is problematic also from the standpoint of an applicative researcher. He/she must navigate this landscape of disconnected methods, with no orientation as to which method is more efficient, effective, or accurate. To make things worse, this applicative researcher may not be fluent in computer programming, ML, or statistics (a social scientist, for example). This goes back to the issue I discussed at the beginning of this book: \textit{we do not need more methods}. If anything, we need a consolidation of the existing methods into a few that do the job, even for a social scientist. I fear that the situation in other parts of network science (e.g. community detection) is similar.

The time is ripe that we take a critical look at the amount of methods we produce and honestly ask if we need them. Network science is a an exciting field, but it might benefit from adjusting its course. Meanwhile, I argue that the best direction for us is towards consolidating the existing methods and applying them to real networks. This book is my modest attempt in that direction.


%
%

\begin{thebibliography}{99.}


\bibitem{Estrada2015}  Estrada, E., Knight, P.A.: A First Course in Network Theory. Oxford University Press, Oxford (2015).

\bibitem{Zweig2016}  Zweig, K. A.: Network Analysis Literacy: A Practical Approach to the Analysis of Networks. Springer, Vienna (2016).

\bibitem{Yang2016}  Yang, S., Keller, F.B., Zheng, L.: Social Network Analysis: Methods and Examples. SAGE Publications, Thousand Oaks (2016).

\bibitem{Latora2017}  Latora, V., Nicosia, V., Russo, G.: Complex Networks: Principles, Methods and Applications. Cambridge University Press, Cambridge (2017).

\bibitem{Newman2018}  Newman, M. E. J.: Networks. 2nd edn. Oxford University Press, Oxford (2018).

\bibitem{Menczer2020} Menczer, F., Fortunato, S., Davis, C. A.: A First Course in Network Science. Cambridge University Press, Cambridge (2020).

\bibitem{Coscia2021}  Coscia, M.: The Atlas for the Aspiring Network Scientist. arXiv:2101.00863 (2021).

\bibitem{Dorogovtsev2022}  Dorogovtsev, S. N., Mendes, J. F. F.: The Nature of Complex Networks. Oxford University Press, Oxford (2022).

\bibitem{Izenman2023}  Izenman, A. J.: Network Models for Data Science: Theory, Algorithms, and Applications. Cambridge University Press, Cambridge (2023).


\bibitem{Hidalgo2016} Hidalgo, C.A.: Disconnected, fragmented, or united? A trans-disciplinary review of network science. Applied Network Science \textbf{1}, 6 (2016)




\bibitem{Levnajic2008} Levnajić, Z., Tadić, B.: Self-organization in Trees and Motifs of Two-dimensional Chaotic Maps with Time Delay. Journal of Statistical Mechanics: Theory and Experiment \textbf{2008}, P03003 (2008)

\bibitem{Levnajic2010a} Levnajić, Z., Tadić, B.: Stability and Chaos in Coupled Two-dimensional Maps on Gene Regulatory Network of Bacterium E. coli. Chaos \textbf{20}, 033115 (2010)

\bibitem{levnajic2010b} Levnajić, Z., Pikovsky, A.: Phase Resetting of Collective Rhythm in Ensembles of Oscillators. Physical Review E \textbf{82}, 056202 (2010)

\bibitem{levnajic2011b} Levnajić, Z.: Emergent Multistability and Frustration in Phase-repulsive Networks of Oscillators. Physical Review E \textbf{84}, 016231 (2011)

\bibitem{faggian2019} Faggian, M., Ginelli, F., Rosas, F., Levnajić, Z.: Synchronization in time-varying random networks with vanishing connectivity. Scientific Reports \textbf{9}, 10207 (2019)

\bibitem{tokuda2019} Tokuda, I., Levnajić, Z., Ishimura, K.: A practical method for estimating coupling functions in complex dynamical systems. Philosophical Transactions of the Royal Society A \textbf{377}, 20190015 (2019)

\bibitem{grau2019a} Grau Leguia, M., Levnajić, Z., Todorovski, L., Ženko, B.: Reconstructing dynamical networks via feature ranking. Chaos \textbf{29}, 093107 (2019)

\bibitem{grau2019b} Grau Leguia, M., Martínez, C.G.B., Malvestio, I., Campo, A.T., Rocamora, R., Levnajić, Z., Andrzejak, R.G.: Inferring directed networks using a rank-based connectivity measure. Physical Review E \textbf{99}, 012319 (2019)

\bibitem{simidjievski2018} Simidjievski, N., Tanevski, J., Ženko, B., Levnajić, Z., Todorovski, L., Džeroski, S.: Decoupling approximation robustly reconstructs directed dynamical networks. New Journal of Physics \textbf{20}, 113003 (2018)

\bibitem{grau2017} Grau Leguia, M., Andrzejak, R.G., Levnajić, Z.: Evolutionary optimization of network reconstruction from derivative-variable correlations. Journal of Physics A: Mathematical and Theoretical \textbf{50}, 334001 (2017)

\bibitem{levnajic2014} Levnajić, Z., Pikovsky, A.: Untangling complex dynamical systems via derivative-variable correlations. Scientific Reports \textbf{4}, 5030 (2014)

\bibitem{levnajic2013} Levnajić, Z.: Derivative-variable correlation reveals the structure of dynamical networks. European Physical Journal B \textbf{86}, 298 (2013)

\bibitem{levnajic2011a} Levnajić, Z., Pikovsky, A.: Network Reconstruction from Random Phase Resetting. Physical Review Letters \textbf{107}, 034101 (2011)

\bibitem{levnajic2012} Levnajić, Z.: Evolutionary Design of Non-frustrated Networks of Phase-repulsive Oscillators. Scientific Reports \textbf{2}, 967 (2012)

\bibitem{crnkic2020} Crnkić, A., Povh, J., Jaćimović, V., Levnajić, Z.: Collective dynamics of phase-repulsive oscillators solves graph coloring problem. Chaos \textbf{30}, 033128 (2020)

\bibitem{tadic2016} Tadić, B., Andjelković, M., Boshkoska, B.M., Levnajić, Z.: Algebraic Topology of Multi-Brain Connectivity Networks Reveals Dissimilarity in Functional Patterns during Spoken Communications. PLoS ONE \textbf{11}, e0166787 (2016)

\bibitem{luzar2014} Lužar, B., Levnajić, Z., Povh, J., Perc, M.: Community Structure and the Evolution of Interdisciplinarity in Slovenia's Scientific Collaboration Network. PLoS ONE \textbf{9}, e94429 (2014)

\bibitem{subelj2015} Šubelj, L., Bajec, M., Kastrin, A., Boshkoska, B.M., Levnajić, Z.: Quantifying the Consistency of Scientific Databases. PLoS ONE \textbf{10}, e0127390 (2015)

\bibitem{Yaveroglu2014}  Yavero\u{g}lu, \"O., Malod-Dognin, N., Davis, D., Levnaji\'c, Z., Janji\'c, V., Karapand\v{z}a, R., Stojmirovi\'c, A., Pr\v{z}ulj, N.: Revealing the hidden language of complex networks. Scientific Reports \textbf{4}, 4547 (2014)

\bibitem{levnajic2010d} Levnajić, Z., Mezić, I.: Ergodic Theory and Visualization. I. Mesochronic Plots for Visualization of Ergodic Partition and Invariant Sets. Chaos \textbf{20}, 033114 (2010)

\bibitem{levnajic2015} Levnajić, Z., Mezić, I.: Ergodic Theory and Visualization. II. Fourier Mesochronic Plots Visualize (Quasi)periodic Sets. Chaos \textbf{25}, 053105 (2015)

\bibitem{levnajic2010c} Levnajić, Z., Prosen, T.: Chaotic Dephasing in a Double-slit Scattering Experiment. Chaos \textbf{20}, 043118 (2010)

\bibitem{Guazzini2015} Guazzini, A., Vilone, D., Donati, C., Nardi, A., Levnajić, Z.: Modeling crowdsourcing as collective problem solving. Scientific Reports \textbf{5}, 16557 (2015)

\bibitem{Ban2017} Ban, K., Perc, M., Levnajić, Z.: Robust clustering of languages across Wikipedia growth. Journal of the Royal Society Open Science \textbf{4}, 171217 (2017)

\bibitem{Guazzini2019} Guazzini, A., Stefanelli, F., Imbimbo, E., Vilone, D., Bagnoli, F., Levnajić, Z.: Humans best judge how much to cooperate when facing hard problems in large groups. Scientific Reports \textbf{9}, 5497 (2019)

\bibitem{Jovic2023} Jović, M., Šubelj, L., Golob, T., Makarovič, M., Yasseri, T., Boberić Krstićev, D., Škrbić, S., Levnajić, Z.: Terrorist attacks sharpen the binary perception of “Us” vs. “Them”. Scientific Reports \textbf{13}, 12451 (2023)

\bibitem{Damij2015} Damij, N., Levnajić, Z., Rejec Skrt, V., Suklan, J.: What motivates us for work? Intricate web of factors beyond money and prestige. PLoS ONE \textbf{10}, e0132641 (2015)

\bibitem{Joksimovic2023} Joksimović, J., Perc, M., Levnajić, Z.: Self-organization in Slovenian public spending. Journal of the Royal Society Open Science \textbf{10}, 221279 (2023)

\bibitem{Damij2023} Damij, N., Levnajić, Z., Modic, D., Suklan, J.: Challenges and Limitations of Pandemic Information Systems: A Literature Review. IEEE Transactions on Engineering Management, doi: 10.1109/TEM.2023.3328894 (2023)

\bibitem{Grote2019} Grote, V., Levnajić, Z., Puff, H., Ohland, T., Goswami, N., Frühwirth, M., Moser, M.: Dynamics of Vagal Activity Due to Surgery and Subsequent Rehabilitation. Frontiers in Neuroscience \textbf{13}, 1116 (2019)

\bibitem{Zorko2020} Zorko, A., Frühwirth, M., Goswami, N., Moser, M., Levnajić, Z.: Heart Rhythm Analyzed via Shapelets Distinguishes Sleep From Awake. Frontiers in Physiology \textbf{10}, 1554 (2020)

\bibitem{metinalista}  Metina lista, ``TOP objave,'' (online). Available: \url{https://metinalista.si/category/znanost-2/top\_objave/}. [Accessed: 28-Jun-2024].




\bibitem{zubarev2024}  Zubarev, V.: Machine Learning for Everyone In simple words. With real-world examples. Yes, again. (Accessed 16 Nov 2024).

\bibitem{shalev2014}  Shalev-Shwartz, S., Ben-David, S.: Understanding Machine Learning: From Theory to Algorithms. Cambridge University Press, Cambridge (2014).

\bibitem{rogers2016} Rogers, S., Girolami, M.: A First Course in Machine Learning. 2nd edn. Chapman and Hall/CRC, Boca Raton (2016).

\bibitem{james2023} James, G., Witten, D., Hastie, T., Tibshirani, R., Taylor, J.: An Introduction to Statistical Learning: with Applications in Python. 1st edn. Springer, Cham (2023).

\bibitem{hastie2009} Hastie, T., Tibshirani, R., Friedman, J.: The Elements of Statistical Learning: Data Mining, Inference, and Prediction. 2nd edn. Springer, New York (2009).





\bibitem{wiki-modelling}  Wikipedia contributors, \textit{``Scientific modelling''}, \textit{Wikipedia, The Free Encyclopedia}, Available at: \url{https://en.wikipedia.org/wiki/Scientific\_modelling} (accessed December 25, 2024).

\bibitem{stanford-encyclopedia}  Frigg, Roman, and Stephan Hartmann, \textit{``Models in Science''}, \textit{The Stanford Encyclopedia of Philosophy (Summer 2009 Edition)}, Edward N. Zalta (ed.), Available at: \url{https://plato.stanford.edu/archives/sum2009/entries/models-science/} (accessed December 25, 2024).

\bibitem{Page2018}  Page, S. E.: The Model Thinker: What You Need to Know to Make Data Work for You. Basic Books, New York (2018).

\bibitem{wiki-wrong}  Wikipedia contributors, \textit{``All models are wrong''}, \textit{Wikipedia, The Free Encyclopedia}, Available at: \url{https://en.wikipedia.org/wiki/All\_models\_are\_wrong} (accessed December 25, 2024).

\bibitem{acebron2005}  Acebrón, J. A., Bonilla, L. L., Pérez Vicente, C. J., Ritort, F., Spigler, R.:  The Kuramoto model: A simple paradigm for synchronization phenomena.  Rev. Mod. Phys. \textbf{77}, 137 (2005).  

\bibitem{rodrigues2016} Rodrigues, F. A., Peron, T. K., Ji, P., Kurths, J.: The Kuramoto model in complex networks. Phys. Rep. \textbf{610}, 1--98 (2016).






\bibitem{barabasi2016book} Barab{\'a}si, A.-L.: Network Science. Cambridge University Press (2016). Available online: \url{https://networksciencebook.com/} (Accessed: January 7, 2025).


\bibitem{goldenberg2010}  Goldenberg, A., Zheng, A. X., Fienberg, S. E., Airoldi, E. M.: A survey of statistical network models. Found. Trends Mach. Learn. \textbf{2}(2), 129--233 (2010).


\bibitem{Bollobas1998}  Bollob\'as, B.: Modern Graph Theory. Graduate Texts in Mathematics, vol. 184. Springer, New York (1998).

\bibitem{Frieze2015}  Frieze, A., Karoński, M.: Introduction to Random Graphs. Cambridge University Press, Cambridge (2015).

\bibitem{Erdos1959}  Erd\H{o}s, P., R\'enyi, A.: On random graphs I. Publ. Math. Debrecen \textbf{6}, 290--297 (1959).

 
\bibitem{michel2019}  Michel, J., Reddy, S., Shah, R., Silwal, S., Movassagh, R.: Directed random geometric graphs. J. Complex Netw. \textbf{7}(5), 792--816 (2019).

\bibitem{fosdick2018}  Fosdick, B.K., Larremore, D.B., Nishimura, J., Ugander, J.: Configuring random graph models with fixed degree sequences. SIAM Rev. \textbf{60}(2), 315--355 (2018).


\bibitem{watts1998}  Watts, D.J., Strogatz, S.H.: Collective dynamics of 'small-world' networks, Nature \textbf{393}, 440–442 (1998).

\bibitem{latora2001}  Latora, V., Marchiori, M.: Efficient behavior of small-world networks. Phys. Rev. Lett. \textbf{87}(19), 198701 (2001).

\bibitem{deSantos2014}  de Santos-Sierra, D., Sendiña-Nadal, I., Leyva, I., Almendral, J.A., Anava, S., Ayali, A., Boccaletti, S.: Emergence of small-world anatomical networks in self-organizing clustered neuronal cultures. PLoS ONE \textbf{9}(1), e85828 (2014).

\bibitem{ferreira2021}  Ferreira, L. N., Hong, I., Rutherford, A., Cebrian, M.: The small-world network of global protests. Sci. Rep. \textbf{11}, 19156 (2021).

\bibitem{humphries2008}  Humphries, M. D., Gurney, K.: Network 'small-world-ness': a quantitative method for determining canonical network equivalence. PLoS ONE \textbf{3}(4), e2051 (2008).

\bibitem{holland1983}  Holland, P.W., Laskey, K.B., Leinhardt, S.: Stochastic blockmodels: First steps. Social Networks \textbf{5}(2), 109–137 (1983).

\bibitem{doreian2004}  Doreian, P., Batagelj, V., Ferligoj, A.: Generalized Blockmodeling. Cambridge University Press, Cambridge (2004).

\bibitem{lee2019}  Lee, C., Wilkinson, D.J.: A review of stochastic block models and extensions for graph clustering. Appl. Netw. Sci. \textbf{4}(1), 122 (2019).

\bibitem{cugmas2020}  Cugmas, M., DeLay, D., Žiberna, A., Ferligoj, A.: Symmetric core-cohesive blockmodel in preschool children's interaction networks. PLoS ONE \textbf{15}(1), e0226801 (2020).

\bibitem{peixoto2014}  Peixoto, T.P.: Hierarchical block structures and high-resolution model selection in large networks. Phys. Rev. X \textbf{4}(1), 011047 (2014).

\bibitem{peixoto2015}  Peixoto, T.P.: Model selection and hypothesis testing for large-scale network models with overlapping groups. Phys. Rev. X \textbf{5}(1), 011033 (2015).

\bibitem{ramirez2022}  Vaca-Ramírez, F., Peixoto, T.P.: Systematic assessment of the quality of fit of the stochastic block model for empirical networks. Phys. Rev. E \textbf{105}(5), 054311 (2022).

\bibitem{snijders2006}  Snijders, T. A. B., Pattison, P .E., Robins, G. L., Handcock, M. S.: New specifications for exponential random graph models. Sociol. Methodol. \textbf{36} (1), 99-153 (2006).

\bibitem{robins2007}  Robins, G., Pattison, P., Kalish, Y., Lusher, D.: An introduction to exponential random graph ($p^*$) models for social networks. Social Networks \textbf{29} (2), 173-191 (2007).

\bibitem{lusher2013}  Lusher, D., Koskinen, J., Robins, G.: Exponential Random Graph Models for Social Networks: Theory, Methods, and Applications. Cambridge University Press, Cambridge (2013).

\bibitem{stivala2021}  Stivala, A., Lomi, A.: Testing biological network motif significance with exponential random graph models. Appl. Netw. Sci. \textbf{6}, 91 (2021).

\bibitem{karkavandi2022}   Karkavandi, M. A., Wang, P., Lusher, D.,  Bastian, B., McKenzie, V., Robins, G.: Perceived friendship network of socially anxious adolescent girls. Social Networks \textbf{68}, 330-345 (2022).

\bibitem{setayesh2022}  Setayesh, A., Zadeh, Z. S. H., Bahrak, B.: Analysis of the global trade network using exponential random graph models. Appl. Netw. Sci. \textbf{7}, 38 (2022).

\bibitem{penrose2003}  Penrose, M.: Random Geometric Graphs. Oxford University Press, Oxford (2003).

\bibitem{dall2002} Dall, J., Christensen, M.: Random geometric graphs. Phys. Rev. E \textbf{66}, 016121 (2002).

\bibitem{fujiwara2011} Fujiwara, N., Kurths, J., Díaz-Guilera, A.: Synchronization in networks of mobile oscillators. Phys. Rev. E \textbf{83}, 025101(R) (2011).

\bibitem{lo2015} Lo, Y.-P., O’Dea, R., Crofts, J.J., Han, C.E., Kaiser, M.: A geometric network model of intrinsic grey-matter connectivity of the human brain. Sci. Rep. \textbf{5}, 15397 (2015).

\bibitem{barthelemy2011} Barthélemy, M.: Spatial networks. Phys. Rep. \textbf{499}, 1–101 (2011).

\bibitem{boguna2021} Boguñá, M., Bonamassa, I., De Domenico, M., Havlin, S., Krioukov, D., Serrano, M.Á.: Network geometry. Nat. Rev. Phys. \textbf{3}, 114–135 (2021).

\bibitem{karrer2011} Karrer, B., Newman, M.E.J.: Stochastic blockmodels and community structure in networks. Phys. Rev. E \textbf{83}, 016107 (2011).

\bibitem{zuev2015} Zuev, K., Boguñá, M., Bianconi, G., Krioukov, D.: Emergence of soft communities from geometric preferential attachment. Sci. Rep. \textbf{5}, 9421 (2015).

\bibitem{cherifi2019} Cherifi, H., Palla, G., Szymanski, B.K., Lu, X.: On community structure in complex networks: challenges and opportunities. Appl. Netw. Sci. \textbf{4}, 117 (2019).

\bibitem{stadtfeld2020} Stadtfeld, C., Takács, K., Vörös, Á.: The emergence and stability of groups in social networks. Soc. Netw. \textbf{60}, 129–145 (2020).

\bibitem{lancichinetti2008} Lancichinetti, A., Fortunato, S., Radicchi, F.: Benchmark graphs for testing community detection algorithms. Phys. Rev. E \textbf{78}, 046110 (2008).

\bibitem{borgatti2000} Borgatti, S.P., Everett, M.G.: Models of core/periphery structures. Soc. Netw. \textbf{21}(4), 375–395 (2000).

\bibitem{verma2016} Verma, T., Russmann, F., Araújo, N.A.M., Nagler, J., Herrmann, H.J.: Emergence of core–peripheries in networks. Nat. Commun. \textbf{7}, 10441 (2016).

\bibitem{csigi2017} Csigi, M., Kőrösi, A., Bíró, J., Heszberger, Z., Malkov, Y., Gulyás, A.: Geometric explanation of the rich-club phenomenon in complex networks. Sci. Rep. \textbf{7}, 1730 (2017).

\bibitem{hebertdufresne2013} Hébert-Dufresne, L., Allard, A., Young, J.-G., Dubé, L.J.: Percolation on random networks with arbitrary $k$-core structure. Phys. Rev. E \textbf{88}, 062820 (2013).

\bibitem{sendina2016} Sendiña-Nadal, I., Danziger, M.M., Wang, Z., Havlin, S., Boccaletti, S.: Assortativity and leadership emerge from anti-preferential attachment in heterogeneous networks. Sci. Rep. \textbf{6}, 21297 (2016).

\bibitem{toivonen2009}  Toivonen, R., Kovanen, L., Kivelä, M., Onnela, J.-P., Saramäki, J., Kaski, K.: A comparative study of social network models: Network evolution models and nodal attribute models. Soc. Netw. \textbf{31}(4), 240–254 (2009).

\bibitem{barabasi1999} Barabási, A.-L., Albert, R.: Emergence of scaling in random networks. Science \textbf{286}(5439), 509–512 (1999).

\bibitem{albert2002} Albert, R., Barabási, A.-L.: Statistical mechanics of complex networks. Rev. Mod. Phys. \textbf{74}, 47–97 (2002).

\bibitem{dorogovtsev2000} Dorogovtsev, S.N., Mendes, J.F.F., Samukhin, A.N.: Structure of growing networks with preferential linking. Phys. Rev. Lett. \textbf{85}, 4633–4636 (2000).

\bibitem{price1976} Price, D.J.D.S.: A general theory of bibliometric and other cumulative advantage processes. J. Am. Soc. Inf. Sci. \textbf{27}(5), 292–306 (1976).

\bibitem{broido2019} Broido, A.D., Clauset, A.: Scale-free networks are rare. Nat. Commun. \textbf{10}, 1017 (2019).

\bibitem{barabasi2018blog} Barabási, A.-L.: Love is all you need. Clauset's fruitless search for scale-free networks. Blog post (2018).

\bibitem{voitalov2019} Voitalov, I., van der Hoorn, P., van der Hofstad, R., Krioukov, D.: Scale-free networks well done. Phys. Rev. Res. \textbf{1}, 033034 (2019).

\bibitem{jacomy2020} Jacomy, M.: Epistemic clashes in network science: Mapping the tensions between idiographic and nomothetic subcultures. Big Data Soc. \textbf{7}(2), 2053951720949577 (2020).

\bibitem{krapivsky2000} Krapivsky, P.L., Redner, S., Leyvraz, F.: Connectivity of growing random networks. Phys. Rev. Lett. \textbf{85}(21), 4629–4632 (2000).

\bibitem{bianconi2001} Bianconi, G., Barabási, A.-L.: Competition and multiscaling in evolving networks. Europhys. Lett. \textbf{54}(4), 436–442 (2001).

\bibitem{fortunato2006} Fortunato, S., Flammini, A., Menczer, F.: Scale-free network growth by ranking. Phys. Rev. Lett. \textbf{96}(21), 218701 (2006).

\bibitem{klemm2002} Klemm, K., Eguíluz, V. M.: Highly clustered scale-free networks. Phys. Rev. E \textbf{65}(3), 036123 (2002).

\bibitem{klimm2014} Klimm, F., Bassett, D.S., Carlson, J.M., Mucha, P.J.: Resolving structural variability in network models and the brain. PLoS Comput. Biol. \textbf{10}(3), e1003491 (2014).

\bibitem{papadopoulos2015} Papadopoulos, F., Psomas, C., Krioukov, D.: Network mapping by replaying hyperbolic growth. IEEE/ACM Trans. Netw. \textbf{23}(1), 198–211 (2015).


\bibitem{vazquez2003} Vázquez, A.: Growing network with local rules: Preferential attachment, clustering hierarchy, and degree correlations. Phys. Rev. E \textbf{67}, 056104 (2003).

\bibitem{saramaki2004} Saramäki, J., Kaski, K.: Scale-free networks generated by random walkers. Physica A \textbf{341}(1-4), 80-86 (2004).

\bibitem{kleinberg1999} Kleinberg, J. M., Kumar, R., Raghavan, P., Rajagopalan, S., Tomkins, A. S.: The Web as a graph: Measurements, models, and methods. Lecture Notes in Computer Science, vol. 1627, pp. 1-17. Springer, Heidelberg (1999).

\bibitem{krapivsky2005} Krapivsky, P.L., Redner, S.: Network growth by copying. Phys. Rev. E 71, 036118 (2005).

\bibitem{wagner1994}  Wagner, G. P.: Evolution of gene networks by gene duplications: a mathematical model and its implications on genome organization. Proc. Natl. Acad. Sci. U. S. A. 91, 4387–4391 (1994). 

\bibitem{sole2002}  Solé, R. V., Pastor-Satorras, R., Smith, E., Kepler, T. B.: A model of large‐scale proteome evolution. Adv. Complex Syst. 5, 43-54 (2002).

\bibitem{bhan2002}  Bhan, A., Galas, D.J., Dewey, T.G.: A duplication growth model of gene expression networks. Bioinformatics 18(11), 1486-1493 (2002).

\bibitem{vazquezFlammini2003}  Vázquez, A., Flammini, A., Maritan, A., Vespignani, A.: Modeling of protein interaction networks. ComPlexUs 1(1), 38-44 (2003).


\bibitem{leskovec2005}  Leskovec, J., Kleinberg, J., Faloutsos, C.: Graphs over time: Densification laws, shrinking diameters and possible explanations. KDD '05: Proceedings of the eleventh ACM SIGKDD international conference on Knowledge discovery in data mining. 177-187 (2005).

\bibitem{bancal2010}  Bancal, J. D., Pastor-Satorras, R.: Steady-state dynamics of the forest fire model on complex networks. Eur. Phys. J. B 76, 109-121 (2010).


\bibitem{leskovec2010} Leskovec, J., Chakrabarti, D., Kleinberg, J., Faloutsos, C., Ghahramani, Z.: Kronecker graphs: An approach to modeling networks. J. Mach. Learn. Res. \textbf{11}, 985-1042 (2010).

\bibitem{seshadhri2013}  Seshadhri, C., Pinar, A., Kolda, T. G.: An in-depth analysis of stochastic Kronecker graphs. J. of the ACM \textbf{60}, 2, 1-32 (2013). 


\bibitem{xie2008} Xie, Y.-B., Zhou, T., Wang, B.-H.: Scale‑free networks without growth. Physica A \textbf{387}, 1683–1688 (2008).

\bibitem{akoglu2009} Akoglu, L., Faloutsos, C.: RTG: A recursive realistic graph generator using random typing. In: Proceedings of ECML‑PKDD 2009, LNCS, vol. 5782, pp. 13–28 (2009). 

\bibitem{smolyarenko2013} Smolyarenko, I. E., Hoppe, K., Rodgers, G. J.: Network growth model with intrinsic vertex fitness. Phys. Rev. E \textbf{88}, 012805 (2013). 

\bibitem{morzy2019} Morzy, M., et al.: Priority attachment: a comprehensive mechanism for generating networks. Sci. Rep. \textbf{9}, 3383 (2019). 

\bibitem{falkenberg2020} Falkenberg, M., et al.: Identifying time dependence in network growth. Phys. Rev. Research \textbf{2}, 023352 (2020). 

\bibitem{levens2022} Levens, W. J., Szorkovszky, A., Sumpter, D. J. T.: Friend of a friend models of network growth. R. Soc. Open Sci. \textbf{9}, 221200 (2022). 

\bibitem{kharel2022} Kharel, S. R., et al.: Degree‑preserving network growth. Nat. Phys. \textbf{18}, 100–106 (2022). 

\bibitem{przulj2006} Pr\v zulj, N., Higham, D. J.: Modelling protein–protein interaction networks via a stickiness index. J. R. Soc. Interface \textbf{3}, 711–716 (2006).


\bibitem{betzel2016}  Betzel, R. F., et al.: Generative models of the human connectome. NeuroImage \textbf{124}, 1054–1064 (2016). 

\bibitem{antal2005}  Antal, T., Krapivsky, P. L., Redner, S.: Dynamics of social balance on networks. Phys. Rev. E \textbf{72}, 036121 (2005). 

\bibitem{talaga2020}  Talaga, S., Nowak, A.: Homophily as a process generating social networks: Insights from social distance attachment model. J. Artif. Soc. Soc. Simul. \textbf{23}(2), 6 (2020).

\bibitem{mattsson2025}   Mattsson, C. E. S.: Network growth under opportunistic attachment. Appl. Netw. Sci. \textbf{10}, 21 (2025).

\bibitem{zhang2015}  Zhang, X., et al.: A biologically inspired network design model. Sci. Rep. \textbf{5}, 10794 (2015).


\bibitem{piva2021}  Piva, G., Ribeiro, F. L., Mata, A. S.: Networks with growth and preferential attachment: Modeling and applications. J. Complex Netw. \textbf{9}(1), cnab008 (2021). 


\bibitem{duvivier2022}  Duvivier, L., Cazabet, R., Robardet, C.: Graph space: using both geometric and probabilistic structure to evaluate statistical graph models. J. Complex Networks \textbf{10}(2), cnac006 (2022).

\bibitem{pham2015}  Pham, T., Sheridan, P., Shimodaira, H.: PAFit: A Statistical Method for Measuring Preferential Attachment in Temporal Complex Networks. PLoS ONE \textbf{10}(9), e0137796 (2015).

\bibitem{pham2021}  Pham, T., Sheridan, P., Shimodaira, H.: Non-parametric estimation of the preferential attachment function from one network snapshot. J. Complex Networks \textbf{9}(5), cnab024 (2021). 

\bibitem{tsiotas2019}  Tsiotas, D.: Detecting different topologies immanent in scale-free networks with the same degree distribution. Proc. Natl. Acad. Sci. U.S.A. \textbf{116}(14), 6701–6706 (2019). 

\bibitem{arnold2021}  Arnold, N. A., Mondragón, R. J., Clegg, R. G.: Likelihood-based approach to discriminate mixtures of network models that vary in time. Sci. Rep. \textbf{11}, 5205 (2021).


\bibitem{burnham2002}  Burnham, K. P., Anderson, D. R.: Model Selection and Multimodel Inference: A Practical Information-Theoretic Approach, 2nd edn. Springer, New York (2002). 

\bibitem{kadane2004}  Kadane, J. B., Lazar, N. A.: Methods and Criteria for Model Selection. J. Amer. Statist. Assoc. \textbf{99}(465), (2004). 

\bibitem{nannen2010}  Nannen, V.: A Short Introduction to Model Selection, Kolmogorov Complexity and Minimum Description Length (MDL). arXiv:1005.2364 (2010). 

\bibitem{ding2018}  Ding, J., Tarokh, V., Yang, Y.: Model Selection Techniques: An Overview. IEEE Signal Process. Mag. \textbf{35}(6), (2018). 






\bibitem{Motallebi2013} Motallebi, S., Aliakbary, S. Habibi, J.: Generative model selection using a scalable and size-independent complex network classifier. Chaos \textbf{23}, 043127 (2013)

\bibitem{Caceres2016} Caceres, R. S., Weiner, L., Schmidt, M. C., Miller, B. A., Campbell, W. M.: Model Selection Framework for Graph-based data. Available via arXiv. \url{https://arxiv.org/abs/1609.04859} (2016)

\bibitem{Nagy2022} Nagy, M., Molontay, R.: Network classification-based structural analysis of real networks and their model-generated counterparts. Network Science \textbf{10}, 2, 146-169 (2022)

\bibitem{Aliakbary2015}  Aliakbary, S., Motallebi, S., Rashidian, S., Habibi, J., Movaghar, A.: Noise-tolerant model selection and parameter estimation for complex networks. Physica A: Statistical Mechanics and its Applications \textbf{427}, 100-112 (2015)

\bibitem{Attar2017}  Attar, N., Aliakbary, S.:  Classification of complex networks based on similarity of topological network features. Chaos \textbf{27}, 091102 (2017)

\bibitem{Singh2021}  Singh, K. V., Verma, A. K., Vig, L.: Deep learning based network similarity for model selection. Data Science \textbf{4}, 63-83 (2021)

\bibitem{Middendorf2005} Middendorf, M., Ziv, E., Wiggins, C. H.: Inferring network mechanisms: The Drosophila melanogaster protein interaction network. PNAS \textbf{102}, 3192-3197 (2005) 

\bibitem{Przulj2007} Pr\v{z}ulj, N.: Biological network comparison using graphlet degree distribution. Vol \textbf{23} ECCB 2006, e177 (2007)

\bibitem{Janssen2012} Janssen, J., Hurshman, M., Kalyaniwalla, N.: Model Selection for Social Networks Using Graphlets. Internet Mathematics \textbf{8}, 338-363 (2012)

\bibitem{Ospina-Forero2018} Ospina-Forero, L., Deane, C. M., Reinert, G.: Assessment of model fit via network comparison methods based on subgraph counts. Journal of Complex Networks \textbf{7}, 2, 226-253 (2018)

\bibitem{Takahashi2012} Takahashi, D. Y., Sato J. R., Ferreira C. E., Fujita, A.: Discriminating Different Classes of Biological Networks by Analyzing the Graphs Spectra Distribution. PLoS ONE \textbf{7}, 12, e49949 (2012)

\bibitem{Santos2021} Santos, S. Fujita, A., Matias, C.: Spectral density of random graphs: convergence properties and application in model fitting. Journal of Complex Networks \textbf{9}, 6, cnab041 (2021)

\bibitem{Domenico2016} De Domenico, M., Biamonte, J.: Spectral Entropies as Information-Theoretic Tools for Complex Network Comparison. Physical Review X \textbf{6}, 041062 (2016)

\bibitem{Middendorf2004} Middendorf, M., Ziv, E., Adams, C., Hom, J., Koytcheff, R., Levovitz, C., Woods, G., Chen, L., Wiggins, C.: Discriminative topological features reveal biological network mechanisms. BMC Bioinformatics \textbf{5}, 181 (2004)

\bibitem{Chen2019} Chen, S., Mira, A., Onnela, J. P.: Flexible model selection for mechanistic network models. Journal of Complex Networks \textbf{8}, 2, (2019)

\bibitem{Goyal2023} Goyal, R., De Gruttola, V., Onnela, J. P.: Framework for converting mechanistic network models to probabilistic models. Journal of Complex Networks \textbf{11}, 5, cnad034 (2023)






\bibitem{zhang2021} Zhang, Y.-J., Yang, K.-C., Radicchi, F.: Systematic comparison of graph embedding methods in practical tasks. Phys. Rev. E \textbf{104}, 044315 (2021). 

\bibitem{gu2021} Gu, W., Tandon, A., Ahn, Y.-Y., Radicchi, F.: Principled approach to the selection of the embedding dimension of networks. Nat. Commun. \textbf{12}, 3772 (2021).

\bibitem{rozemberczki2021} Rozemberczki, B., Allen, C., Sarkar, R.: Multi-scale attributed node embedding. J. Complex Netw. \textbf{9}(2), cnab014 (2021).

\bibitem{dehghan2022} Dehghan-Kooshkghazi, A., Kamiński, B., Kraiński, Ł., Prałat, P., Théberge, F.: Evaluating node embeddings of complex networks. J. Complex Netw. \textbf{10}(4), cnac030 (2022).

\bibitem{baptista2023} Baptista, A., Sánchez-García, R. J., Baudot, A., Bianconi, G.: Zoo guide to network embedding. J. Phys. Complex. \textbf{4}, 042001 (2023).

\bibitem{dong2024} Dong, T., Sun, Y., Liang, F.: Deep network embedding with dimension selection. Neural Netw. \textbf{179}, 106512 (2024).

\bibitem{milocco2024} Milocco, R., Jansen, F., Garlaschelli, D.: Multi-scale node embeddings for graph modeling and generation. arXiv preprint arXiv:2412.04354 (2024).



\bibitem{schieber2017} Schieber, T. A., Carpi, L., Díaz-Guilera, A., Pardalos, P. M., Masoller, C., Ravetti, M. G.: Quantification of network structural dissimilarities. Nat. Commun. \textbf{8}, 13928 (2017). 

\bibitem{bento2019} Bento, J., Ioannidis, S.: A family of tractable graph metrics. Appl. Netw. Sci. \textbf{4}, 107 (2019). 

\bibitem{bagrow2019} Bagrow, J. P., Bollt, E. M.: An information-theoretic, all-scales approach to comparing networks. Appl. Netw. Sci. \textbf{4}, 45 (2019).

\bibitem{hartle2020} Hartle, H., Klein, B., McCabe, S., Daniels, A., St-Onge, G., Murphy, C., Hébert-Dufresne, L.: Network comparison and the within-ensemble graph distance. Proc. R. Soc. A \textbf{476}(2243), 20190744 (2020). 

\bibitem{wills2020} Wills, P., Meyer, F. G.: Metrics for graph comparison: A practitioner’s guide. PLoS One \textbf{15}(2), e0228728 (2020).

\bibitem{felippe2024} Felippe, H., Battiston, F., Kirkley, A.: Network mutual information measures for graph similarity. Commun. Phys. \textbf{7}, 335 (2024).



\bibitem{schreiber2000} Schreiber, T., Schmitz, A.: Surrogate time series. Physica D \textbf{142}(3–4), 346–382 (2000).

\bibitem{luo2005} Luo, X., Nakamura, T., Small, M.: Surrogate test to distinguish between chaotic and pseudoperiodic time series. Phys. Rev. E \textbf{71}, 026230 (2005).

\bibitem{lucio2012} Lucio, J. H., Valdés, R., Rodríguez, L. R.: Improvements to surrogate data methods for nonstationary time series. Phys. Rev. E \textbf{85}, 056202 (2012).

\bibitem{lancaster2018} Lancaster, G., Iatsenko, D., Pidde, A., Ticcinelli, V., Stefanovska, A.: Surrogate data for hypothesis testing of physical systems. Phys. Rep. \textbf{748}, 1–60 (2018).

\bibitem{moore2022} Moore, J. M., Yan, G., Altmann, E. G.: Nonparametric power-law surrogates. Phys. Rev. X \textbf{12}, 021056 (2022). 

\bibitem{hobson2021} Hobson, E. A., Silk, M. J., Fefferman, N. H., Larremore, D. B., Rombach, P., Shai, S., Pinter-Wollman, N.: A guide to choosing and implementing reference models for social network analysis. Biol. Rev. \textbf{96}(6), 2716--2734 (2021).



\bibitem{kudelka2017} Kudelka, M., Ochodkova, E., Zehnalova, S.: Around average behavior: 3-lambda network model. arXiv preprint arXiv:1701.01274 (2017).

\bibitem{bunimovich2019} Bunimovich, L. A., Smith, D. C., Webb, B. Z.: Specialization models of network growth. J. Complex Netw. \textbf{7}(3), 375--392 (2019).

\bibitem{beckage2019} Beckage, N. M., Colunga, E.: Network growth modeling to capture individual lexical learning. Hindawi Complexity \textbf{2019}, 7690869 (2019).

\bibitem{villegas2020} Villegas, P., Mu\~{n}oz, M. A., Bonachela, J. A.: Evolution in the Debian GNU/Linux software network: analogies and differences with gene regulatory networks. J. R. Soc. Interface \textbf{17}(163), 20190845 (2020).

\bibitem{molinero2020} Molinero, C., Hernando, A.: A model for the generation of road networks. arXiv preprint arXiv:2001.08180 (2020).

\bibitem{arora2022} Arora, V., Amico, E., Go\~{n}i, J., Ventresca, M.: Investigating cognitive ability using action-based models of structural brain networks. J. Complex Netw. \textbf{10}(4), cnac037 (2022).

\bibitem{mcmillan2022} McMillan, C., Kreager, D. A., Veenstra, R.: Keeping to the code: How local norms of friendship and dating inform macro-structures of adolescents’ romantic networks. Soc. Netw. \textbf{70}, 126--137 (2022).



\bibitem{valles2018} Vallès‑Català, T., Peixoto, T. P., Sales‑Pardo, M., Guimerà, R.: Consistencies and inconsistencies between model selection and link prediction in networks. Phys. Rev. E \textbf{97}, 062316 (2018).

\bibitem{valdano2019} Valdano, E., Arenas, A.: Exact rank-reduction of network models. Phys. Rev. X \textbf{9}, 031050 (2019).

\bibitem{airoldi2011} Airoldi, E. M., Bai, X., Carley, K. M.: Network sampling and classification: An investigation of network model representations. Decis. Support Syst. \textbf{51}(3), 506--518 (2011).

\bibitem{alves2020} Alves, L. G. A., Aleta, A., Rodrigues, F. A., Moreno, Y., Amaral, L. A. N.: Centrality anomalies in complex networks as a result of model over-simplification. New J. Phys. \textbf{22}, 013043 (2020).

\bibitem{arora2020} Arora, V., Guo, D., Dunbar, K. D., Ventresca, M.: Examining the variability in network populations and its role in generative models. Network Sci. \textbf{8}(S1), 43-64 (2020). 

\bibitem{sikdar2023} Sikdar, S., Cedre, D. G., Ford, T. W., Weninger, T.: The infinity mirror test for graph models. IEEE Trans. Knowl. Data Eng. \textbf{35}(4), 4281--4292 (2023).



\bibitem{ubaldi2016} Ubaldi, E., Perra, N., Karsai, M., Vezzani, A., Burioni, R., Vespignani, A.: Asymptotic theory of time‑varying social networks with heterogeneous activity and tie allocation. Sci. Rep. \textbf{6}, 35724 (2016). https://doi.org/10.1038/srep35724

\bibitem{young2019} Young, J.-G., St‑Onge, G., Laurence, E., Murphy, C., Hébert‑Dufresne, L., Desrosiers, P.: Phase transition in the recoverability of network history. Phys. Rev. X \textbf{9}, 041056 (2019). 



\bibitem{menezes2014} Menezes, T., Roth, C.: Symbolic regression of generative network models. Sci. Rep. \textbf{4}, 6284 (2014).

\bibitem{medland2016} Medland, M.R., Harrison, K.R., Ombuki‑Berman, B.M.: Automatic inference of graph models for directed complex networks using genetic programming. In: 2016 IEEE Congress on Evolutionary Computation (CEC), pp. 2337–2344. IEEE (2016). 

\bibitem{tran2019} Tran, Q.H., Vo, V.T., Hasegawa, Y.: Scale‑variant topological information for characterizing the structure of complex networks. Phys. Rev. E \textbf{100}, 032308 (2019). 

\bibitem{xiao2021} Xiao, X., Chen, H., Bogdan, P.: Deciphering the generating rules and functionalities of complex networks. Sci. Rep. \textbf{11}, 22964 (2021). 

\bibitem{staudt2017} Staudt, C.L., Hamann, M., Gutfraind, A., Safro, I., Meyerhenke, H.: Generating realistic scaled complex networks. Appl. Netw. Sci. \textbf{2}, 36 (2017).

\bibitem{arora2017} Arora, V., Ventresca, M.: Action‑based modeling of complex networks. Sci. Rep. \textbf{7}, 6673 (2017).
 
\bibitem{attar2020}  Attar, N., Aliakbary, S., Hosseini Nezhad, Z.: Automatic generation of adaptive network models based on similarity to the desired complex network. Physica A \textbf{545}, 123353 (2020).

\bibitem{fujita2020} Fujita, A., Lira, E. S., Santos, S. S., Bando, S. Y., Soares, G. E., Takahashi, D. Y.: A semi-parametric statistical test to compare complex networks. J. Complex Netw. \textbf{8}(2), cnz028 (2020).

\bibitem{wang2012} Wang, W.-Q., Zhang, Q.-M., Zhou, T.: Evaluating network models: A likelihood analysis. EPL \textbf{98}, 28004 (2012). 

\bibitem{reeves2020} Reeves, T., Damle, A., Benson, A.R.: Network interpolation. SIAM J. Math. Data Sci. \textbf{2}(2), 505–528 (2020).

\end{thebibliography}



\backmatter


\end{document}